\documentclass[a4paper,11pt]{article}
\usepackage{jcappub}
\usepackage[utf8]{inputenc}
\usepackage{amsmath,mathtools}
\usepackage{tensor}
\usepackage{amssymb}
\usepackage{graphicx,epsfig,amssymb}
\usepackage{physics}
\usepackage{lipsum}  
\usepackage{bigints}
\usepackage{verbatim}
\usepackage{hyperref}
\usepackage{color}
\usepackage{multirow}
\usepackage{float}
\usepackage{array}
\usepackage{orcidlink}
\usepackage{booktabs}
\newcolumntype{P}[1]{>{\centering\arraybackslash}p{#1}}

\usepackage{caption}
\usepackage{subcaption}

\begin{document}


\title{Intertwined Constraints in Extended Cosmologies: Dark Energy, Curvature, Neutrinos, and Inflation}

\author[a]{William Giar\`e \orcidlink{0000-0002-4012-9285},}
\author[b]{Dong Ha Lee \orcidlink{0009-0000-2992-3157},}
\author[b]{Eleonora Di Valentino \orcidlink{0000-0001-8408-6961},}

\affiliation[a]{Department of Physics and Astronomy, University of Hawai‘i, Honolulu, HI 96822, USA}

\affiliation[b]{School of Mathematical and Physical Sciences, University of Sheffield, Hounsfield Road, Sheffield S3 7RH, United Kingdom}

\emailAdd{giare@hawaii.edu}
\emailAdd{dhlee1@sheffield.ac.uk}
\emailAdd{e.divalentino@sheffield.ac.uk}

\abstract{We present a systematic reassessment of cosmological constraints beyond $\Lambda$CDM by progressively relaxing the assumptions underlying Dark Energy (DE), Curvature, Neutrinos, and Inflation. Using the latest CMB data together with DESI BAO and different SN catalogues, we show that the preference for dynamical DE persists across all the extended cosmologies considered. $\Omega_k$ remains compatible with flatness, despite a mild $2.2\sigma$ preference for $\Omega_k>0$ that is substantially degraded in dynamical DE extensions. Constraints on $N_{\rm eff}$ are broadly consistent with $N_{\rm eff}=3.04$, while cosmological upper limits on the total neutrino mass vary substantially across the cosmologies explored, ranging from $\sum m_\nu\lesssim 0.06$ eV to $\lesssim 0.2$ eV. We quantify both the preference for the mass ordering and the apparent tension between cosmology and oscillation experiments, showing that they are strongly framework dependent. We find no evidence for inflationary tensor modes, with $r\lesssim 0.035$. Constraints on the spectral index $n_s$ show significant model dependence. Allowing for the scalar runnings produces a mild shift toward $\alpha_s>0$ and $\beta_s>0$ that can reabsorb the preference for larger $n_s$ found in small-scale CMB data, although both $\alpha_s$ and $\beta_s$ remain consistent with zero at $\sim 1.5\sigma$. We highlight the implications for slow-roll inflation and benchmark models. None of the extensions considered here can resolve the $H_0$ tension. We discuss the implications for $\Omega_m$ and $S_8$. Overall, dynamical DE is the only significant deviation from $\Lambda$CDM and has the strongest impact on the inferred conclusions in the other sectors of the model.}

\makeatletter
\gdef\@fpheader{}
\makeatother
\maketitle
\clearpage


\section{Introduction}
\label{sec.introduction}

The standard $\Lambda$CDM model of cosmology provides the most successful description of the large-scale Universe, accounting for observations across widely separated cosmic epochs and physical scales, from primordial Big Bang Nucleosynthesis (BBN)~\cite{Peebles:1966zz,Wagoner:1966pv,Steigman:1977kc,Yang:1978ge,Walker:1991ap,Copi:1994ev,Schramm:1997vs,Steigman:2007xt,Iocco:2008va,Cyburt:2015mya,Cooke:2017cwo,Fields:2019pfx} and the Cosmic Microwave Background (CMB)~\cite{Fixsen:1996nj,Bennett:1996ce,Boomerang:2000jdg,WMAP:2012fli,WMAP:2012nax,Planck:2018nkj,Planck:2018vyg,ACT:2023dou,ACT:2023kun,AtacamaCosmologyTelescope:2025blo,SPT-3G:2024atg,SPT-3G:2025bzu} to late-time probes such as baryon acoustic oscillations (BAO)~\cite{SDSS:2005xqv,2dFGRS:2005yhx,Jones:2009yz,Beutler:2011hx,Beutler:2012px,WiggleZ:2012sek,BOSS:2013rlg,BOSS:2012gof,BOSS:2012dmf,BOSS:2014hhw,BOSS:2016wmc,eBOSS:2019dcv,eBOSS:2020yzd,DESI:2024uvr,DESI:2024mwx,DESI:2025zgx}, galaxy surveys~\cite{DES:2016jjg,KiDS:2020suj,DES:2021wwk,Wright:2025xka,DES:2026fyc}, and Type Ia supernovae (SNIa)~\cite{SupernovaSearchTeam:1998fmf,SupernovaCosmologyProject:1998vns,Pan-STARRS1:2017jku,Scolnic:2021amr,Brout:2022vxf,DES:2025sig,Rubin:2026qdt}.

At its core, the model rests upon two pillars of well-established physics: General Relativity (GR), which governs gravitational interactions, and the Standard Model (SM) of particle physics, which describes the known constituents of matter and radiation, together with all other non-gravitational forces. Remarkably, however, GR and the SM, taken alone, are insufficient to account for all phenomena inferred from data, compelling the inclusion of three additional components: dark matter (DM), dark energy (DE), and an early phase of inflationary expansion.

Within $\Lambda$CDM, these components enter as minimal phenomenological additions required to match cosmological observations~\cite{Peebles:1982ff,Blumenthal:1984bp,Davis:1985rj,Peebles:2002gy}. DM is modeled as a cold, non-baryonic, pressureless fluid interacting only gravitationally, while DE is described by a positive cosmological constant $\Lambda$, corresponding to a constant vacuum energy density and asymptotic de Sitter expansion. Inflation~\cite{Guth:1980zm,Starobinsky:1980te,Linde:1981mu,Albrecht:1982wi,Mukhanov:1981xt,Hawking:1982cz,Starobinsky:1982ee,Guth:1982ec,Bardeen:1983qw} is encoded through a nearly scale-invariant spectrum of adiabatic perturbations, characterized by a scalar amplitude and spectral tilt.

Increasingly precise and diverse cosmological measurements allow us to test these structural assumptions with progressively greater stringency, both directly and indirectly. If DM consists of new non-baryonic particles beyond the SM, it may leave signatures at microphysical scales, accessible to laboratory experiments, collider searches, or indirect non-cosmological probes~\cite{Goodman:1984dc,Drukier:1986tm,Bertone:2004pz,Clowe:2006eq,Feng:2010gw,Bertone:2016nfn,Bertone:2018krk,DiLuzio:2020wdo,Cirelli:2024ssz}. Despite sustained and increasingly sensitive efforts, however, no conclusive detection has yet been achieved~\cite{Arcadi:2017kky,Schumann:2019eaa,Cirelli:2024ssz}. Similarly, inflation provides a second direct target. The super-adiabatic amplification of quantum fluctuations during the accelerated phase is expected to generate a stochastic background of primordial gravitational waves (PGWs), whose amplitude is set by the inflationary energy scale~\cite{Starobinsky:1979ty,Rubakov:1982df,Fabbri:1983us,Abbott:1984fp}. If sufficiently large, this primordial background would imprint B-mode polarization in the CMB at large angular scales~\cite{Kamionkowski:1996ks,Seljak:1996gy,Zaldarriaga:1996xe}, providing direct evidence for inflation. At the time of writing, no definitive primordial signal has been reported~\cite{Planck:2018jri,BICEP:2021xfz}. Direct tests of DE are instead more challenging: the energy scale associated with cosmic acceleration is extraordinarily small~\cite{Weinberg:1988cp,Carroll:2000fy}, making controlled laboratory probes extremely difficult despite several proposed mechanisms and experimental strategies~\cite{Khoury:2003aq,Khoury:2003rn,Brax:2004qh,Burrage:2014oza,Vagnozzi:2021quy,Yuan:2025twx}.

The intrinsic challenges and continued lack of direct detections of the fundamental constituents of $\Lambda$CDM motivate a shift toward indirect tests of the framework. These operate at two distinct levels. The first consists in comparing cosmological parameters inferred within $\Lambda$CDM with independent determinations of the same quantities that do not rely on its assumptions. While this does not directly probe the microphysical nature of DM, DE, or inflation, it provides a stringent end-to-end test of the framework. The Hubble constant is the clearest example. Within $\Lambda$CDM, $H_0$ is inferred with remarkable precision from CMB observations~\cite{Planck:2018vyg,ACT:2023kun,SPT-3G:2025bzu}. At the same time, it can also be measured directly from local distance indicators~\cite{Freedman:2020dne,Birrer:2020tax,Riess:2021jrx,Anderson:2023aga,Scolnic:2023mrv,Jones:2022mvo,Anand:2021sum,Freedman:2021ahq,Uddin:2023iob,Huang:2023frr,Li:2024yoe,Pesce:2020xfe,Kourkchi:2020iyz,Schombert:2020pxm,Blakeslee:2021rqi,deJaeger:2022lit,Murakami:2023xuy,Breuval:2024lsv,Freedman:2024eph,Riess:2024vfa,Vogl:2024bum,Scolnic:2024hbh,Said:2024pwm,Boubel:2024cqw,Scolnic:2024oth,Li:2025ife,Jensen:2025aai,Riess:2025chq,Benisty:2025tct,Newman:2025gwg,Stiskalek:2025ktq,H0DN:2025lyy,Agrawal:2025tuv,Bhardwaj:2025kbw}. Recent combinations of the local distance ladder, including SH0ES~\cite{Riess:2021jrx} within the $H_0$ Distance Network (H0DN)~\cite{H0DN:2025lyy}, find a local expansion rate in tension with early-Universe inferences at more than seven standard deviations~\cite{H0DN:2025lyy}. The long-standing Hubble tension has therefore evolved from a mild anomaly into one of the central challenges for $\Lambda$CDM cosmology~\cite{Verde:2019ivm,DiValentino:2020zio,DiValentino:2021izs,Perivolaropoulos:2021jda,Schoneberg:2021qvd,Shah:2021onj,Abdalla:2022yfr,DiValentino:2022fjm,Kamionkowski:2022pkx,Giare:2023xoc,Hu:2023jqc,Verde:2023lmm,DiValentino:2024yew,Efstathiou:2024dvn,CosmoVerseNetwork:2025alb,Cai:2026swf}.\footnote{Unresolved systematic effects remain an actively investigated possibility; see, e.g., Refs.~\cite{Efstathiou:2020wxn,Mortsell:2021nzg,Mortsell:2021tcx,Riess:2021jrx,Sharon:2023ioz,Murakami:2023xuy,Riess:2023bfx,Bhardwaj:2023mau,Brout:2023wol,Dwomoh:2023bro,Uddin:2023iob,Riess:2024ohe,Freedman:2024eph,Riess:2024vfa}.}

A second class of indirect tests concerns the internal consistency of the model itself. As observations become increasingly precise, constraints on the baseline $\Lambda$CDM scenario tighten, while its defining assumptions can be tested more stringently. By promoting fixed parameters to free quantities in extended analyses, one may ask whether a single minimal parameter set can still provide a statistically consistent description of all datasets. If $\Lambda$CDM captures the underlying physics, these additional degrees of freedom should remain compatible with the baseline scenario within reasonable uncertainties. Conversely, persistent deviations may indicate that one or more structural assumptions require revision. DE provides a particularly important example. Although a simple cosmological-constant description has shown remarkable agreement with observations~\cite{Escamilla:2023oce}, recent analyses of high-precision BAO data from the Dark Energy Spectroscopic Instrument (DESI)~\cite{DESI:2025zgx}, combined with CMB and SNIa information, have reported growing $3.2$--$3.4\sigma$ preferences for a dynamical dark-energy equation of state (EoS)~\cite{DESI:2025fii}, introducing a new challenge for $\Lambda$CDM.\footnote{The post-DESI literature discussing the preference for dynamical DE, its robustness, and possible theoretical interpretations is extensive. Without claiming completeness, see, e.g., Refs.~\cite{Cortes:2024lgw,Shlivko:2024llw,Luongo:2024fww,Gialamas:2024lyw,Wang:2024dka,Ye:2024ywg,Tada:2024znt,Carloni:2024zpl,Chan-GyungPark:2024mlx,Efstathiou:2024xcq,Bhattacharya:2024hep,Andriot:2024jsh,Li:2024qus,Reboucas:2024smm,Giare:2024gpk,Ramadan:2024kmn,Yang:2024kdo,Jiang:2024xnu,RoyChoudhury:2024wri,Giare:2024oil,Giare:2025pzu,Efstathiou:2025tie,Kessler:2025kju,RoyChoudhury:2025dhe,Scherer:2025esj,Wolf:2025jlc,Santos:2025wiv,Specogna:2025guo,Cheng:2025lod,Cheng:2025hug,Ozulker:2025ehg,Li:2025vuh,Lee:2025pzo,Fazzari:2025lzd,Teixeira:2025czm,Smith:2025icl,Cai:2025mas,Herold:2025hkb,Cheng:2025yue,Chudaykin:2025aux,Chudaykin:2025lww,Chudaykin:2025vdh,Ivanov:2026dvl,Chudaykin:2026nls,Gokcen:2026pkq,Ishak:2025cay,Najafi:2026kxs,Yang:2026yaq,Dinda:2026ktu,Dainotti:2026enb,Kessler:2026dbi,Li:2026xaz,Giare:2026tyk,Li:2026asg,Ren:2026jyw,Andriot:2026lac}.}

Taken together, direct and indirect tests invite a careful reassessment of the present status of the $\Lambda$CDM framework. While the model continues to provide an overall statistically successful description of a remarkably broad range of observations, the increasing precision and diversity of cosmological data have also brought to light a number of external tensions and internal inconsistencies that have grown in significance. Thus, it would be misleading to conclude that it has emerged unambiguously strengthened. In our view, $\Lambda$CDM is not overturned; yet it appears less robust and less empirically secure than it was often perceived to be in earlier stages of precision cosmology.

It is important to stress that this uncertainty carries implications beyond the internal consistency of the model itself. If cosmology is to function as a high-precision laboratory for testing fundamental physics, confidence in its foundational assumptions is essential. When those assumptions come under strain, it becomes increasingly difficult to disentangle genuine signals of new microphysics on the largest observable scales from artifacts of an incomplete cosmological framework. A particularly illustrative example is provided by neutrinos. Oscillation experiments have established the existence of at least three neutrino flavors and at least two massive states, but they do not determine the absolute mass scale or the mass ordering~\cite{Super-Kamiokande:1998kpq,SNO:2002tuh,KamLAND:2002uet,deSalas:2020pgw,Esteban:2024eli,Capozzi:2025wyn}. Cosmology has long offered one of the most promising, and arguably the most sensitive, avenues for constraining the total neutrino mass and potentially clarifying the ordering~\cite{Lesgourgues:2006nd,Lesgourgues:2012uu,TopicalConvenersKNAbazajianJECarlstromATLee:2013bxd,Vagnozzi:2017ovm}. In recent years, however, cosmological analyses allowing the total neutrino mass to vary within minimal $\Lambda$CDM have yielded bounds that approach, and in some cases appear in tension with, the lower limits implied by oscillation data, for both the inverted and normal orderings~\cite{Jiang:2024viw,Capozzi:2025wyn}. Crucially, these bounds depend sensitively on the assumed cosmological framework~\cite{Loverde:2024nfi,Craig:2024tky,Naredo-Tuero:2024sgf,RoyChoudhury:2024wri,Green:2024xbb,Elbers:2024sha,Elbers:2025vlz,Cozzumbo:2025ewt,Sharma:2025iux,Elbers:2025vlz,Giare:2025ath,Lynch:2025ine,RoyChoudhury:2025dhe,Graham:2025dqn,Pulido-Hernandez:2026hcs,Jhaveri:2025neg,Feng:2026pzs,Yang:2026yaq,Kibris:2026cqq}. The apparent discrepancies can be significantly alleviated, and sometimes removed, by relaxing some of the underlying assumptions, for instance by considering dynamical DE models. Conversely, they may become even more pronounced in alternative scenarios motivated by the Hubble tension.

In this landscape, marked by sub-percent precision data on the one hand and tensions of growing significance on the other, extracting parameter constraints within a fixed baseline scenario is no longer sufficient. It becomes important to reassess the structural assumptions underlying the concordance model in a systematic way: quantifying the robustness of its constraints, examining how they shift under controlled extensions, and clarifying how tensions propagate across different sectors of the framework. This provides the basis for a more careful assessment of the present empirical status of $\Lambda$CDM and of its role as a framework for testing fundamental physics.

In this work, we pursue this reassessment in a systematic, sector-by-sector manner. We provide updated and consolidated constraints on the parameters defining the $\Lambda$CDM framework and its extensions, examine the network of parameter correlations, and quantify how inferences shift when baseline assumptions are relaxed. We then assess how these shifts affect conclusions about fundamental physics and clarify several open questions surrounding recently emerged internal and external inconsistencies. We organize our reassessment around five interconnected sectors of the $\Lambda$CDM framework.

\begin{itemize}
\item \textbf{The Dark Energy Sector:} We reassess the emerging preference for a dynamical DE component in light of DESI DR1 and DR2. Particular attention is given to whether the apparent quintessence-to-phantom evolution persists when other sectors, such as curvature and neutrino mass, are relaxed simultaneously, and to how the statistical evidence changes in enlarged parameter spaces.
\item \textbf{Spatial Curvature:} We revisit current constraints on $\Omega_{k}$, assessing whether the Universe can be regarded as spatially flat once high- and low-redshift probes are combined within different extensions of the cosmological model. Particular attention is given to the interplay between curvature and the DE sector, since both $\Omega_k$ and a dynamical DE component
affect the late-time distance-redshift relation probed by BAO and SN data.
\item \textbf{The Neutrino Sector:} We update constraints on $N_{\rm eff}$ and $\sum m_\nu$, and reassess their dependence on cosmological assumptions. We focus in particular on the consistency between cosmological upper bounds and the lower limits from oscillation experiments, and on how conclusions about the neutrino mass ordering change in extended cosmologies.
\item \textbf{Cosmological Inflation:} We reassess constraints on the primordial sector beyond the minimal power-law spectrum, allowing for running, running-of-the-running, and tensor modes. We quantify the stability of constraints on $n_s$, $r$, and slow-roll parameters when late-time and neutrino-sector assumptions are relaxed, and reinterpret the results in terms of inflationary model viability.
\item \textbf{Cosmic Tensions:} We examine how structural extensions propagate into the main cosmological tensions, including those involving $H_0$, $S_8$, and $\Omega_m$. We assess whether these tensions point toward coherent extensions of $\Lambda$CDM or instead reflect sector-dependent shifts that cannot be reconciled within a unified framework.
\end{itemize}

This work is structured as follows. In Sec.~\ref{sec.methods}, we outline the observational and statistical framework of our analysis. Sec.~\ref{sec.DE} is devoted to the DE sector, while Sec.~\ref{sec.curvature} revisits current constraints on the spatial geometry of the Universe. In Sec.~\ref{sec.nu}, we reassess the neutrino sector, and in Sec.~\ref{sec.inflation}, we examine implications for inflationary physics. After scrutinizing each sector individually and in combination, Sec.~\ref{sec.tensions} provides an updated assessment of internal and external cosmological tensions, while Sec.~\ref{sec.model_comparison} summarizes the relative goodness of fit and information-criterion performance of the different cosmological models. Our conclusions are presented in Sec.~\ref{sec.conclusions}. The manuscript is complemented by two appendices: Appendix~\ref{appendix.Tables} reports the full numerical constraints for all parameters and dataset combinations, together with the corresponding $\chi^2$ values, while Appendix~\ref{appendix.Plots} provides additional posterior distributions and parameter contours beyond those discussed in the main text.
\section{Observational and Statistical Framework}
\label{sec.methods}

In this section we describe the observational and statistical framework adopted in our analysis. We begin by summarizing the cosmological parameter spaces considered and by outlining the numerical pipeline used for theoretical predictions and parameter inference. We then specify the observational datasets and likelihoods employed in the analysis.

\subsection{Numerical and Statistical Framework}
The cosmological model space explored in this work is organized around two broad late-time frameworks. As a baseline, we adopt the standard spatially flat $\Lambda$CDM model, described by the six free parameters $\{\omega_b\,,\,\omega_c\,,\,\theta_s\,,\,\tau\,,\,A_s\,,\,n_s\}$, corresponding to the physical baryon density, the physical cold-DM density, the angular size of the sound horizon at recombination, the optical depth to reionization, the amplitude of the primordial scalar power spectrum, and its spectral tilt. In the first framework, DE is described by a cosmological constant, and the baseline model is progressively enlarged by opening different sectors of the cosmological parameter space, namely the neutrino sector, through the total neutrino mass $\sum m_\nu$\footnote{In all runs involving free neutrino mass, we adopt the standard implementation with \texttt{num$\_$massive$\_$neutrinos: 3} and \texttt{neutrino$\_$hierarchy: degenerate}, corresponding to three massive neutrinos with equal masses. Since the present analysis is primarily sensitive to the total neutrino mass $\sum m_\nu$, this approximation is sufficient for the purposes of this work.} and the effective number of relativistic species $N_{\rm eff}$, spatial curvature, through $\Omega_k$, and the inflationary sector, through the tensor-to-scalar ratio $r$, the running of the scalar spectral index $\alpha_s$, and the running of the running of the scalar spectral index $\beta_s$. In the second framework, the same sequence of extensions is repeated after replacing the cosmological constant with a dynamical DE component described by the CPL parametrization, $w(a)=w_0+w_a(1-a)$, where $w_0$ and $w_a$ denote the present-day value and first-order time variation of the DE equation of state~\cite{Chevallier:2000qy,Linder:2002et}. These parameters are allowed to vary in different combinations throughout the work, including the fully extended case in which all these sectors are opened simultaneously.

\begin{table}[t!]
\centering
\renewcommand{\arraystretch}{1.4}
\begin{tabular}{l@{\hspace{1.5 cm}}l@{\hspace{1.5 cm}}l}
\hline\hline
\textbf{Parameter} & \textbf{Prior} & \textbf{Description} \\
\hline
$\Omega_b h^2 \equiv \omega_b$ & $[0.005\,,\,0.1]$ & Physical baryon density \\
$\Omega_c h^2 \equiv \omega_c$ & $[0.001\,,\,0.99]$ & Physical cold-DM density \\
$100\theta_{s}$ & $[0.5\,,\,10]$ & Angular size of the sound horizon at recombination \\
$\tau$ & $[0.01\,,\,0.8]$ & Optical depth to reionization \\
$\ln(10^{10}A_s)$ & $[1.61\,,\,3.91]$ & Amplitude of the primordial scalar power spectrum \\
$n_s$ & $[0.8,\,1.2]$ & Scalar spectral index \\
$\sum m_\nu$ [eV] & $[0\,,\,5]$ & Total neutrino mass \\
$N_{\rm eff}$ & $[0.05\,,\,10]$ & Effective number of relativistic species \\
$\Omega_k$ & $[-0.3\,,\,0.3]$ & Spatial curvature \\
$r$ & $[0\,,\,0.5]$ & Tensor-to-scalar ratio \\
$\alpha_s$ & $[-1\,,\,1]$ & Running of $n_s$ \\
$\beta_s$ & $[-1\,,\,1]$ & Running of the running \\
$w_0$ & $[-3\,,\,1]$ & Present-day DE equation of state \\
$w_a$ & $[-3\,,\,2]$ & First-order DE time variation \\
\hline\hline
\end{tabular}
\caption{Flat priors adopted for the cosmological parameters varied in this work.}
\label{tab:priors}
\end{table}

Theoretical predictions for the different cosmological models considered in this work are computed with the Boltzmann solver \texttt{CAMB}~\cite{Lewis:1999bs,Lewis:2002ah,Howlett:2012mh}, while Bayesian parameter inference is performed through Markov Chain Monte Carlo (MCMC) sampling using \texttt{Cobaya}~\cite{Torrado:2020dgo,Lewis:2013hha}. For each model, the posterior distribution is sampled until convergence is achieved according to the Gelman-Rubin criterion~\cite{Gelman:1992zz}, requiring $R-1 \lesssim 0.02$. The priors adopted for the parameters varied in the analysis are summarized in Table~\ref{tab:priors}.

The resulting MCMC chains are then post-processed to derive marginalized constraints and posterior distributions for the cosmological and derived parameters. Standard numerical summaries and posterior visualizations are obtained using \texttt{GetDist}~\cite{Lewis:2019xzd} together with additional custom analysis and visualization tools.\footnote{Some of the custom tools used in this work are publicly available at \url{https://github.com/williamgiare/wgcosmo}.} Specific statistical diagnostics and numerical pipelines adopted in the different sectors of the analysis are introduced on a case-by-case basis in the corresponding sections.

\subsection{Datasets}

Our baseline datasets and likelihoods are the following.
\begin{itemize}
    \item \textbf{Planck low-T}: Planck-2018 low-$\ell$ temperature \texttt{Commander} likelihood~\cite{Planck:2019nip}, accounting for measurements of the TT spectrum at $\ell<30$. 
    \item \textbf{Planck low-E}: Planck-2018 low-$\ell$ E-mode polarization measurements based on the \texttt{SRoll2} likelihood~\cite{Delouis:2019bub}, accounting for measurements of the EE spectrum at $\ell<30$. 
    \item \textbf{Planck (TT-TE-EE)}: Planck-2018 intermediate-multipole temperature and polarization measurements of the TT, TE, and EE spectra~\cite{Planck:2019nip}, with cuts at $\ell_{\max}=(1000,600,600)$, respectively.
    \item \textbf{ACT (TT-TE-EE)}: ACT DR6 high-multipole temperature and polarization measurements of the TT, TE, and EE spectra~\cite{AtacamaCosmologyTelescope:2025blo} over the multipole range $\ell \geq 600$. 
    \item \textbf{SPT (TT-TE-EE)}: SPT-3G D1 high-multipole temperature and polarization measurements of the TT, TE, and EE spectra~\cite{SPT-3G:2025bzu}, considered over the multipole range $\ell \geq 400$, with TT cut at $\ell_{\max}=3000$ and TE/EE cut at $\ell_{\max}=4000$.
    \item \textbf{Planck-ACT-SPT Lensing}: Combined CMB lensing information obtained from the ACT DR6 lensing likelihood~\cite{ACT:2023dou,ACT:2023kun}, together with Planck lensing~\cite{Carron:2022eyg} and the SPT-3G \texttt{MUSE} lensing likelihood~\cite{SPT-3G:2024atg}.
    \item \textbf{BICEP-Keck}: Large-scale B-mode polarization information, corresponding to the BK18 data release, based on measurements of the BB spectrum~\cite{BICEP:2021xfz} over the multipole range $20\lesssim \ell \lesssim 330$.
    \item \textbf{BAO}: DESI DR2 BAO distance measurements from galaxies, quasars, and Ly$\alpha$ forest data~\cite{DESI:2025qqy,DESI:2025zpo} in 14 redshift bins, with the full covariance matrix between the measurements included in the likelihood~\cite{DESI:2025zgx}.
    \item \textbf{Pantheon-Plus}: Pantheon-Plus (PP) compilation of Type Ia supernovae, consisting of 1701 light curves from 1550 distinct objects and spanning redshifts up to $z\simeq 2.26$~\cite{Brout:2022vxf,Scolnic:2021amr}.  
    \item \textbf{DES-Dovekie}: DES-Dovekie (DD) supernova likelihood~\cite{DES:2025sig}, corresponding to the recalibrated DES 5-year Type Ia supernova sample~\cite{DES:2024jxu,DES:2024upw,DES:2024hip} based on the Dovekie cross-calibration program, built from approximately 1600 likely DES supernovae and about 200 low-redshift supernovae from external surveys~\cite{DES:2025sig}.
    
\end{itemize}

These likelihoods are combined into two baseline dataset combinations used throughout this work. The first is denoted by \textbf{CMB+DESI+DD}. It includes the full CMB likelihood, namely Planck, ACT, SPT, the combined Planck-ACT-SPT lensing component, and BK18, together with DESI DR2 BAO and the DES-Dovekie supernova sample. The second is denoted by \textbf{CMB+DESI+PP}. It is defined identically, with the only difference being the replacement of DES-Dovekie with the Pantheon-Plus compilation.

\clearpage

\section{The Dark Energy Sector}
\label{sec.DE}
As originally reported in the first data release (DR1)~\cite{DESI:2024mwx} and subsequently reinforced in the second (DR2)~\cite{DESI:2025zgx}, DESI’s precise measurements of the transverse comoving distance $D_M(z)$, the Hubble rate $H(z)$, and their combinations, all relative to the sound horizon ($r_d$) at the baryon-drag epoch, provide statistically non-negligible evidence for an evolving DE component~\cite{DESI:2025fii}. 

In the widely used CPL parametrization, joint analyses of DESI BAO, CMB data, and different SN samples identify a region of parameter space that departs from the cosmological-constant limit $(w_0,w_a)=(-1,0)$, pointing to a present-day quintessence-like EoS, $w_0>-1$, with evolution toward phantom values in the past, $w_a<0$~\cite{DESI:2025fii}.

Following the DESI releases, the preference for dynamical DE has been extensively tested against dataset choices~\cite{Giare:2025pzu,Ishak:2025cay,DES:2026jmi}, BAO-bin dependence~\cite{Liu:2024gfy,DESI:2024aqx,Dinda:2024kjf}, SN calibration systematics~\cite{Efstathiou:2024xcq,Notari:2024zmi,Huang:2025som,Gialamas:2024lyw}, CMB likelihood choices~\cite{Giare:2024oil,Li:2025vuh}, and alternative EoS parametrizations~\cite{Giare:2024gpk,Wolf:2025jlc}. These studies do not identify a single dataset, redshift bin, or likelihood component as solely responsible for the signal, although its statistical significance can vary across data combinations and analysis choices. 

Here we quantify to what extent the preference for dynamical DE depends on the assumptions adopted in other cosmological sectors and whether it can be absorbed or diluted when curvature, neutrino-sector parameters, and inflationary parameters are allowed to vary. We also test whether any such reduction reflects genuine physical degeneracies or simply a loss of constraining power in enlarged parameter spaces.

\begin{table*}[tpb!]
\centering
\renewcommand{\arraystretch}{1}
\resizebox{\textwidth}{!}{
\begin{tabular}{l @{\hspace{0.5 cm}} l @{\hspace{0.5 cm}}c@{\hspace{0.5 cm}}c@{\hspace{0.5 cm}}c@{\hspace{0.5 cm}}cccccc}
\hline\hline
\\
\textbf{Model} & \textbf{Dataset} & \boldmath{$w_0$} & \boldmath{$w_a$} & \boldmath{$\rho_{w_0w_a}$} &  \boldmath{$z_p$} & \boldmath{$w(z_p)$} & \boldmath{$z_{\rm cross}$} & \textbf{\#}\boldmath{$\sigma$} & \textbf{FoM}\\
\\
\hline
&&&&&\\
$w_0 w_a\mathrm{CDM}$
& CMB+DESI+PP & $-0.839\pm 0.055$ & $-0.60\pm 0.20$ & $-0.894$ & $0.324$ & $-0.986^{+0.025}_{-0.024}$ & $0.367^{+0.090}_{-0.072}$ & $2.741$ & $208.463$ \\[2ex]
& CMB+DESI+DD & $-0.808\pm 0.054$ & $-0.70\pm 0.21$ & $-0.910$ & $0.318$ & $-0.978^{+0.022}_{-0.022}$ & $0.375^{+0.074}_{-0.058}$ & $3.178$ & $216.828$ \\[4ex]

$w_0 w_a\mathrm{CDM} + \sum m_\nu$  
& CMB+DESI+PP & $-0.845\pm 0.055$ & $-0.56^{+0.22}_{-0.19}$ & $-0.884$ & $0.300$ & $-0.975^{+0.025}_{-0.025}$ & $0.381^{+0.119}_{-0.080}$ & $2.086$ & $188.564$ \\[2ex]
& CMB+DESI+DD & $-0.812\pm 0.056$ & $-0.68^{+0.24}_{-0.20}$ & $-0.905$ & $0.276$ & $-0.959^{+0.023}_{-0.024}$ & $0.383^{+0.096}_{-0.064}$ & $2.866$ & $187.266$ \\[4ex]

$w_0 w_a\mathrm{CDM} + N_\mathrm{eff}$
& CMB+DESI+PP & $-0.831\pm 0.054$ & $-0.68\pm 0.20$ & $-0.882$ & $0.312$ & $-0.992^{+0.025}_{-0.025}$ & $0.333^{+0.078}_{-0.065}$ & $2.783$ & $191.002$ \\[2ex]
& CMB+DESI+DD & $-0.796\pm 0.057$ & $-0.80^{+0.24}_{-0.21}$ & $-0.908$ & $0.306$ & $-0.983^{+0.024}_{-0.024}$ & $0.343^{+0.064}_{-0.051}$ & $3.507$ & $185.387$ \\[4ex]

$w_0 w_a\mathrm{CDM} + \sum m_\nu + N_\mathrm{eff}$
& CMB+DESI+PP & $-0.836\pm 0.056$ & $-0.63^{+0.23}_{-0.20}$ & $-0.883$ & $0.288$ & $-0.978^{+0.026}_{-0.026}$ & $0.348^{+0.096}_{-0.071}$ & $2.065$ & $172.602$ \\[2ex]
& CMB+DESI+DD & $-0.800\pm 0.057$ & $-0.77^{+0.25}_{-0.22}$ & $-0.904$ & $0.282$ & $-0.969^{+0.024}_{-0.025}$ & $0.353^{+0.079}_{-0.056}$ & $2.839$ & $171.307$ \\[4ex]

$w_0 w_a\mathrm{CDM} + \Omega_k$  
& CMB+DESI+PP & $-0.856\pm 0.057$ & $-0.51\pm 0.22$ & $-0.902$ & $0.318$ & $-0.980^{+0.024}_{-0.024}$ & $0.389^{+0.141}_{-0.085}$ & $1.743$ & $190.256$ \\[2ex]
& CMB+DESI+DD & $-0.822\pm 0.059$ & $-0.63\pm 0.23$ & $-0.922$ & $0.306$ & $-0.971^{+0.022}_{-0.023}$ & $0.390^{+0.101}_{-0.067}$ & $2.524$ & $190.678$ \\[4ex]

$w_0 w_a\mathrm{CDM} + r$ 
& CMB+DESI+PP & $-0.840\pm 0.054$ & $-0.59^{+0.20}_{-0.18}$ & $-0.890$ & $0.342$ & $-0.992^{+0.025}_{-0.024}$ & $0.366^{+0.093}_{-0.074}$ & $2.646$ & $209.747$ \\[2ex]
& CMB+DESI+DD & $-0.811\pm 0.055$ & $-0.69\pm 0.21$ & $-0.915$ & $0.330$ & $-0.983^{+0.022}_{-0.022}$ & $0.376^{+0.075}_{-0.057}$ & $3.197$ & $214.080$ \\[4ex]

$w_0 w_a\mathrm{CDM} + r + \alpha_s$
& CMB+DESI+PP & $-0.841\pm 0.054$ & $-0.59\pm 0.19$ & $-0.894$ & $0.324$ & $-0.986^{+0.024}_{-0.024}$ & $0.366^{+0.090}_{-0.072}$ & $2.656$ & $213.431$ \\[2ex]
& CMB+DESI+DD & $-0.811\pm 0.055$ & $-0.69\pm 0.21$ & $-0.914$ & $0.318$ & $-0.978^{+0.023}_{-0.022}$ & $0.375^{+0.077}_{-0.058}$ & $3.222$ & $213.099$ \\[4ex]

$w_0 w_a\mathrm{CDM} + r + \alpha_s + \beta_s$ 
& CMB+DESI+PP & $-0.840\pm 0.054$ & $-0.60^{+0.20}_{-0.18}$ & $-0.895$ & $0.330$ & $-0.989^{+0.024}_{-0.024}$ & $0.363^{+0.089}_{-0.074}$ & $2.780$ & $211.143$ \\[2ex]
& CMB+DESI+DD & $-0.810\pm 0.054$ & $-0.70^{+0.21}_{-0.19}$ & $-0.911$ & $0.312$ & $-0.977^{+0.022}_{-0.022}$ & $0.370^{+0.072}_{-0.057}$ & $3.133$ & $219.756$ \\[4ex]

$w_0 w_a\mathrm{CDM} + r + \alpha_s + \beta_s + \Omega_k + \sum m_\nu + N_\mathrm{eff}$
& CMB+DESI+PP & $-0.855\pm 0.059$ & $-0.55^{+0.26}_{-0.23}$ & $-0.882$ & $0.276$ & $-0.975^{+0.027}_{-0.027}$ & $0.356^{+0.152}_{-0.085}$ & $1.937$ & $147.738$ \\[2ex]
& CMB+DESI+DD & $-0.821\pm 0.062$ & $-0.69^{+0.29}_{-0.25}$ & $-0.908$ & $0.264$ & $-0.965^{+0.026}_{-0.025}$ & $0.350^{+0.106}_{-0.064}$ & $2.343$ & $143.168$ \\[4ex]
\hline
\bottomrule
\end{tabular}
}
\caption{Constraints on the CPL dark energy parameters obtained for the different cosmological models considered in this work. For each model we report the 68\% confidence level constraints on $w_0$ and $w_a$, their correlation coefficient $\rho_{w_0 w_a}$, the pivot redshift $z_p$, the corresponding pivot equation of state $w(z_p)$, and the redshift $z_{\rm cross}$ at which $w(z)$ crosses $w=-1$. The last two columns report the statistical preference for dynamical dark energy (in units of $\sigma$) and the Figure of Merit (FoM) for the $(w_0,w_a)$ constraints. Results are shown for two different SN samples combined with CMB and DESI BAO data. The various rows correspond to progressively extended cosmological parameter spaces in which additional sectors of the baseline $\Lambda$CDM model are allowed to vary.}
\label{tab:results.DE}
\end{table*}

\begin{figure}[htp!]
    \centering
    \includegraphics[width=0.8\linewidth]{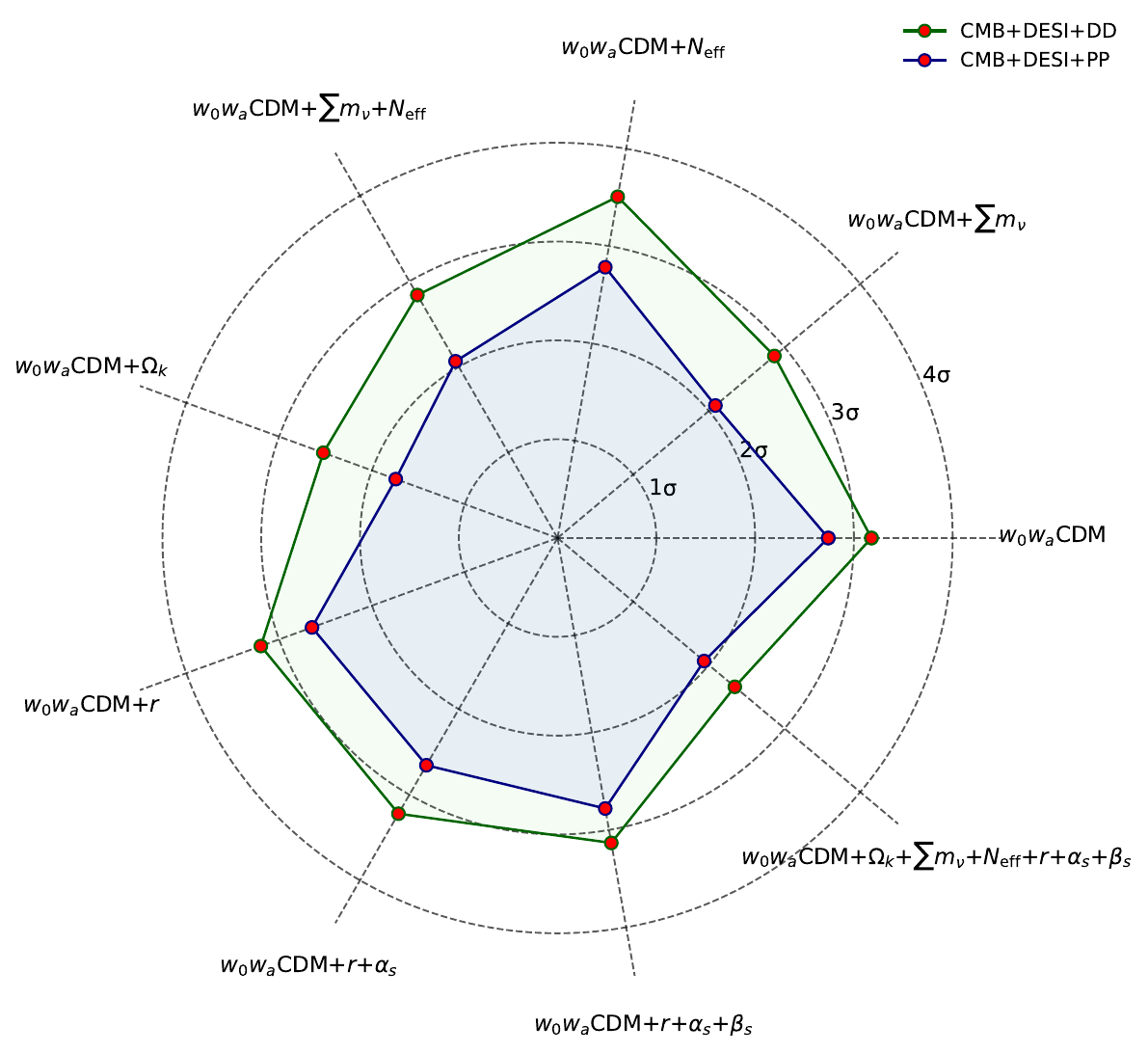}
    \caption{Statistical preference for dynamical DE, quantified in units of Gaussian-equivalent $\sigma$, obtained for different extensions of the baseline cosmological model. Each radial axis corresponds to a different parameter-space configuration in which additional cosmological sectors are progressively relaxed beyond $w_0w_a$CDM. The green and blue curves show the results obtained using two different SN samples combined with CMB and DESI BAO data. Moving anticlockwise, the extensions include neutrino-sector parameters ($\sum m_\nu$, $N_{\rm eff}$), spatial curvature ($\Omega_k$), and inflationary parameters ($r$, $\alpha_s$, $\beta_s$).}
    \label{fig:DDE_sigmas}
\end{figure}

\subsection{Preference for Dynamical Dark Energy in Extended Cosmologies}
To quantify the preference for dynamical DE in extended parameter spaces, we construct two parallel families of cosmological models. In the first, the DE sector is fixed to a cosmological constant, while the baseline $\Lambda$CDM framework is progressively enlarged by allowing additional parameters to vary. In particular, we consider extensions involving the total neutrino mass $\sum m_\nu$, the effective number of relativistic species $N_{\rm eff}$, the spatial curvature parameter $\Omega_k$, the tensor-to-scalar ratio $r$, the running of the scalar spectral index $\alpha_s$, and the running of the running of the scalar spectral index $\beta_s$. These extensions are studied individually, in selected combinations, and in a fully extended setup. In the second family, we consider the same set of extensions but promote DE to a dynamical component described by the CPL parametrization.

This construction yields two matched sets of models: one with DE fixed to $\Lambda$, and one with DE described by $(w_0,w_a)$, while all other extensions are kept identical. The preference for dynamical DE can then be quantified by comparing each CPL model with its corresponding $\Lambda$ counterpart.\footnote{In practice this amounts to comparing pairs of models that differ only by the two additional CPL parameters, for example $w_0w_a{\rm CDM}+\sum m_\nu$ versus $\Lambda{\rm CDM}+\sum m_\nu$, or $w_0w_a{\rm CDM}+\sum m_\nu+N_{\rm eff}$ versus $\Lambda{\rm CDM}+\sum m_\nu+N_{\rm eff}$, and so on for the full set of extensions considered.} For each pair we compute
\begin{equation}
\Delta\chi^2=\chi^2_{\min}(\Lambda{\rm CDM}+\theta)-\chi^2_{\min}(w_0w_a{\rm CDM}+\theta),
\end{equation}
where $\theta$ denotes the additional set of parameters defining the extended cosmology. Since the two models are nested (i.e., $\Lambda$CDM+$\theta$ is recovered from $w_0w_a$CDM+$\theta$ for $w_0=-1$ and $w_a=0$), the distribution of $\Delta\chi^2$ can be approximated using Wilks' theorem, which states that in the asymptotic large-sample limit the likelihood-ratio statistic follows a $\chi^2$ distribution with degrees of freedom equal to the difference in the number of free parameters between the models. In our case this corresponds to $\Delta k = 2$, associated with the additional CPL parameters $(w_0,w_a)$. The probability of obtaining a value at least as extreme as the observed one is therefore given by the $p$-value
$p = 1 - F_{\chi^2}(|\Delta\chi^2|,\Delta k)$, where $F_{\chi^2}(x,k)$ denotes the cumulative distribution function of the $\chi^2$ distribution with $k$ degrees of freedom:
\begin{equation}
F_{\chi^2}(x,k) =
\frac{1}{2^{k/2}\Gamma(k/2)}
\int_{0}^{x} t^{k/2-1} e^{-t/2}\,dt.
\end{equation}
Finally, for ease of interpretation we convert the $p$-value into an equivalent Gaussian significance expressed in units of standard deviations by computing $\#\sigma=\Phi^{-1}(1-p/2)$, where
\begin{equation}
\Phi(x)=\frac{1}{\sqrt{2\pi}}\int_{-\infty}^{x} e^{-t^2/2}\,dt .
\label{eq:Phi}
\end{equation}
denotes the cumulative distribution function of the standard normal distribution.

The statistical significance of the preference for dynamical DE, together with the main results of this analysis, is summarized in Table~\ref{tab:results.DE}, with the corresponding Gaussian-equivalent significances reported in the column labeled \#$\sigma$. A complementary visual summary is provided in Fig.~\ref{fig:DDE_sigmas}, where the same information is displayed in the form of a radar plot.

Across all extensions involving the inflationary parameters $r$, $\alpha_s$, and $\beta_s$, the preference for dynamical DE remains essentially unchanged with respect to the minimal $w_0w_a$CDM scenario. The baseline CPL model gives a significance of $\sim2.8\sigma$ for CMB+DESI+PP and $\sim3.2\sigma$ for CMB+DESI+DD, and very similar values are obtained when tensor modes, running, and the running of the running are allowed to vary. This stability is physically expected. The parameters $\alpha_s$ and $\beta_s$ modify the scale dependence of the primordial scalar spectrum and are primarily constrained by the detailed shape of the CMB temperature and polarization spectra, especially at intermediate and high multipoles. The tensor-to-scalar ratio $r$ mainly affects large-scale CMB anisotropies and polarization, including the B-mode signal. By contrast, dynamical DE affects the CMB only weakly through the late-time ISW effect, while the preference for DDE is mainly driven by late-time geometric probes, especially BAO and SN distances anchored by the CMB-calibrated sound horizon. As a result, freedom in the inflationary sector does not significantly mimic or absorb the preference for late-time dynamical DE.

A second set of extensions concerns the neutrino sector. When the effective number of relativistic species $N_{\rm eff}$ is allowed to vary, the preference for dynamical DE remains essentially stable and shows a slight increase in both dataset combinations, reaching $2.78\sigma$ for CMB+DESI+PP and $3.51\sigma$ for CMB+DESI+DD. $N_{\rm eff}$ changes the amount of radiation before recombination and therefore controls the expansion rate of the Universe in the pre-recombination era. Higher (lower) values of $N_{\rm eff}$ lead to a faster (slower) expansion and consequently to a smaller (larger) sound horizon. In the CMB anisotropy spectra, this also affects the diffusion scale and the epoch of matter-radiation equality, which together modify the detailed structure of the acoustic peaks. For this reason, $N_{\rm eff}$ is constrained mainly by the high-$\ell$ temperature and polarization spectra, where the CMB is most sensitive to pre-recombination physics. The fits show a mild shift toward values of $N_{\rm eff}$ below the value assumed in $\Lambda$CDM, where this parameter is fixed to $N_{\rm eff}=3.04$. Such a shift corresponds to a slightly larger sound horizon. To preserve the angular scale of the acoustic peaks, this increase must be compensated by a corresponding increase in the angular-diameter distance to the last-scattering surface. At fixed cosmological parameters, a CPL evolution in which the EoS becomes phantom-like at $z \gtrsim 0.3$ implies that the DE density decreases more rapidly toward higher redshift than in the $\Lambda$CDM case, reducing the total energy density at intermediate redshifts and lowering the expansion rate $H(z)$. Since $D_A(z_{\star})$ depends on the integral of $1/H(z)$ along the line of sight, a small reduction of $H(z)$ leads to a modest increase in the integrated distance to last scattering, helping to compensate for the larger sound horizon associated with smaller values of $N_{\rm eff}$.

As for the total neutrino mass $\sum m_\nu$, once massive neutrinos become non-relativistic, their large thermal velocities cause them to free-stream over cosmological scales, suppressing the growth of matter perturbations below the corresponding free-streaming scale. At the same time, they contribute to the total matter density $\Omega_m$ in proportion to their total mass. Allowing $\sum m_\nu$ to vary therefore introduces an additional degeneracy direction with the dynamical-DE parameters $w_0$ and $w_a$, reducing the nominal statistical significance of the preference for dynamical DE to $2.09\sigma$ for CMB+DESI+PP and $2.87\sigma$ for CMB+DESI+DD.

When both neutrino-sector parameters are allowed to vary simultaneously, the preference remains intermediate between these two cases, with significances of $2.07\sigma$ and $2.84\sigma$ for PP and DD, respectively. This again supports the interpretation that the dilution is driven mainly by $\sum m_\nu$, whereas $N_{\rm eff}$ may even slightly reinforce the dynamical-DE preference.

Allowing spatial curvature to vary produces the largest reduction among the single-sector extensions, lowering the preference for dynamical DE to $1.74\sigma$ for CMB+DESI+PP and $2.52\sigma$ for CMB+DESI+DD. Curvature alters cosmological distances both through the expansion history and through the geometrical projection relating radial and transverse directions, acting directly on the same observables from which the preference for dynamical DE originates. For a given line-of-sight comoving distance $\chi=\int_0^z dz'/H(z')$, non-zero curvature changes the transverse distance through $D_M(z)=S_k(\chi)$, where $S_k(\chi)$ encodes the curvature-dependent mapping between radial and transverse separations. In addition, it modifies the expansion rate through the term $\Omega_k(1+z)^2$. These effects absorb part of the geometrical shift that would otherwise be attributed to dynamical DE, producing the largest reduction in significance among the single-parameter extensions considered here. The overlap is only partial, however, because the redshift scaling of the two effects is different. The curvature contribution scales as $(1+z)^2$ and therefore affects the geometry smoothly over a wider redshift range, while DE modifies the expansion history mainly at low and intermediate redshifts, when it becomes dynamically relevant.

In the fully extended cosmology, the preference for dynamical DE decreases to $1.94\sigma$ for CMB+DESI+PP and $2.34\sigma$ for CMB+DESI+DD when $\Omega_k$, $\sum m_\nu$, $N_{\rm eff}$, $r$, $\alpha_s$, and $\beta_s$ are allowed to vary together with the CPL parameters $(w_0,w_a)$. This reduction may suggest that the preference is largely absorbed in a sufficiently enlarged parameter space, but it also raises the question of whether the effect reflects a genuine weakening of the signal or simply a loss of constraining power.

\begin{figure}[htp!]
    \centering
    \includegraphics[width=0.7\linewidth]{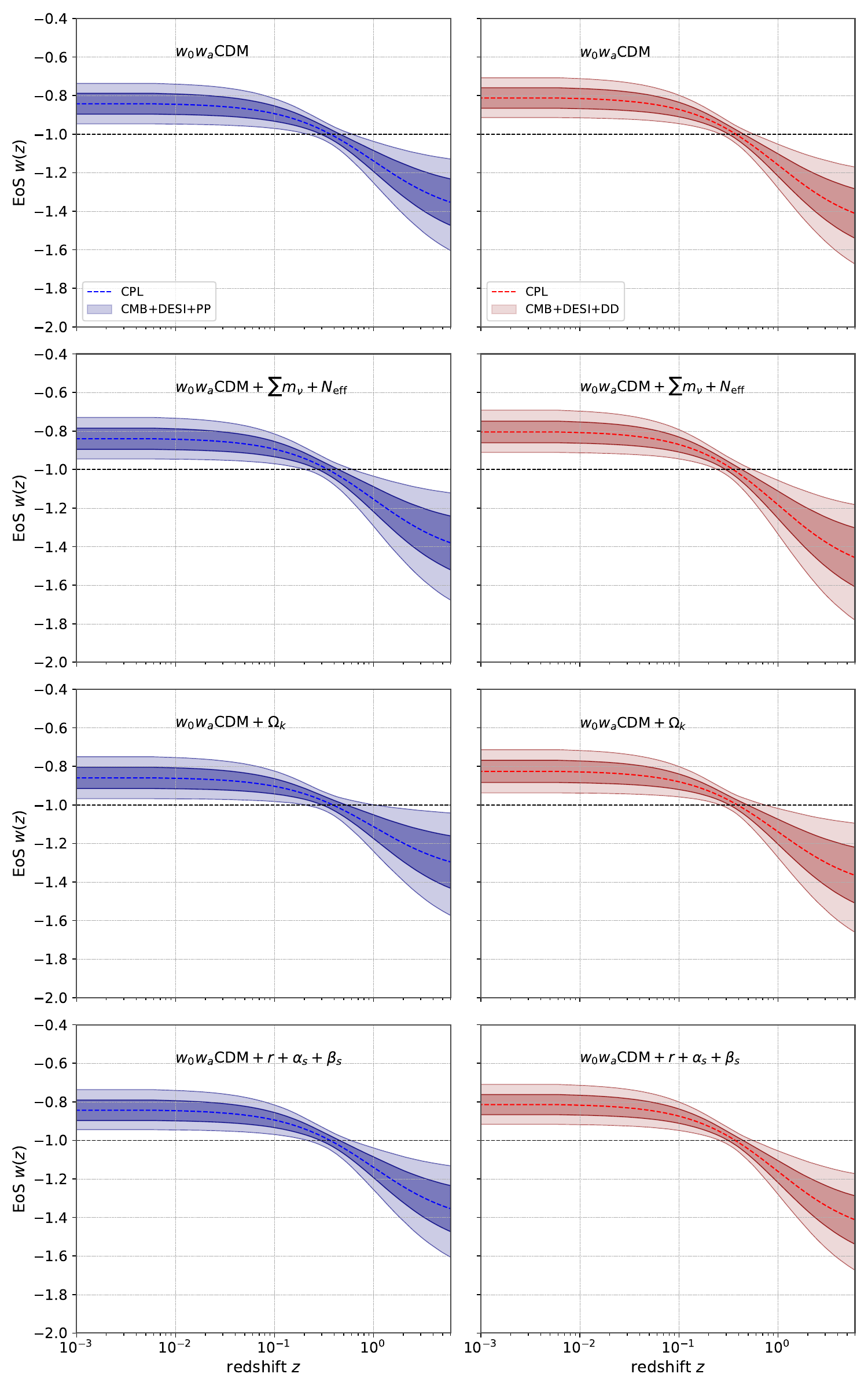}
    \caption{Reconstruction of the DE EoS $w(z)$ obtained using the CPL parametrization for different extensions of the baseline cosmological model. The left (blue) and right (red) columns show the results obtained using two different SN samples combined with CMB and BAO data. From top to bottom, we progressively relax additional sectors of the cosmological model beyond $\Lambda$CDM: the baseline CPL extension, the inclusion of neutrino-sector parameters ($\sum m_\nu$, $N_{\rm eff}$), spatial curvature ($\Omega_k$), and inflationary parameters ($r$, $\alpha_s$, $\beta_s$). The shaded regions represent the 68\% and 95\% confidence intervals. The horizontal dashed line indicates the $\Lambda$CDM value $w=-1$.}
    \label{fig:EoS}
\end{figure}

\begin{figure}[htp!]
    \centering
    \includegraphics[width=0.9\linewidth]{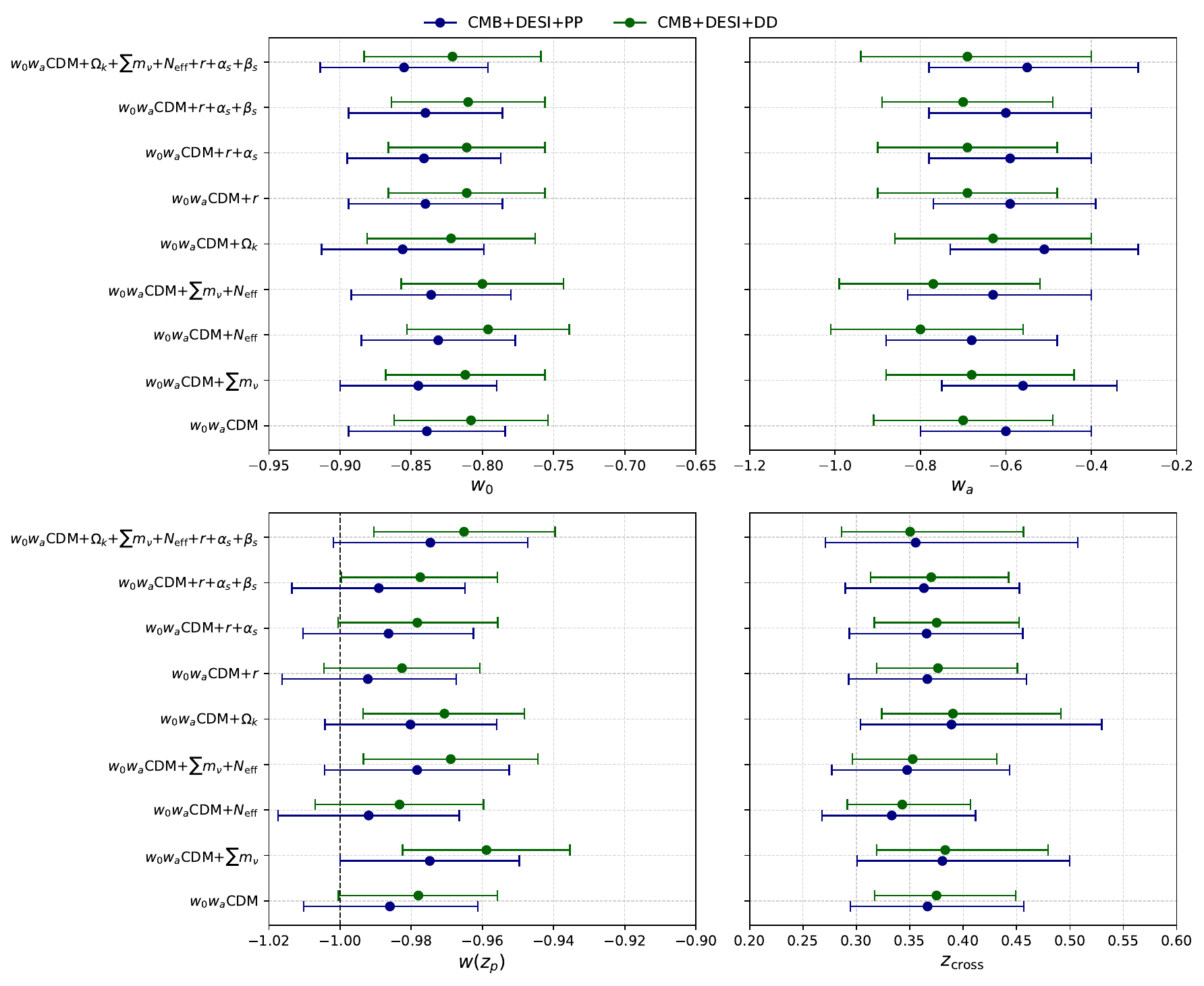}
    \caption{Summary of the constraints obtained for the CPL parameters and derived quantities across the different cosmological models considered in this work. The panels show the marginalized constraints on $w_0$ (top left), $w_a$ (top right), the pivot equation of state $w(z_p)$ (bottom left), and the crossing redshift $z_{\rm cross}$ at which $w(z)$ crosses $w=-1$ (bottom right). Results are shown for the two SN samples combined with CMB and DESI BAO data, corresponding to the same model extensions reported in Table~\ref{tab:results.DE}. Each row corresponds to a different cosmological parameter-space configuration in which additional sectors of the baseline model are progressively relaxed.}
    \label{fig:summary_plot_DDE}
\end{figure}

\subsection{Constraining Power in Extended Cosmologies}

To determine whether the reduced preference for dynamical DE reflects a genuine weakening of the signal or simply a loss of constraining power in enlarged parameter spaces, we compute, for each extension discussed above, the Figure of Merit (FoM), defined as
\begin{equation}
{\rm FoM} \equiv \frac{1}{\sqrt{\det \mathbb{C}_{w_0w_a}}},
\end{equation}
where $\mathbb{C}_{w_0w_a}$ denotes the $2\times2$ covariance submatrix of the parameters $(w_0,w_a)$. Geometrically, this quantity is inversely proportional to the area of the confidence ellipse in the $(w_0,w_a)$ plane and therefore provides a direct measure of how tightly the DE EoS is constrained. A larger FoM corresponds to a smaller allowed region in parameter space and hence to stronger constraining power. 

Table~\ref{tab:results.DE} shows a clear correlation between the decrease in statistical significance and the degradation of the FoM. The minimal $w_0w_a$CDM model gives FoM values of order $\sim 210$. Extensions involving $N_{\rm eff}$ leave the FoM close to $\sim 190$-$200$, consistent with the stability of the dynamical-DE preference in those cases. Allowing curvature to vary lowers the FoM to $\sim 190$, indicating a moderate degradation of the $(w_0,w_a)$ constraints, accompanied by a noticeable reduction in significance. When multiple sectors are relaxed simultaneously, the FoM drops more substantially. In the fully extended model, it decreases to $\sim 147$ for CMB+DESI+PP and $\sim 143$ for CMB+DESI+DD, indicating a substantial enlargement of the allowed region in the $(w_0,w_a)$ plane. The weaker preference in very large parameter spaces is therefore primarily a dilution effect, reflecting the cumulative impact of several degeneracy directions, even when the preferred region remains broadly similar.

\subsection{$w_0$-$w_a$ Correlations and Pivot Redshift in Extended Cosmologies}

Table~\ref{tab:results.DE} also clarifies the geometric structure of the constraints in the $(w_0,w_a)$ plane. A first notable feature is the strong correlation between the two CPL parameters, quantified by $\rho_{w_0w_a}$. Across all model extensions and for both dataset combinations, we find $\rho_{w_0w_a}\simeq -0.9$, indicating a highly elongated allowed region with negative slope. This reflects the well-known degeneracy between the present-day value of the EoS and its redshift evolution. Many combinations of $(w_0,w_a)$ can produce similar expansion histories over the redshift range probed by the data: a larger present-day value of $w_0$ can be compensated by a more negative $w_a$, leading to comparable integrated distances. Importantly, this degeneracy direction remains remarkably stable when the cosmological model is extended. Additional sectors enlarge the allowed region, as reflected by the degradation of the FoM, but they do not significantly rotate the direction in which the constraints are most sensitive. This is particularly informative for spatial curvature. Since $\Omega_k$ directly modifies the distance-redshift relation, it is the most natural candidate for rotating the $(w_0,w_a)$ degeneracy. However, even when curvature is allowed to vary, $\rho_{w_0w_a}$ remains essentially unchanged. Curvature therefore introduces an additional degeneracy that broadens the constraints, but does not alter the underlying direction selected by the late-time geometric data. This is consistent with the fact that curvature produces a smoother geometric shift over a broad redshift interval, whereas the CPL preference is tied to the specific low- and intermediate-redshift evolution probed by BAO and SN data.

The same geometry also determines the pivot redshift $z_p$, defined as the redshift at which the EoS is maximally constrained. In the CPL parametrization, many combinations of $w_0$ and $w_a$ along the principal degeneracy direction produce very similar expansion histories, and therefore nearly identical distance-redshift relations. By contrast, variations orthogonal to this degeneracy modify the expansion history more efficiently and are much more tightly constrained. Since the EoS at a given redshift corresponds to a specific linear combination of $w_0$ and $w_a$, each redshift effectively selects a direction in the $(w_0,w_a)$ plane. The pivot redshift is the value of $z$ for which this combination aligns with the best-constrained direction of the likelihood, so that the uncertainty on $w(z)$ is minimized and $w_0$ and $w_a$ become effectively decorrelated. In this sense, $z_p$ identifies the epoch at which the data most directly constrain the effective value of the EoS.

As shown in Table~\ref{tab:results.DE}, the pivot redshift remains confined to a relatively narrow interval, $z_p\simeq0.25$--$0.34$, across all model extensions and dataset combinations. The fact that the pivot consistently falls around $z\sim0.3$ indicates that the dominant leverage arises from the redshift range where BAO and SN measurements provide the strongest constraints on the late-time distance-redshift relation.

\subsection{Phantom crossing in Extended Cosmologies}
A useful way to further characterize the reconstructed EoS is to compare the pivot redshift $z_p$ with the phantom-crossing redshift $z_{\rm cross}$, defined by $w(z_{\rm cross})=-1$. As reported in Table~\ref{tab:results.DE}, the two redshifts lie remarkably close to each other: while $z_p\simeq0.25-0.34$, the crossing occurs at $z_{\rm cross}\simeq0.33-0.39$, with only mild variations across model extensions and dataset combinations. Since $z_p$ identifies the redshift at which the EoS is most tightly constrained, this implies that the best-constrained portion of the reconstruction lies close to the point where the preferred trajectory crosses the $\Lambda$CDM value. Consistently, the pivot EoS remains very close to $-1$ in all cases, typically $w(z_p)\in[-0.97,-0.99]$. The preference for dynamical DE therefore does not manifest itself as a large local departure from $w=-1$ at the redshift where the constraints are strongest. Rather, the reconstruction remains anchored near the $\Lambda$ value around $z\sim0.3$, while the evidence for dynamics appears through the slope of $w(z)$ away from this point, with the preferred trajectory evolving toward more phantom-like values at higher redshift. This also explains the stability of $z_{\rm cross}$: even when the statistical significance changes or the allowed $(w_0,w_a)$ region broadens, the crossing remains confined to a narrow redshift interval.

The overall behaviour of the reconstructed EoS is illustrated in Fig.~\ref{fig:EoS}. Across all extensions, the preferred evolution remains quintessence-like at low redshift and moves toward phantom-like values at higher redshift. Although the width of the allowed region depends on the cosmological parameter space, the shape of the reconstruction is remarkably stable. The different trajectories remain closely aligned and cross $w=-1$ within a narrow redshift interval, consistent with the values of $z_{\rm cross}$ reported in Table~\ref{tab:results.DE}. This provides a direct visual confirmation of the interpretation above. Enlarging the cosmological parameter space mainly broadens the allowed region around an otherwise stable trajectory, rather than shifting the reconstruction toward a qualitatively different behaviour. The same conclusion is reinforced by Fig.~\ref{fig:summary_plot_DDE}, which summarizes the constraints on $w_0$, $w_a$, $w(z_p)$, and $z_{\rm cross}$. While the uncertainties increase as additional parameters are introduced, the central values show little variation across extensions. Both dataset combinations consistently favour $w_0>-1$ and $w_a<0$, corresponding to an evolution from a mildly quintessence-like regime today toward phantom-like values at higher redshift, with $z_{\rm cross}$ clustered around $z\sim0.35$.

Overall, DESI late-time distances, combined with CMB and SN data, select a remarkably stable pattern in the redshift dependence of the expansion history. Within the CPL framework, this translates into a characteristic DE evolution: mildly quintessence-like today, crossing near $z\sim0.35$, and becoming phantom-like at higher redshift. Additional cosmological freedom broadens the allowed region and lowers the nominal significance, but does not substantially alter the underlying geometric pattern preferred by the data.
\section{Spatial Curvature}
\label{sec.curvature}

In $\Lambda$CDM, the Universe is assumed to be spatially flat. This assumption is well motivated by inflationary expectations and remains broadly consistent with the current observational landscape. Most late-time geometric probes do not provide robust evidence for $\Omega_k \neq 0$. DESI BAO distance measurements are compatible with spatial flatness~\cite{DESI:2025zgx}, in agreement with previous SDSS BAO results~\cite{eBOSS:2020yzd}. Weak-lensing and galaxy-clustering data point in the same direction: DES, when combined with CMB and SN, tightly constrains $\Omega_k\sim 0$~\cite{DES:2022ccp}, while KiDS-Legacy remains compatible with flatness within its broader uncertainties~\cite{KiDS:2020ghu}. High-resolution CMB observations also lead to somewhat consistent conclusions. ACT, in combination with Planck CMB and DESI BAO, finds no statistically significant evidence for curvature~\cite{ACT:2025tim}, while SPT primary CMB spectra alone constrain $\Omega_k$ to remain within $1\sigma$ of zero~\cite{SPT-3G:2025bzu}.

The historical outlier is Planck. Analyses of the PR3 temperature and polarization spectra, when based on primary CMB data alone and without lensing reconstruction, found a preference for closed geometry, with $\Omega_k<0$ at more than $3\sigma$~\cite{Planck:2018vyg,DiValentino:2019qzk,Handley:2019tkm}. This indication is closely tied to the well-known lensing anomaly~\cite{DiValentino:2019qzk}, since a closed geometry can partly mimic the enhanced smoothing of the acoustic peaks.\footnote{In Planck PR4 the preference for $\Omega_k<0$ is reduced, although residual likelihood dependence remains: \texttt{CamSpec} still finds a mild preference for a closed Universe~\cite{Rosenberg:2022sdy}, while \texttt{HiLLiPoP} is broadly compatible with flatness~\cite{Tristram:2023haj}.} The Planck curvature anomaly has therefore motivated a broad literature assessing whether this signal reflects a genuine departure from the baseline cosmological model, or instead unmodelled systematics~\cite{Park:2017xbl,Handley:2019tkm,Ryan:2019uor,DiValentino:2019qzk,Efstathiou:2020wem,DiValentino:2020hov,Benisty:2020otr,Vagnozzi:2020rcz,Vagnozzi:2020dfn,DiValentino:2020kpf,Yang:2021hxg,Bargiacchi:2021hdp,Cao:2021ldv,Dhawan:2021mel,Dinda:2021ffa,Zuckerman:2021kgm,Gonzalez:2021ojp,Akarsu:2021max,DiValentino:2022oon,DiValentino:2022rdg,Cao:2022ugh,Glanville:2022xes,Bel:2022iuf,Yang:2022kho,Stevens:2022evv,Vigneron:2022bgr,Jimenez:2022asc,Liu:2022mpj,Banerjee:2023rvg,Favale:2023lnp,Zhang:2023eup,Blachier:2023ooc,Qi:2023oxv,Giare:2023ejv,Foidl:2024xlv,Deng:2024uuz,Jensko:2024bee,Gariazzo:2024sil,Sanz-Wuhl:2024uvi,Amendola:2024gkz,Bhattacharya:2024hep,Shimon:2024mbm,Kuzmichev:2025fpm,Yadav:2025wbc,Du:2025csv,Fortunato:2025qxc,Chudaykin:2025lww,Favale:2025mgk,Deng:2025uou,Chen:2025mlf,Wang:2025hvo,Specogna:2025ufe,Millard:2026wnd,Kumar:2026kbo,Pulido-Hernandez:2026hcs,Wu:2026qog,Comini:2026nsj}.

In this section we revisit curvature constraints within the same sector-by-sector framework adopted throughout this work, using the updated and comprehensive dataset combinations described in Sec.~\ref{sec.methods}. The interest in $\Omega_k$ goes beyond establishing whether the Universe is exactly spatially flat. Spatial curvature is also an important geometric degree of freedom in extended cosmologies, since it modifies both the expansion history and the mapping between radial and transverse distances. As a result, $\Omega_k$ can absorb part of the geometric freedom that would otherwise be assigned to other sectors, especially late-time DE. This makes curvature particularly relevant when comparing $\Lambda$CDM-like extensions with dynamical-DE models, where both sectors can reshape the distance-redshift relation probed by BAO and SN data.

\subsection{Spatial Curvature in Extended Cosmological Models}

For each cosmological model, we quantify the statistical preference for non-zero spatial curvature using the full marginalized posterior distribution of $\Omega_k$ and computing the posterior probabilities on the two sides of the flat limit, $\mathcal{P}(\Omega_k<0)$ and $\mathcal{P}(\Omega_k>0)$, corresponding to closed and open geometry, respectively. We then define the cumulative posterior probability up to the flat limit as $F(0)\equiv \mathcal{P}(\Omega_k\leq 0)$. From this, we construct a two-sided posterior tail probability,
\begin{equation}
p = 2\,\min\left[F(0),\,1-F(0)\right],
\end{equation}
which quantifies how far the posterior is displaced from $\Omega_k=0$. With this definition, small values of $p$ indicate that the flat limit lies in the tail of the posterior and is therefore disfavored by the data, while values of order unity indicate no significant preference away from flatness. For ease of interpretation, we also express this result in terms of an equivalent Gaussian significance, defined as $\#\sigma = \Phi^{-1}\left(1-p/2\right)$, where $\Phi^{-1}$ denotes the inverse cumulative distribution function of a unit Gaussian defined in Eq.~\eqref{eq:Phi}.\footnote{We stress that this procedure does not assume that the posterior distribution of $\Omega_k$ is exactly Gaussian. The probabilities $\mathcal{P}(\Omega_k<0)$, $\mathcal{P}(\Omega_k>0)$, and $p$ are computed directly from the weighted posterior samples, while $\#\sigma$ is only used as a convenient Gaussian-equivalent summary statistic.}

\begin{table*}[t!]
\centering
\renewcommand{\arraystretch}{1}
\resizebox{\textwidth}{!}{
\begin{tabular}{l @{\hspace{0.5 cm}} l @{\hspace{0.5 cm}}c@{\hspace{0.5 cm}}c@{\hspace{0.5 cm}}c@{\hspace{0.5 cm}}cccccc}
\hline\hline
\\
\textbf{Model} & \textbf{Dataset} & \boldmath{$\Omega_k$} & \boldmath{$\mathcal P(\Omega_k<0)$} & \boldmath{$\mathcal P(\Omega_k>0)$} & \boldmath{$p$}& \textbf{\#}\boldmath{$\sigma$}\\
\\
\hline
&&&&&\\
$\Lambda\mathrm{CDM}+\Omega_k$
& CMB+DESI+PP & $0.0024\pm 0.0011$ & $0.014$ & $0.986$ & $0.027$ & $2.205$ \\[2ex]
& CMB+DESI+DD & $0.0024\pm 0.0011$ & $0.014$ & $0.986$ & $0.028$ & $2.200$ \\[4ex]

$\Lambda\mathrm{CDM} +r + \alpha_s + \beta_s + \Omega_k + \sum m_\nu + N_\mathrm{eff}$
& CMB+DESI+PP & $0.0031\pm 0.0015$ & $0.014$ & $0.986$ & $0.028$ & $2.197$ \\[2ex]
& CMB+DESI+DD & $0.0032\pm 0.0015$ & $0.011$ & $0.989$ & $0.022$ & $2.286$ \\[4ex]

$w_0 w_a\mathrm{CDM} + \Omega_k$  
& CMB+DESI+PP & $0.0011\pm 0.0013$ & $0.193$ & $0.807$ & $0.387$ & $0.865$ \\[2ex]
& CMB+DESI+DD & $0.0008\pm 0.0013$ & $0.256$ & $0.744$ & $0.513$ & $0.655$ \\[4ex]

$w_0 w_a\mathrm{CDM} + r + \alpha_s + \beta_s + \Omega_k + \sum m_\nu + N_\mathrm{eff}$
& CMB+DESI+PP & $0.0021\pm 0.0016$ & $0.095$ & $0.905$ & $0.190$ & $1.310$ \\[2ex]
& CMB+DESI+DD & $0.0020\pm 0.0016$ & $0.108$ & $0.892$ & $0.217$ & $1.236$ \\[4ex]

\hline
\bottomrule
\end{tabular}
}
\caption{Marginalized constraints on $\Omega_k$ and corresponding posterior probabilities for the four curvature extensions considered in this work. For each model and dataset combination, we report the marginalized constraint on $\Omega_k$, the posterior probabilities $\mathcal P(\Omega_k<0)$ and $\mathcal P(\Omega_k>0)$, the associated two-sided posterior tail probability $p$, and the equivalent Gaussian significance.}
\label{tab:results.omk}
\end{table*}

\begin{figure}[t!]
    \centering
    \includegraphics[width=\linewidth]{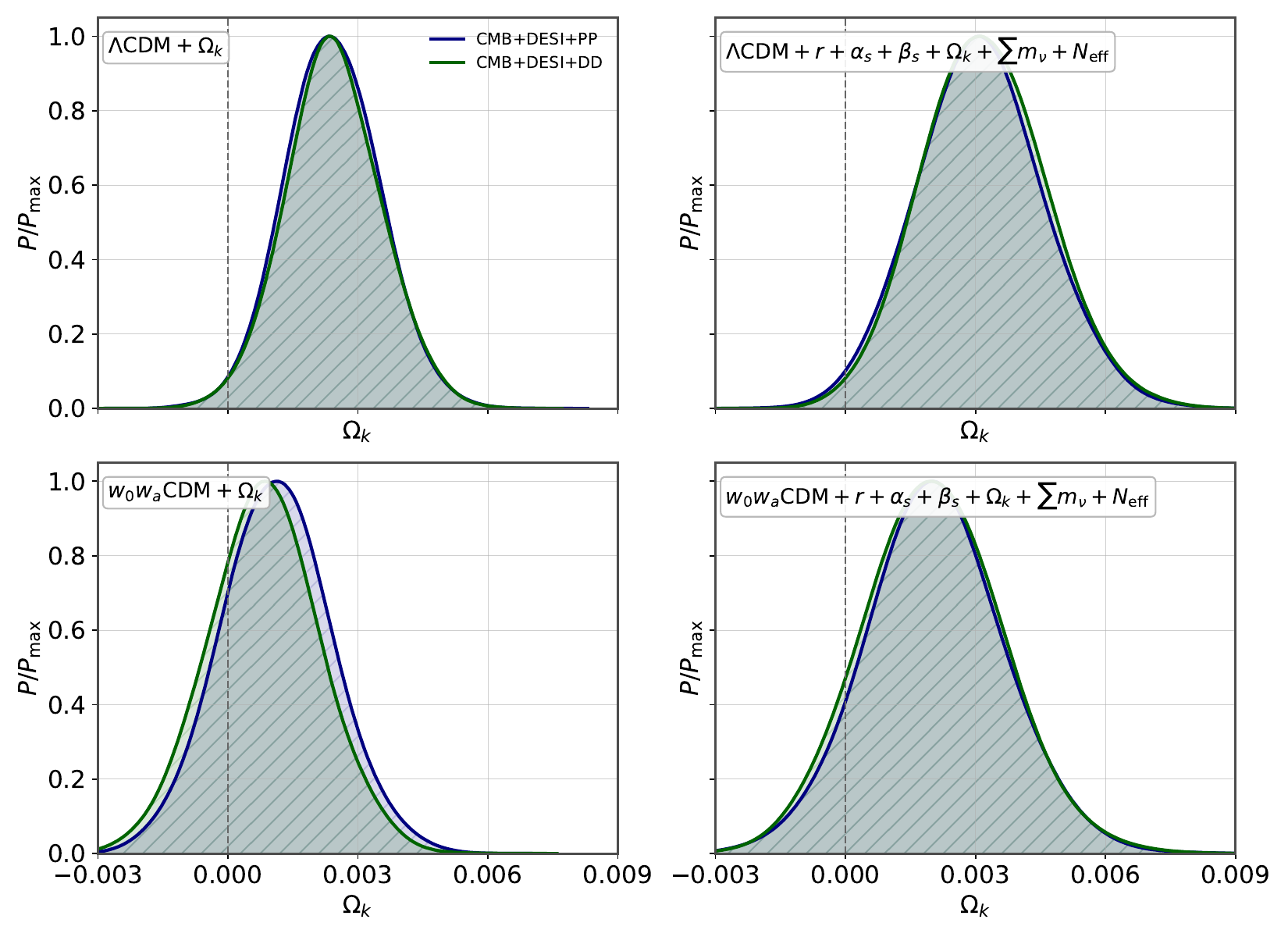}
    \caption{Marginalized one-dimensional posterior distributions of $\Omega_k$ for the four curvature extensions considered in this work, obtained from the CMB+DESI+PP and CMB+DESI+DD dataset combinations.}
    \label{fig:omk_1D}
\end{figure}

The corresponding numerical results are summarized in Table~\ref{tab:results.omk}, while the marginalized one-dimensional posterior distributions of $\Omega_k$ are shown in Fig.~\ref{fig:omk_1D}.

Within the $\Lambda$CDM-like branch, we observe a mild preference for positive $\Omega_k$. In the baseline $\Lambda$CDM+$\Omega_k$ case, both CMB+DESI+PP and CMB+DESI+DD yield $\Omega_k=0.0024\pm0.0011$, corresponding to a posterior probability $\mathcal P(\Omega_k>0)=0.986$ and a Gaussian-equivalent significance of about $2.2\sigma$. This indication is not washed out when additional sectors are opened simultaneously. In the fully extended $\Lambda$CDM-like model, the preference for $\Omega_k>0$ remains essentially unchanged, and even increases slightly for the DD combination, reaching $2.3\sigma$.

In the baseline $w_0w_a$CDM+$\Omega_k$ model, the posterior shifts significantly toward the flat limit, with $\Omega_k=0.0011\pm0.0013$ for CMB+DESI+PP and $\Omega_k=0.0008\pm0.0013$ for CMB+DESI+DD. The corresponding Gaussian-equivalent significances drop to only $0.87\sigma$ and $0.66\sigma$, respectively, fully consistent with a spatially flat Universe. Even in the fully extended dynamical-DE branch, we do not observe any significant preference for curvature, and the constraints remain substantially compatible with spatial flatness at $1.31\sigma$ for CMB+DESI+PP and $1.24\sigma$ for CMB+DESI+DD.

The mild preference for positive $\Omega_k$ is therefore present only when DE is fixed to a cosmological constant, but largely disappears once the late-time expansion history is allowed to vary dynamically. This suggests that in $\Lambda$CDM-like models, an open geometry is at least partly acting as an effective surrogate for missing late-time DE freedom. Although a small positive $\Omega_k$ modifies the expansion history with a different redshift dependence from dynamical DE, both affect the late-time background geometry and hence induce partially similar shifts in the distance-redshift relation over the redshift range most relevant for the geometric probes considered here. As a result, positive $\Omega_k$ can partly mimic the same broad geometric adjustment that, once the DE sector is opened, is more naturally absorbed by a dynamical component favored by the data.

Overall, our results do not provide a robust case for either an open or a closed Universe. Instead, they show that the mild preference for $\Omega_k>0$ found in $\Lambda$CDM-like models, and more broadly the inferred constraints on spatial curvature, are strongly dependent on the assumptions made in the DE sector.

\clearpage
\section{The Neutrino Sector}
\label{sec.nu}

Flavour oscillations require non-degenerate neutrino mass eigenstates and non-trivial leptonic mixing~\cite{Pontecorvo:1957cp,Pontecorvo:1967fh,Maki:1962mu,Gribov:1968kq,Bilenky:1978nj,Bilenky:1980cx,Kayser:1981ye,Giunti:1991ca}. Neutrino oscillation experiments precisely measure differences between the squared masses of the propagation eigenstates~\cite{Bilenky:1978nj,Esteban:2024eli}. Solar neutrino experiments, together with KamLAND, determine the smaller splitting in the $(\nu_1,\nu_2)$ sector, $\Delta m^2_{21} \simeq 7.4 \times 10^{-5}\,\mathrm{eV}^2$~\cite{SNO:2002tuh,KamLAND:2008dgz,Esteban:2024eli}, while atmospheric, accelerator, and reactor data probe the larger splitting involving the third mass eigenstate, $|\Delta m^2_{32}| \simeq 2.5 \times 10^{-3}\,\mathrm{eV}^2$~\cite{Super-Kamiokande:1998kpq,T2K:2011ypd,MINOS:2013utc,DayaBay:2012fng,Esteban:2024eli}. These measurements imply that at least two neutrino mass eigenstates are non-zero. However, since the atmospheric splitting is determined through $|\Delta m^2_{32}|$, oscillation data do not directly distinguish normal ordering (NO), $m_1 < m_2 < m_3$, from inverted ordering (IO), $m_3 < m_1 < m_2$, nor do they determine the absolute mass scale $\sum m_\nu=m_1+m_2+m_3$. Within each ordering, the measured splittings imply only a minimum value for $\sum m_\nu$, obtained by setting the lightest neutrino mass to zero, namely $m_1=0$ in NO and $m_3=0$ in IO:
\begin{equation}
\sum m_\nu \gtrsim
\begin{cases}
0.059\,\mathrm{eV} \,, & \text{for NO} \\
0.10\,\mathrm{eV} \,, & \text{for IO}.
\end{cases}
\end{equation}

Because neutrino properties affect both the expansion history and the growth of structure across widely separated epochs and scales, cosmology represents one of our best opportunities, and arguably the most precise one, to measure the total neutrino mass and independently test the existence of three light relativistic species~\cite{Lesgourgues:2006nd,Lesgourgues:2012uu,TopicalConvenersKNAbazajianJECarlstromATLee:2013bxd,Vagnozzi:2017ovm}. 

Somewhat paradoxically, however, this very precision has turned neutrino cosmology into a new arena where cosmological inference can come into tension with particle-physics expectations. Within a $\Lambda$CDM cosmology, DESI BAO measurements show mild shifts relative to the region of parameter space favoured by Planck~\cite{DESI:2025zgx}. The best-fit parameters inferred from the two datasets are slightly displaced, leading to an overall $2$--$2.5\sigma$ disagreement~\cite{DESI:2025zgx,Ye:2025ark}. Such disagreement is largely driven by a mismatch in the present-day matter density parameter $\Omega_m$~\cite{Loverde:2024nfi,Colgain:2024mtg,Elbers:2025vlz,Lynch:2025ine,Sailer:2025lxj,Jhaveri:2025neg}, with DESI favouring smaller values than Planck~\cite{DESI:2025zgx,DESI:2025fii}. This directly affects the neutrino-mass inference because once neutrinos become non-relativistic, they contribute to the matter density proportionally to their total mass~\cite{Lesgourgues:2006nd},
\begin{equation}
\Omega_\nu h^2 \simeq \frac{\sum m_\nu}{93.12\,\mathrm{eV}} \, .
\end{equation}
Within a minimal $\Lambda$CDM+$\sum m_\nu$ analysis of Planck and DESI data, the preference for lower $\Omega_m$ pushes cosmological neutrino-mass bounds close to, and in some cases into tension with, the lower limits from oscillation experiments discussed above~\cite{Loverde:2024nfi,Jiang:2024viw,Lynch:2025ine,Capozzi:2025wyn}.\footnote{Some of us had previously pointed out that large-scale temperature and polarization anomalies can also propagate into parameters strongly correlated with $\sum m_{\nu}$, such as $\tau$~\cite{Giare:2023ejv}. Following the DESI results, independent analyses confirmed that relaxing the large-scale $E$-mode polarization constraint can substantially weaken the quoted neutrino-mass bounds~\cite{Allali:2025wwi,Allali:2025yvp,Sailer:2025lxj,Jhaveri:2025neg,Sullivan:2026tas}.} An even more puzzling result emerges when the well-motivated physical prior $\sum m_\nu > 0$ is relaxed and the formally unphysical region $\sum m_\nu < 0$ is allowed. In this case, several independent analyses have found that the likelihood peaks at negative values of the neutrino-mass parameter~\cite{Craig:2024tky,Naredo-Tuero:2024sgf,Green:2024xbb,Elbers:2024sha}. In models with additional freedom relative to $\Lambda$CDM, such as dynamical DE, these bounds can be relaxed and become consistent again with the lower limits from particle physics~\cite{DESI:2025zgx,DESI:2025fii,Giare:2025ath,Kibris:2026cqq,Yang:2026yaq,Li:2026ldf}. Here we quantify how the constraints on $\sum m_\nu$ and $N_{\rm eff}$ change as the assumptions of $\Lambda$CDM are relaxed, and compare the resulting bounds with the lower limits implied by oscillation data across the extended cosmological models considered in this work.

\begin{table*}[t!]
\centering
\renewcommand{\arraystretch}{1}
\resizebox{\textwidth}{!}{
\begin{tabular}{l @{\hspace{0.5 cm}} l @{\hspace{0.5 cm}} c @{\hspace{0.5 cm}} c @{\hspace{0.5 cm}} c @{\hspace{0.5 cm}} c @{\hspace{0.5 cm}} c @{\hspace{0.5 cm}} c @{\hspace{0.5 cm}} c c @{\hspace{0.5 cm}}}
\hline\hline
\\[-1.5ex]
\textbf{Model} & \textbf{Dataset} & \boldmath{$N_{\rm eff}$} & \boldmath{$\Delta N_{\rm eff} / \sigma_{N_{\rm eff}}$} & \boldmath{$\sum m_{\nu}$} \textbf{[eV]} & \boldmath{$\ln\!\left(\mathcal B_{\rm NO/IO}\right)$} &
\multicolumn{2}{c}{\boldmath{$\mathcal P_{\sum m_{\nu}<\rm{osc}}$}} &
\multicolumn{2}{c}{\boldmath{$\mathrm{OVL}$}} \\
\cmidrule(lr){7-8}\cmidrule(lr){9-10}
& & & & & &
\textbf{NO} & \textbf{IO} &
\textbf{NO} & \textbf{IO} \\
\\[-0.5ex]
\hline
& & & & & & & & \\
$\Lambda\mathrm{CDM}+\sum m_{\nu}$
&  CMB+DESI+PP    &  $3.04$ (fixed) & -- & $<0.0608$ & $3.329 \pm 0.154$ & $0.942$ & $0.999$ & $0.013$ & $0.001$ \\[2ex]
&  CMB+DESI+DD &  $3.04$ (fixed) & -- & $<0.0613$ & $3.169 \pm 0.131$ & $0.941$ & $0.998$ & $0.014$ & $0.002$ \\[4ex]

$\Lambda\mathrm{CDM}+N_{\rm eff}$
&  CMB+DESI+PP & $3.00\pm 0.10$ & $-0.393$ &  $0.06$ (fixed) & -- & -- & -- & -- & -- \\[2ex]
&  CMB+DESI+DD & $3.00\pm 0.11$ & $-0.456$ &  $0.06$ (fixed) & -- & -- & -- & -- & -- \\[4ex]

$\Lambda\mathrm{CDM}+\sum m_{\nu}+N_{\rm eff}$ 
&  CMB+DESI+PP & $2.97\pm 0.10$ & $-0.734$ & $<0.0587$ & $3.068 \pm 0.127$ & $0.950$ & $0.998$ & $0.013$ & $0.001$ \\[2ex]
&  CMB+DESI+DD & $2.97\pm 0.11$ & $-0.728$ & $<0.0592$ & $3.139 \pm 0.135$ & $0.949$ & $0.998$ & $0.014$ & $0.002$ \\[4ex]

$\Lambda\mathrm{CDM}+r+\alpha_s+\beta_s+\Omega_k+\sum m_{\nu}+N_{\rm eff}$
& CMB+DESI+PP & $2.93\pm 0.18$ & $-0.655$ & $<0.122$ & $0.936 \pm 0.012$ & $0.679$ & $0.900$ & $0.036$ & $0.031$ \\[2ex]
&  CMB+DESI+DD & $2.92\pm 0.18$ & $-0.725$ & $<0.122$ & $0.936 \pm 0.013$ & $0.676$ & $0.900$ & $0.035$ & $0.030$ \\[4ex]

$w_0w_a\mathrm{CDM}+\sum m_{\nu}$
&  CMB+DESI+PP    &  $3.04$ (fixed) & -- & $<0.113$ & $1.048 \pm 0.012$ & $0.709$ & $0.921$ & $0.031$ & $0.026$ \\[2ex]
&  CMB+DESI+DD &  $3.04$ (fixed) & -- & $<0.123$ & $0.915 \pm 0.009$ & $0.644$ & $0.889$ & $0.035$ & $0.029$ \\[4ex]

$w_0w_a\mathrm{CDM}+N_{\rm eff}$ 
&  CMB+DESI+PP & $2.90\pm 0.11$ & $-1.286$ &  $0.06$ (fixed) & -- & -- & -- & -- & -- \\[2ex]
&  CMB+DESI+DD & $2.88\pm 0.11$ & $-1.442$ &  $0.06$ (fixed) & -- & -- & -- & -- & -- \\[4ex]

$w_0w_a\mathrm{CDM}+\sum m_{\nu}+N_{\rm eff}$ 
&  CMB+DESI+PP & $2.89\pm 0.11$ & $-1.337$ & $<0.0992$ & $1.360 \pm 0.020$ & $0.769$ & $0.953$ & $0.029$ & $0.022$ \\[2ex]
&  CMB+DESI+DD & $2.88\pm 0.11$ & $-1.447$ & $<0.113$ & $1.054 \pm 0.014$ & $0.703$ & $0.920$ & $0.032$ & $0.026$ \\[4ex]

$w_0w_a\mathrm{CDM}+r+\alpha_s+\beta_s+\Omega_k+\sum m_{\nu}+N_{\rm eff}$
&  CMB+DESI+PP & $2.90^{+0.18}_{-0.20}$ & $-0.757$ & $<0.171$ & $0.477 \pm 0.007$ & $0.502$ & $0.752$ & $0.050$ & $0.046$ \\[2ex]
&  CMB+DESI+DD & $2.88^{+0.18}_{-0.20}$ & $-0.882$ & $<0.184$ & $0.387 \pm 0.006$ & $0.443$ & $0.696$ & $0.052$ & $0.049$ \\[4ex]
\hline
\bottomrule
\end{tabular}
}
\caption{Summary of the neutrino-sector constraints obtained for each model and dataset combination considered in this work. We report the marginalized constraint on $N_{\rm eff}$, its normalized displacement $\Delta N_{\rm eff}/\sigma_{N_{\rm eff}}$ relative to $N_{\rm eff}=3.04$, the 95\% CL upper bound on the total neutrino mass $\sum m_\nu$, and the logarithmic Bayes factor $\ln(\mathcal B_{\rm NO/IO})$ quantifying the relative preference for normal over inverted ordering. We also report two complementary diagnostics of the tension between cosmological neutrino-mass constraints and oscillation-informed expectations, namely the tail probability $\mathcal P_{\sum m_\nu<{\rm osc}}$ and the overlap coefficient ${\rm OVL}$, each evaluated separately for NO and IO.}
\label{tab:results.neutrinos}
\end{table*}

\begin{figure}[t]
    \centering  \includegraphics[width=0.6\linewidth]{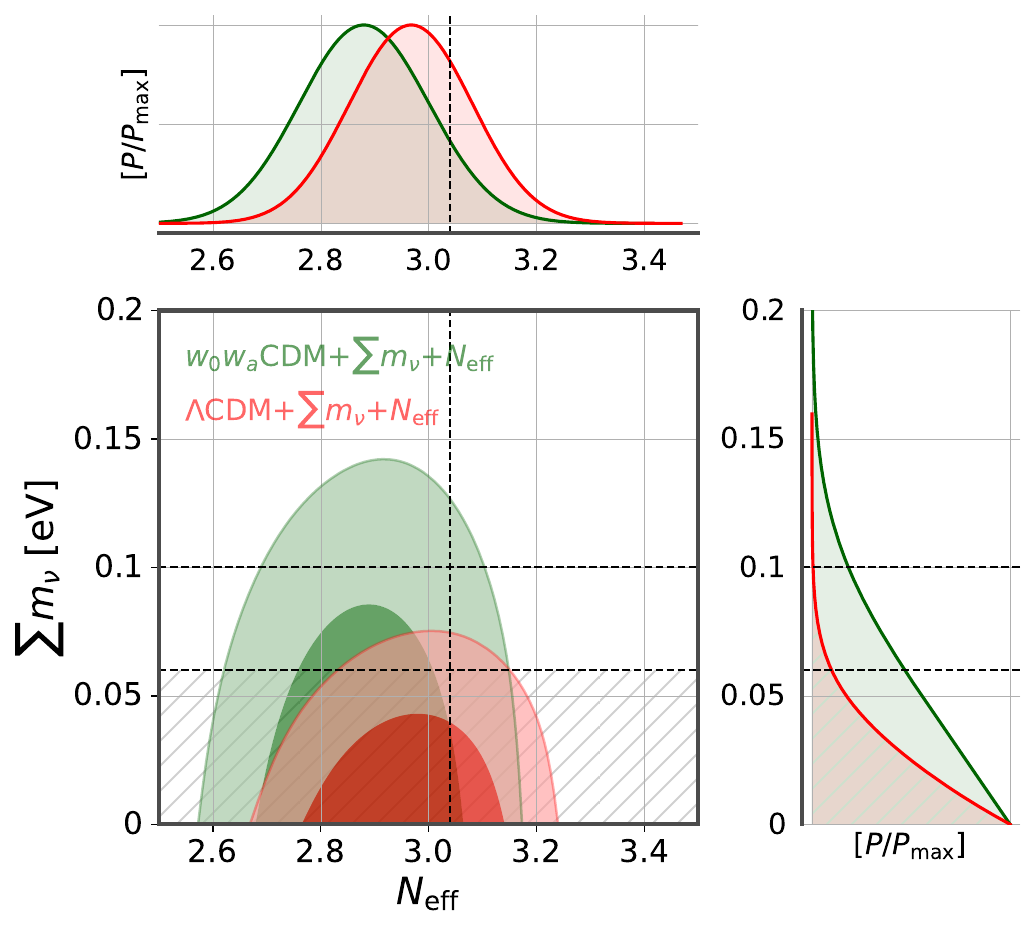}
    \caption{Joint constraints in the ($\sum m_\nu$, $N_{\rm eff}$) plane for $\Lambda$CDM+$\sum m_\nu$+$N_{\rm eff}$ and $w_0w_a$CDM+$\sum m_\nu$+$N_{\rm eff}$ for the dataset combination CMB+DESI+DD. The contours show the 68\% and 95\% CL regions, while the upper and right panels show the corresponding one-dimensional marginalized posteriors, normalized to their maximum value. The vertical dashed line marks the standard prediction $N_{\rm eff}=3.044$. The hatched region below $\sum m_\nu\simeq 0.06\,{\rm eV}$ lies below the minimal value allowed by oscillation data for NO, while the dashed horizontal line at $\sum m_\nu\simeq 0.10\,{\rm eV}$ indicates the approximate minimal value for IO.}
    \label{fig:mnu_neff}
\end{figure}

\subsection{Neutrino Mass Bounds in Extended Cosmologies}

The 95\% confidence level upper bounds on the total neutrino mass $\sum m_\nu$ are reported in Table~\ref{tab:results.neutrinos}. Across all model extensions, CMB+DESI+PP typically gives slightly tighter bounds than CMB+DESI+DD. The difference, however, is small, and the two SN dataset combinations lead to broadly consistent conclusions.

In the minimal $\Lambda$CDM+$\sum m_\nu$ model, the bound $\sum m_\nu\simeq 0.061\,{\rm eV}$ lies extremely close to the minimum value allowed by oscillation data in NO, and well below the minimum value allowed in IO. Allowing $N_{\rm eff}$ and $\sum m_\nu$ to vary together makes the limits slightly tighter, $\sum m_\nu<0.0587\,{\rm eV}$ for CMB+DESI+PP and $\sum m_\nu<0.0592\,{\rm eV}$ for CMB+DESI+DD, pushing the upper limit even closer to the NO floor and worsening the mismatch with oscillation experiments. This is seen directly in Fig.~\ref{fig:mnu_neff}, where the $\Lambda$CDM+$\sum m_\nu+N_{\rm eff}$ posterior in the $(\sum m_\nu,N_{\rm eff})$ plane remains compressed toward very small neutrino masses, with the allowed region piling up close to the lower edge set by oscillation data.

Relaxing the assumption of a cosmological constant, the bounds instead weaken substantially. In $w_0w_a$CDM+$\sum m_\nu$, we find $\sum m_\nu<0.113\,{\rm eV}$ for CMB+DESI+PP and $\sum m_\nu<0.123\,{\rm eV}$ for CMB+DESI+DD. These values are no longer compressed against the NO floor and become compatible with both hierarchies, NO and IO. A similar conclusion holds when $N_{\rm eff}$ is also allowed to vary within dynamical DE scenarios; see Fig.~\ref{fig:mnu_neff}. The relaxation is therefore driven primarily by the additional late-time background freedom introduced by the dynamical DE sector. This is consistent with the DE analysis in Sec.~\ref{sec.DE}, where we showed that allowing $\sum m_\nu$ to vary introduced an additional degeneracy direction with the CPL parameters, reducing the nominal significance of the dynamical-DE preference. Here the same correlation acts in the opposite direction, relaxing the cosmological upper bound on $\sum m_\nu$ once the DE sector is allowed to evolve.

Dynamical DE, however, is not the only way to relax the neutrino-mass bound. Inflationary parameters such as $\alpha_s$ and $\beta_s$ modify the scale dependence of the primordial spectrum and can correlate with the small-scale CMB information that also constrains $N_{\rm eff}$ through the damping tail. Most importantly, spatial curvature acts directly at the background level by modifying cosmological distances and adding geometrical freedom. Since the neutrino-mass bound is inferred from the combination of CMB and late-time information, additional freedom in both sectors can weaken the constraint even without modifying DE. In $\Lambda$CDM+$r+\alpha_s+\beta_s+\Omega_k+\sum m_\nu+N_{\rm eff}$, where the cosmological constant is kept fixed but several other sectors are enlarged simultaneously, the bound relaxes to $\sum m_\nu<0.122\,{\rm eV}$ for both CMB+DESI+PP and CMB+DESI+DD. Agreement with the oscillation lower limits can therefore also be recovered within a $\Lambda$-based late-time cosmology, provided that other assumptions of the minimal model are relaxed.

Finally, when all extensions are allowed simultaneously, including dynamical DE, curvature, neutrino-sector freedom, and inflationary-sector freedom, the bounds weaken further to $\sum m_\nu<0.171\,{\rm eV}$ for CMB+DESI+PP and $\sum m_\nu<0.184\,{\rm eV}$ for CMB+DESI+DD, as expected from the cumulative effect of several degeneracy directions.

Overall, adding background-level freedom through dynamical DE or spatial curvature, and more generally enlarging the cosmological parameter space, relaxes the neutrino mass bounds and restores compatibility with the particle-physics lower limits. None of the extensions considered here, however, leads to a preference for non-zero neutrino mass.

\begin{figure}[t]
    \centering  \includegraphics[width=\linewidth]{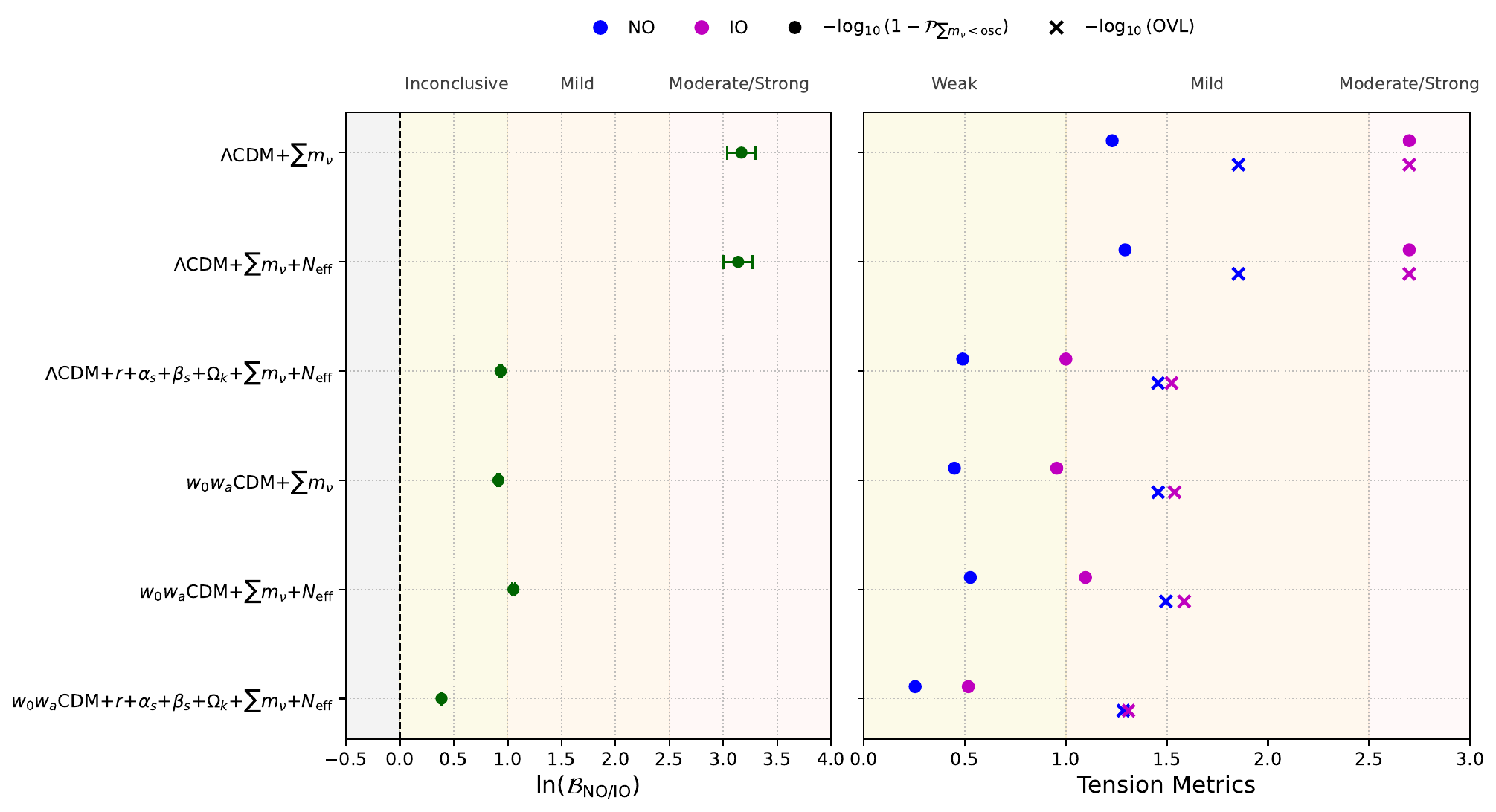}
    \caption{Summary of the preference for the neutrino mass ordering and of the tension with particle-physics expectations across the cosmological models considered in this work, for the dataset combination CMB+DESI+DD. The left panel shows $\ln(\mathcal{B}_{\rm NO/IO})$, with positive values indicating a preference for normal ordering over inverted ordering. The right panel summarizes the level of tension with oscillation-based mass information for both orderings, using two complementary estimators: $-\log_{10}(1-\mathcal P_{\sum m_\nu<{\rm osc}})$ and $-\log_{10}(\mathrm{OVL})$. Blue and magenta markers denote NO and IO, respectively. The dataset combination CMB+DESI+PP yields fully analogous results.}
    \label{fig:mnu_summary}
\end{figure}

\subsection{Ordering Preference in Extended Cosmologies}

The remarkable model dependence of the mass bounds implies that both the inferred preference for a given mass ordering and the possible tension with oscillation data are themselves framework dependent.

Here we quantify the relative preference for NO over IO across the same extended cosmologies. To this end, we define the logarithmic Bayes factor
\begin{equation}
\ln(\mathcal{B}_{\rm NO/IO}) = \ln Z_{\rm NO}-\ln Z_{\rm IO},
\end{equation}
where $Z_{\rm NO}$ and $Z_{\rm IO}$ are the ordering-conditioned evidences for NO and IO, respectively. Each $Z_H$, with $H=\{{\rm NO},{\rm IO}\}$, measures the compatibility between the cosmological posterior on $\sum m_\nu$ and the total-mass distribution implied by oscillation data under the ordering $H$. To construct the latter, we build two oscillation-informed reference distributions, one for NO and one for IO. Once a mass ordering is assumed, the full neutrino spectrum is fixed by the value of the lightest mass. For NO, the lightest state is $m_1$, so that
\begin{equation}
\sum m_\nu^{\rm NO}(m_1)=
m_1+\sqrt{m_1^2+\Delta m_{21}^2}
+\sqrt{m_1^2+\Delta m_{31}^2}.
\end{equation}
For IO, the lightest state is $m_3$, so that
\begin{equation}
\sum m_\nu^{\rm IO}(m_3)=
m_3+\sqrt{m_3^2+|\Delta m_{31}^2|}
+\sqrt{m_3^2+|\Delta m_{31}^2|+\Delta m_{21}^2}.
\end{equation}
By sampling the lightest mass and mapping it into $\sum m_\nu$ separately for NO and IO, we obtain two normalized distributions, $P_{\rm osc}^{\rm NO}(\sum m_\nu)$ and $P_{\rm osc}^{\rm IO}(\sum m_\nu)$.\footnote{In constructing these reference distributions, we adopt a prior uniform in the lightest neutrino mass over $m_{\rm lightest}\in[0,2]\,{\rm eV}$, fix the mass-squared splittings to $\Delta m^2_{21}=7.42\times10^{-5}\,{\rm eV}^2$, $\Delta m^2_{31}=2.517\times10^{-3}\,{\rm eV}^2$ for NO, and $|\Delta m^2_{31}|=2.498\times10^{-3}\,{\rm eV}^2$ for IO, and normalize the resulting distributions over the same total-mass interval $\Sigma\in[0,5]\,{\rm eV}$ adopted in the cosmological analysis.} The ordering-conditioned evidence for each ordering is then computed as
\begin{equation}
Z_H =
\int_{0}^{\infty} d\Sigma\,
P_{\rm cosmo}(\Sigma)\,
P_{\rm osc}^{H}(\Sigma),
\end{equation}
where $\Sigma\equiv\sum m_{\nu}$ and $P_{\rm cosmo}$ is the cosmological posterior distribution of the total neutrino mass obtained within each extended model. This quantity is the overlap evidence between the cosmological posterior and the oscillation-informed mass distribution for ordering $H$.\footnote{In practice, $Z_H$ is computed by evaluating the oscillation-informed weight of each sampled value of $\sum m_\nu$ in the cosmological chain and averaging it over the posterior. The uncertainty on $\ln(\mathcal B_{\rm NO/IO})$ is estimated by bootstrap resampling the weighted cosmological samples and taking the standard deviation of the resulting $\ln(\mathcal B_{\rm NO/IO})$ distribution.} If the cosmological posterior is concentrated close to the NO floor and below the minimum mass allowed by IO, the overlap evidence $Z_{\rm IO}$ is suppressed and $\ln(\mathcal{B}_{\rm NO/IO})$ becomes positive. If, instead, the posterior has substantial support above the IO threshold, the two ordering-conditioned evidences become more similar and $\ln(\mathcal{B}_{\rm NO/IO})$ moves closer to zero. Thus, $\ln(\mathcal{B}_{\rm NO/IO})$ measures the compatibility of the cosmological mass posterior with the NO spectrum relative to the IO spectrum, with positive values indicating a preference for NO over IO.

We report the values of $\ln(\mathcal{B}_{\rm NO/IO})$ in Table~\ref{tab:results.neutrinos} and summarize them in the left panel of Fig.~\ref{fig:mnu_summary}. As a qualitative guide, we use the conventional Jeffreys/Trotta scale~\cite{Jeffreys:1939xee,Trotta:2008qt}, where $|\ln\mathcal B|<1$ is usually regarded as inconclusive, $1\leq|\ln\mathcal B|<2.5$ as weak evidence, $2.5\leq|\ln\mathcal B|<5$ as moderate-to-strong evidence, and $|\ln\mathcal B|\geq5$ as strong or very strong evidence.

The behaviour of $\ln(\mathcal B_{\rm NO/IO})$ closely tracks the strength of the cosmological upper bound on $\sum m_\nu$. In the minimal $\Lambda$CDM+$\sum m_\nu$ model, where the bounds lie close to the NO floor and almost entirely below the minimum total mass allowed in IO, the evidence for NO is substantial, with $\ln(\mathcal B_{\rm NO/IO})\simeq3.3$. It decreases once late-time background freedom is introduced. In $w_0w_a$CDM+$\sum m_\nu$, the logarithmic Bayes factor drops to $\ln(\mathcal B_{\rm NO/IO})\simeq 1.0$. In both the $\Lambda$CDM-based and dynamical-DE cases, allowing $N_{\rm eff}$ to vary does not change the qualitative conclusion.

A similar reduction is obtained in the fully extended $\Lambda$CDM-based model. Even without allowing DE to evolve, the simultaneous variation of $r$, $\alpha_s$, $\beta_s$, $\Omega_k$, $\sum m_\nu$, and $N_{\rm eff}$ relaxes the mass bound and lowers the ordering preference to $\ln(\mathcal B_{\rm NO/IO})\simeq0.94$. This confirms that the apparent preference for NO is not specific to the DE sector, but depends more generally on the freedom allowed in the cosmological model. In fact, the weakest ordering preference is found in the fully extended dynamical-DE model. When all sectors are allowed to vary simultaneously, $\ln(\mathcal B_{\rm NO/IO})$ drops below $0.5$ for both dataset combinations, corresponding to inconclusive evidence.

Overall, the cosmological preference for NO over IO is highly model dependent. It is substantial in minimal $\Lambda$CDM-based analyses, where the total-mass posterior is compressed close to the NO floor and below the IO threshold, but becomes weak or inconclusive once additional cosmological freedom is allowed.

\subsection{Consistency with Oscillation Data in Extended Cosmologies}

We now assess the absolute consistency between the cosmological posterior on $\sum m_\nu$ and the mass ranges implied by oscillation data. Unlike $\ln(\mathcal B_{\rm NO/IO})$, which compares NO and IO relative to each other, the diagnostics introduced here test whether a given cosmological posterior is compatible with the total-mass range allowed under each ordering separately.

The agreement between cosmology and oscillation data conditional on NO or IO is quantified through two complementary diagnostics.

\begin{itemize}
\item The first diagnostic is the marginal posterior probability that the cosmological total mass lies below the minimum value allowed by oscillation data for a given ordering
\begin{equation}
\mathcal P_{\sum m_\nu<{\rm osc}}^{H}
=
\int_0^{\Sigma_{\rm min}^{H}}
d\Sigma \,
P_{\rm cosmo}(\Sigma),
\end{equation}
where, as usual, $\Sigma\equiv\sum m_{\nu}$ while $\Sigma_{\rm min}^{H}\equiv (\sum m_\nu)_{\rm min}^{H}$ is the lower bound implied by oscillation data: $\simeq0.059\,{\rm eV}$ for $H={\rm NO}$ and $\simeq0.10\,{\rm eV}$ for $H={\rm IO}$.\footnote{In practice, this is computed directly from the MCMC chain as the total posterior weight of samples satisfying $\sum m_\nu < (\sum m_\nu)_{\rm min}^{H}$.}
This quantity measures the fraction of the cosmological posterior that falls in the region forbidden by oscillation data under the assumed ordering. Values close to zero indicate consistency with the oscillation experiments, while values close to one indicate that most of the posterior lies below the minimum total mass required by that ordering. It therefore isolates the lower-bound problem directly.

\item The second diagnostic is the overlap coefficient,
\begin{equation}
{\rm OVL}_{H}
=
\int_{0}^{\infty} d\Sigma\,
\min\left[
P_{\rm cosmo}(\Sigma),
P_{\rm osc}^{H}(\Sigma)
\right].
\end{equation}
Since both distributions are normalized over the same range of $\sum m_\nu$, we have $0\leq {\rm OVL}_{H}\leq1$ by construction.\footnote{In the numerical implementation, $P_{\rm cosmo}$ is reconstructed as a weighted histogram from the MCMC samples, while $P_{\rm osc}^{H}$ is evaluated on the same grid. The integral is then computed by summing the common area shared by the two distributions.} Values ${\rm OVL}_{H}\sim 1$ indicate that the cosmological posterior and the oscillation-informed distribution occupy the same region of total-mass space, while values ${\rm OVL}_{H}\sim 0$ indicate little common support. The overlap coefficient therefore provides a simple geometric measure of agreement between cosmology and oscillation data under the assumed ordering.
\end{itemize}

The two quantities are complementary. The tail probability measures how much cosmological posterior weight lies below the minimum total mass allowed by oscillation data. The overlap coefficient quantifies the probability area shared by the cosmological and oscillation-informed distributions.

In Table~\ref{tab:results.neutrinos}, we report both the diagnostics separately for NO and IO. They are also summarized graphically in the right panel of Fig.~\ref{fig:mnu_summary}, where, for visual clarity, the estimators are shown in logarithmic form so as to place them on a common qualitative scale ranging from weak to moderate-to-strong tension. They give a coherent picture: in the minimal $\Lambda$CDM+$\sum m_\nu$ model, the cosmological posterior is strongly compressed below the oscillation lower bounds. About $94\%$ of the posterior lies below the NO floor, and essentially all of it, $\simeq99.9\%$, lies below the IO floor. The overlaps are correspondingly tiny, ${\rm OVL}_{\rm NO}\simeq0.013$ and ${\rm OVL}_{\rm IO}\simeq0.001$. Thus, the models that give the strongest apparent preference for NO also produce the strongest absolute mismatch with the oscillation-informed mass ranges.

Allowing $N_{\rm eff}$ to vary together with $\sum m_\nu$ does not improve the situation, since the mass bound becomes even slightly tighter. The mismatch is instead reduced in $w_0w_a$CDM+$\sum m_\nu$. The posterior weight below the NO floor drops to $\simeq71\%$ for CMB+DESI+PP and $\simeq64\%$ for CMB+DESI+DD. IO remains more strained, with $\simeq92\%$ and $\simeq89\%$ of the posterior still below the IO floor. The overlaps increase by a factor of two to three relative to minimal $\Lambda$CDM. Dynamical DE therefore improves the absolute consistency with oscillation data, especially for NO. When $N_{\rm eff}$ is also varied within dynamical-DE models, the improvement is partially reduced for CMB+DESI+PP and broadly similar for CMB+DESI+DD.

A comparable improvement appears in the fully extended $\Lambda$CDM-based model. Even with DE fixed to a cosmological constant, allowing $r$, $\alpha_s$, $\beta_s$, $\Omega_k$, $\sum m_\nu$, and $N_{\rm eff}$ to vary reduces the posterior weight below the NO floor to $\simeq68\%$, and below the IO floor to $\simeq90\%$, for both dataset combinations. The overlaps increase to ${\rm OVL}_{\rm NO}\simeq0.035$ and ${\rm OVL}_{\rm IO}\simeq0.030$.

The best consistency is obtained in the fully extended dynamical-DE model. The posterior weight below the NO floor drops to $\simeq50\%$ for CMB+DESI+PP and $\simeq44\%$ for CMB+DESI+DD, while the fraction below the IO floor decreases to $\simeq75\%$ and $\simeq70\%$. The overlaps reach their largest values, ${\rm OVL}_{\rm NO}\simeq0.050$ and ${\rm OVL}_{\rm IO}\simeq0.049$. 

Overall, the apparent tension with oscillation data is highly model dependent. Enlarging the cosmological parameter space progressively improves consistency with the oscillation-informed mass ranges, although IO remains significantly more difficult to accommodate even in extended parameter spaces.

\subsection{Effective Number of Relativistic Species in Extended Cosmologies}

\begin{figure}[t]
    \centering  \includegraphics[width=0.7\linewidth]{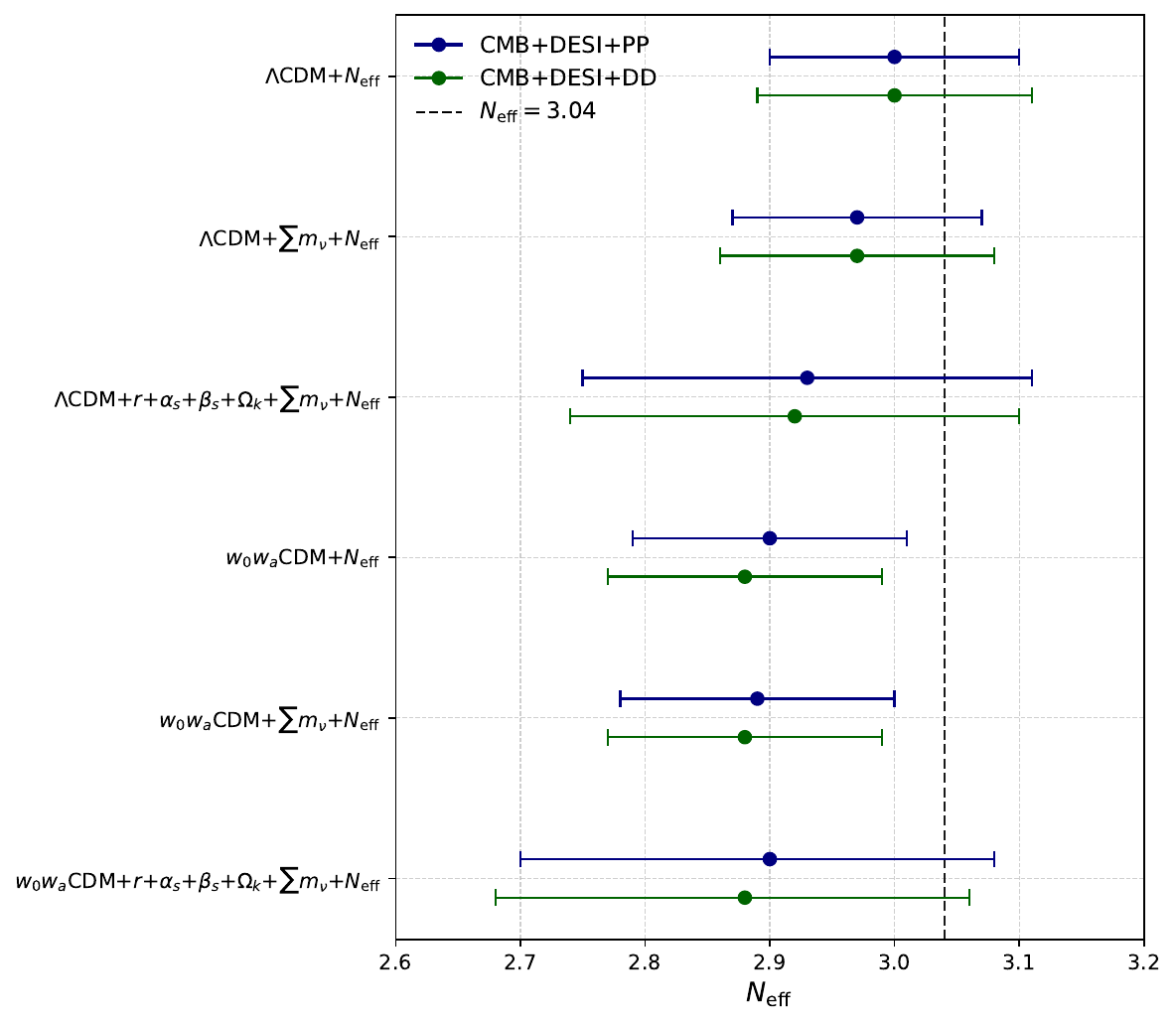}
    \caption{Summary of the marginalized constraints on $N_{\rm eff}$ across the cosmological models in which the effective number of relativistic species is allowed to vary. Results are shown for the two dataset combinations, CMB+DESI+PP and CMB+DESI+DD, corresponding to the same model extensions reported in Table~\ref{tab:results.neutrinos}. Each row denotes a different cosmological parameter-space configuration, while the vertical dashed line marks the standard expectation $N_{\rm eff}=3.04$.}
    \label{fig:Neff}
\end{figure}

We finally consider the constraints on $N_{\rm eff}$, which parameterizes the contribution of all relativistic species beyond photons to the radiation density of the Universe, including neutrinos and any additional dark-radiation component. Accounting for non-instantaneous neutrino decoupling, residual energy transfer during $e^\pm$ annihilation, finite-temperature QED effects, and flavour oscillations, the SM prediction is $N_{\rm eff}\simeq 3.04$~\cite{Mangano:2005cc,Bennett:2019ewm,Akita:2020szl,Froustey:2020mcq,Bennett:2020zkv,Drewes:2024wbw}.

The corresponding constraints are reported in Table~\ref{tab:results.neutrinos} and summarized in Fig.~\ref{fig:Neff}, together with the normalized displacement $\Delta N_{\rm eff}/\sigma_{N_{\rm eff}}$ relative to $N_{\rm eff}=3.04$. Across all extensions, the shifts are small but consistently negative. In $\Lambda$CDM+$N_{\rm eff}$, the data prefer $N_{\rm eff}\simeq3.00$, corresponding to a displacement of only $\simeq0.4\sigma$ below the SM value. Allowing $\sum m_\nu$ to vary simultaneously shifts the result slightly further down, $N_{\rm eff}\simeq2.97$, corresponding to $\simeq0.7\sigma$. In the fully extended $\Lambda$CDM-based model, the uncertainty increases and the shift remains below $1\sigma$, with $N_{\rm eff}\simeq2.92$.

Dynamical-DE models show the largest downward shifts. In $w_0w_a$CDM+$N_{\rm eff}$, we find $N_{\rm eff}\simeq2.88$, corresponding to $\simeq 1.4\sigma$, with a similar displacement when $\sum m_\nu$ is also varied. In the fully extended dynamical-DE model, however, the larger parameter space increases the uncertainty and reduces the normalized shift below $1\sigma$. 

Overall, the data show a mild and persistent tendency toward values of $N_{\rm eff}$ below $3.04$, but the significance remains modest in all cases, see Fig.~\ref{fig:Neff}. 

\clearpage
\section{Cosmological Inflation}
\label{sec.inflation}

At linear order, inflationary perturbations split into decoupled scalar and tensor sectors~\cite{Mukhanov:1990me,Kodama:1984ziu}. Scalar modes describe rotationally invariant curvature and density fluctuations, which provide the primordial seeds of the CMB temperature and polarization anisotropies after horizon re-entry~\cite{Mukhanov:1990me,Hu:1997hv}. Their amplitude and tilt are therefore tightly constrained by CMB observations~\cite{Planck:2018jri}. The amplitude sets the overall normalization of the CMB temperature, polarization, and lensing spectra once the optical depth to reionization is constrained. Its residual model dependence is driven mainly by the $A_s e^{-2\tau}$ degeneracy, with subleading correlations involving the matter density, late-time growth, and lensing normalization~\cite{Gariazzo:2024sil}. The scalar tilt is instead more sensitive to the assumed cosmological framework. In extended models, it can move closer to scale invariance when compensating for changes in the pre-recombination expansion history or in the calibration of the CMB damping tail~\cite{DiValentino:2018zjj,Keeley:2020rmo,Ye:2021nej,Takahashi:2021bti,Forconi:2021que,Braglia:2021sun,Giare:2022rvg,Ye:2022efx,Jiang:2022uyg,Jiang:2022qlj,Lin:2022gbl,Hazra:2022rdl,Giare:2023kiv,Giare:2023wzl,Jiang:2023bsz,Peng:2023bik,Forconi:2023hsj,Fu:2023tfo,Giare:2024akf,Giare:2024sdl,Wang:2024tjd,Cecchini:2024xoq,Peng:2025tqt,Forconi:2025zzu,Wolf:2025ecy,SPT-3G:2025vtb,McDonough:2025lzo,Balkenhol:2025wms,Yuan:2026xcg,Garny:2026gcs,Sabogal:2026qvy}. 

The tensor sector instead describes transverse-traceless metric perturbations, corresponding to a stochastic background of PGWs~\cite{Grishchuk:1974ny,Starobinsky:1979ty,Rubakov:1982df}. Their cleanest observational signature is a contribution to large-scale CMB $B$-mode polarization~\cite{Kamionkowski:1996ks,Seljak:1996gy,Zaldarriaga:1996xe}. So far, no tensor signal has been detected, and current data only provide upper limits on the amplitude of PGWs~\cite{BICEP:2021xfz,SPT-3G:2025vtb}. These limits are generally stable against changes in other sectors of the cosmological model, because they are driven primarily by large-scale $B$-mode measurements from BICEP/Keck~\cite{BICEP:2021xfz}. However, recent studies have suggested that indirect correlations may still emerge in enlarged cosmologies, particularly in dynamical-DE extensions~\cite{Wang:2024sgo,Wang:2025ljj}.

Here we quantify how much primordial-sector constraints depend on assumptions made elsewhere in the cosmological model. Using the same extended cosmologies considered throughout this work, we first constrain the scalar and tensor spectral parameters within increasingly flexible parametrizations of the primordial spectra. We then map these constraints onto potential slow-roll parameters and Hubble-flow parameters. The former directly constrain the inflationary potential and its derivatives, while the latter characterize the evolution of the Hubble rate during inflation and quantify departures from exact de Sitter expansion.

\subsection{Spectral parameters in Extended Cosmological Models}

\begin{table*}[htpb!]
\centering
\renewcommand{\arraystretch}{1}
\resizebox{\textwidth}{!}{
\begin{tabular}{l @{\hspace{0.5 cm}} l @{\hspace{0.5 cm}}c@{\hspace{0.5 cm}}c@{\hspace{0.5 cm}}c@{\hspace{0.5 cm}}c@{\hspace{0.5 cm}}c@{\hspace{0.5 cm}}c@{\hspace{0.5 cm}}c@{\hspace{0.5 cm}}c@{\hspace{0.5 cm}}c}
\hline\hline
\\
\textbf{Model} & \textbf{Dataset} & \boldmath{$\ln\left(10^{10}\,A_s\right)$} & \boldmath{$n_s$} & \boldmath{$\alpha_s$} &  \boldmath{$\beta_s$}  & \boldmath{$r$} & \boldmath{$n_t$} & \boldmath{$\alpha_t \cdot 10^4 $} & \boldmath{$\beta_t \cdot 10^6$}\\
\\
\hline
&&&&&\\

$\Lambda\mathrm{CDM}+r$ 
&  CMB+DESI+PP & $3.0662^{+0.010}_{-0.011}$ & $0.9738\pm 0.0029$ & $--$ & $--$ & $<0.0352$ & $>-0.00440$ & $-0.483 \pm 0.275$ & $--$ \\[2ex]
&  CMB+DESI+DD & $3.065\pm 0.011$ & $0.9736\pm 0.0029$ & $--$ & $--$ & $<0.0354$ & $>-0.00443$ & $-0.489 \pm 0.278$ & $--$ \\[4ex]

$\Lambda\mathrm{CDM}+r+\alpha_s$  
&  CMB+DESI+PP & $3.063^{+0.011}_{-0.012}$ & $0.9736\pm 0.0030$ & $0.0039\pm 0.0051$ & $--$ & $<0.0353$ & $>-0.00441$ & $-0.477 \pm 0.280$ & $6.506 \pm 12.872$ \\[2ex]
&  CMB+DESI+DD & $3.063^{+0.010}_{-0.012}$ & $0.9736\pm 0.0029$ & $0.0038\pm 0.0050$ & $--$ & $<0.0343$ & $>-0.00429$ & $-0.471 \pm 0.275$ & $6.230 \pm 12.442$ \\[4ex]

$\Lambda\mathrm{CDM}+r+\alpha_s+\beta_s$  
&  CMB+DESI+PP & $3.067^{+0.011}_{-0.012}$ & $0.9702\pm 0.0037$ & $0.0077\pm 0.0055$ & $0.0124\pm 0.0086$ & $<0.0351$ & $>-0.00439$ & $-0.561 \pm 0.329$ & $14.476 \pm 15.907$ \\[2ex]
&  CMB+DESI+DD & $3.066^{+0.011}_{-0.012}$ & $0.9702\pm 0.0037$ & $0.0074\pm 0.0056$ & $0.0123\pm 0.0088$ & $<0.0342$ & $>-0.00428$ & $-0.558 \pm 0.322$ & $13.609 \pm 15.199$ \\[4ex]

$\Lambda\mathrm{CDM}+r+\alpha_s+\beta_s+\Omega_k+\sum m_{\nu}+N_{\rm eff}$  
& CMB+DESI+PP & $3.060^{+0.012}_{-0.014}$ & $0.9595\pm 0.0087$ & $0.0046\pm 0.0085$ & $0.0145\pm 0.0096$ & $<0.0362$ & $>-0.00453$ & $-0.796 \pm 0.520$ & $5.866 \pm 22.353$ \\[2ex]
&  CMB+DESI+DD & $3.059^{+0.012}_{-0.014}$ & $0.9588\pm 0.0087$ & $0.0043\pm 0.0086$ & $0.0143\pm 0.0098$ & $<0.0351$ & $>-0.00439$ & $-0.788 \pm 0.516$ & $5.265 \pm 21.473$ \\[4ex]

$w_0w_a\mathrm{CDM}+r$  
&  CMB+DESI+PP & $3.0552^{+0.010}_{-0.012}$ & $0.9716\pm 0.0030$ & $--$ & $--$ & $<0.0345$ & $>-0.00431$ & $-0.510 \pm 0.295$ & $--$ \\[2ex]
&  CMB+DESI+DD & $3.054^{+0.010}_{-0.011}$ & $0.9714\pm 0.0031$ & $--$ & $--$ & $<0.0350$ & $>-0.00438$ & $-0.532 \pm 0.303$ & $--$ \\[4ex]

$w_0w_a\mathrm{CDM}+r+\alpha_s$  
&  CMB+DESI+PP & $3.052^{+0.010}_{-0.012}$ & $0.9716\pm 0.0031$ & $0.0043\pm 0.0050$ & $--$ & $<0.0345$ & $>-0.00431$ & $-0.507 \pm 0.301$ & $6.954 \pm 12.218$ \\[2ex]
&  CMB+DESI+DD & $3.051^{+0.010}_{-0.012}$ & $0.9714\pm 0.0030$ & $0.0045^{+0.0052}_{-0.0046}$ & $--$ & $<0.0343$ & $>-0.00429$ & $-0.509 \pm 0.299$ & $7.331 \pm 12.597$ \\[4ex]

$w_0w_a\mathrm{CDM}+r+\alpha_s+\beta_s$  
&  CMB+DESI+PP & $3.055^{+0.011}_{-0.012}$ & $0.9681\pm 0.0039$ & $0.0082\pm 0.0056$ & $0.0125\pm 0.0087$ & $<0.0348$ & $>-0.00435$ & $-0.591 \pm 0.347$ & $14.889 \pm 15.883$ \\[2ex]
&  CMB+DESI+DD & $3.054^{+0.010}_{-0.012}$ & $0.9679\pm 0.0039$ & $0.0082\pm 0.0056$ & $0.0123\pm 0.0087$ & $<0.0352$ & $>-0.00440$ & $-0.595 \pm 0.360$ & $14.680 \pm 16.139$ \\[4ex]

$w_0w_a\mathrm{CDM}+r+\alpha_s+\beta_s+\Omega_k+\sum m_{\nu}+N_{\rm eff}$  
& CMB+DESI+PP & $3.053^{+0.013}_{-0.014}$ & $0.9591\pm 0.0094$ & $0.0032\pm 0.0090$ & $0.0121\pm 0.0097$ & $<0.0355$ & $>-0.00444$ & $-0.792 \pm 0.525$ & $2.932 \pm 22.571$ \\[2ex]
& CMB+DESI+DD & $3.052^{+0.013}_{-0.015}$ & $0.9580\pm 0.0094$ & $0.0025\pm 0.0090$ & $0.012\pm 0.010$ & $<0.0357$ & $>-0.00446$ & $-0.815 \pm 0.544$ & $1.077 \pm 22.604$ \\[4ex]
\hline
\bottomrule
\end{tabular}
}
\caption{Constraints on the scalar and tensor spectral parameters obtained after imposing the slow-roll consistency relations. We report results for different choices of the primordial parametrization and extended cosmological models, using the CMB+DESI+PP and CMB+DESI+DD dataset combinations. The tensor tilt $n_t$ and its runnings $\alpha_t$ and $\beta_t$ are derived from $r$, $n_s$, and $\alpha_s$ through the consistency relations. Constraints are quoted at 68\% CL, except for the upper limits on $r$ (and the corresponding lower limits on $n_T$), which are reported at 95\% CL.}
\label{tab:results.inflation}
\end{table*}

The simplest dynamical realizations of inflation are based on Einstein gravity and a single weakly coupled degree of freedom, $\phi$, evolving on a sufficiently flat potential $V(\phi)$ that can sustain a phase of slow-roll accelerated expansion~\cite{Riotto:2002yw,Kinney:2009vz,Baumann:2009ds,Martin:2013tda,Senatore:2016aui}. Within this framework, primordial scalar and tensor perturbations are expected to be nearly Gaussian~\cite{Maldacena:2002vr,Acquaviva:2002ud,Bartolo:2004if}, so that their leading statistical information is captured by the two-point correlation functions, or equivalently by the primordial power spectra. In a quasi-de Sitter background, the spectra of quantum fluctuations are nearly scale invariant and, at leading order, are commonly described by power-law expansions specified by amplitudes and constant tilts, which quantify departures from exact scale invariance~\cite{Mukhanov:1981xt,Hawking:1982cz,Starobinsky:1982ee,Bardeen:1983qw,Lidsey:1995np,Kosowsky:1995aa}. Here we go beyond this leading-order description by adopting higher-order parametrizations that allow for the scale dependence of both the scalar and tensor tilts. 

For the scalar spectrum, we consider the following extended parametrization
\begin{equation}
\ln \mathcal{P}_{s}(k)
=
\ln A_s
+
(n_s-1)\ln\left(\frac{k}{k_\star}\right)
+
\frac{\alpha_s}{2}\ln^2\left(\frac{k}{k_\star}\right)
+
\frac{\beta_s}{6}\ln^3\left(\frac{k}{k_\star}\right),
\end{equation}
where $k_\star=0.05\,{\rm Mpc}^{-1}$ is the pivot scale, $A_s\equiv \mathcal{P}_s(k_\star)$ is the scalar amplitude, $n_s-1\equiv \left.d\ln\mathcal P_s/d\ln k\right|_{k_\star}$ is the scalar tilt, $\alpha_s\equiv \left.dn_s/d\ln k\right|_{k_\star}$ is its running, and $\beta_s\equiv \left.d\alpha_s/d\ln k\right|_{k_\star}$ is the running of the running.

Analogously, for the tensor spectrum we use
\begin{equation}
\ln \mathcal{P}_{T}(k)
=
\ln\left(r A_s\right)
+
n_T\ln\left(\frac{k}{k_\star}\right)
+
\frac{\alpha_T}{2}\ln^2\left(\frac{k}{k_\star}\right)
+
\frac{\beta_T}{6}\ln^3\left(\frac{k}{k_\star}\right),
\end{equation}
where $r\equiv \mathcal{P}_T(k_\star)/\mathcal{P}_s(k_\star)$ is the tensor-to-scalar ratio, $n_T\equiv \left.d\ln\mathcal{P}_T/d\ln k\right|_{k_\star}$ is the tensor tilt, $\alpha_T\equiv \left.dn_T/d\ln k\right|_{k_\star}$ is its running, and $\beta_T\equiv \left.d\alpha_T/d\ln k\right|_{k_\star}$ is the tensor running of the running.\footnote{For representative studies of higher-order parametrizations of the primordial spectra, including spectral runnings and associated consistency relations, with applications to inflationary observables, PGWs, and parameter inference, see, e.g., Refs.~\cite{Liddle:1994cr,Turner:1995ge,Lidsey:1995np,Hui:2001ce,Bartolo:2001rt,Kinney:2002qn,Ashoorioon:2005ep,Gong:2007ha,Cheng:2012je,Palma:2014faa,Boyle:2014kba,Cabass:2015jwe,Huang:2015gca,Meerburg:2015zua,Brooker:2016imi,Broy:2016zik,Giare:2019snj,Cano:2020oaa,Capurri:2020qgz,Giare:2020plo,Giare:2020vhn,Giare:2020vss,Vagnozzi:2020gtf,Forconi:2021que,Kinney:2021nje,Li:2021nqa,Benetti:2021uea,Giare:2022wxq,Giovannini:2022etg,German:2023euc,Giare:2023wzl,Bianchi:2024qyp,Cecchini:2024xoq,Giare:2024sdl,Ballardini:2024ado,Ballardini:2024irx,Balkenhol:2025wms,Brax:2025osk,Haddad:2026hsf,Sabogal:2026qvy} and references therein.}

In single-field slow-roll inflation, the scalar and tensor spectral parameters are determined by the same background dynamics. This leads to the following consistency relations
~\cite{Cortes:2006ap,Kuroyanagi:2011iw,Giare:2019snj}:
\begin{equation}
n_T=-\frac{r}{8},
\end{equation}
\begin{equation}
\alpha_T=
\frac{r}{8}(n_s-1)+\frac{r^2}{64},
\end{equation}
\begin{equation}
\beta_T=
\frac{r}{8}\left[\alpha_s-\left(n_s-1\right)^2\right]
-\frac{3r^2}{64}\left(n_s-1\right)
-\frac{r^3}{256}.
\end{equation}
Imposing these consistency relations, we derive constraints on the scalar and tensor parameters across the extended cosmologies analyzed in this work. The results are summarized in Table~\ref{tab:results.inflation}.

The tensor sector remains remarkably stable across the full set of models considered. The 95\% CL upper bounds on $r$ vary only between $r\lesssim 0.034$ and $r\lesssim 0.036$, corresponding to a full spread of about $6\%$ around the $\Lambda$CDM+$r$ value $r\lesssim 0.0352$. This variation is small compared with the enlargement of the parameter space, and shows no systematic dependence on whether the late-time background is described by $\Lambda$CDM or by $w_0w_a$CDM. Overall, we find no preference for a non-vanishing primordial tensor amplitude at the 95\% CL. The stability of $r$ directly propagates to the derived tensor parameters: the tensor tilt is fixed by $n_T=-r/8$, and therefore remains essentially unchanged as long as the bound on $r$ is stable. The higher-order tensor quantities are further slow-roll suppressed, with $\alpha_T$ depending on $r(n_s-1)$ and $r^2$, and $\beta_T$ controlled by combinations involving $r\,\alpha_s$, $r\,(n_s-1)^2$, $r^2\,(n_s-1)$, and $r^3$. As a result, even moderate shifts in the scalar running parameters have only a limited impact on the tensor runnings, which remain small and consistent with the expected slow-roll hierarchy across all the models explored. The difference between dataset combinations involving the PP and DD supernova samples is also negligible.

\begin{figure}[t!]
    \centering    \includegraphics[width=\linewidth]{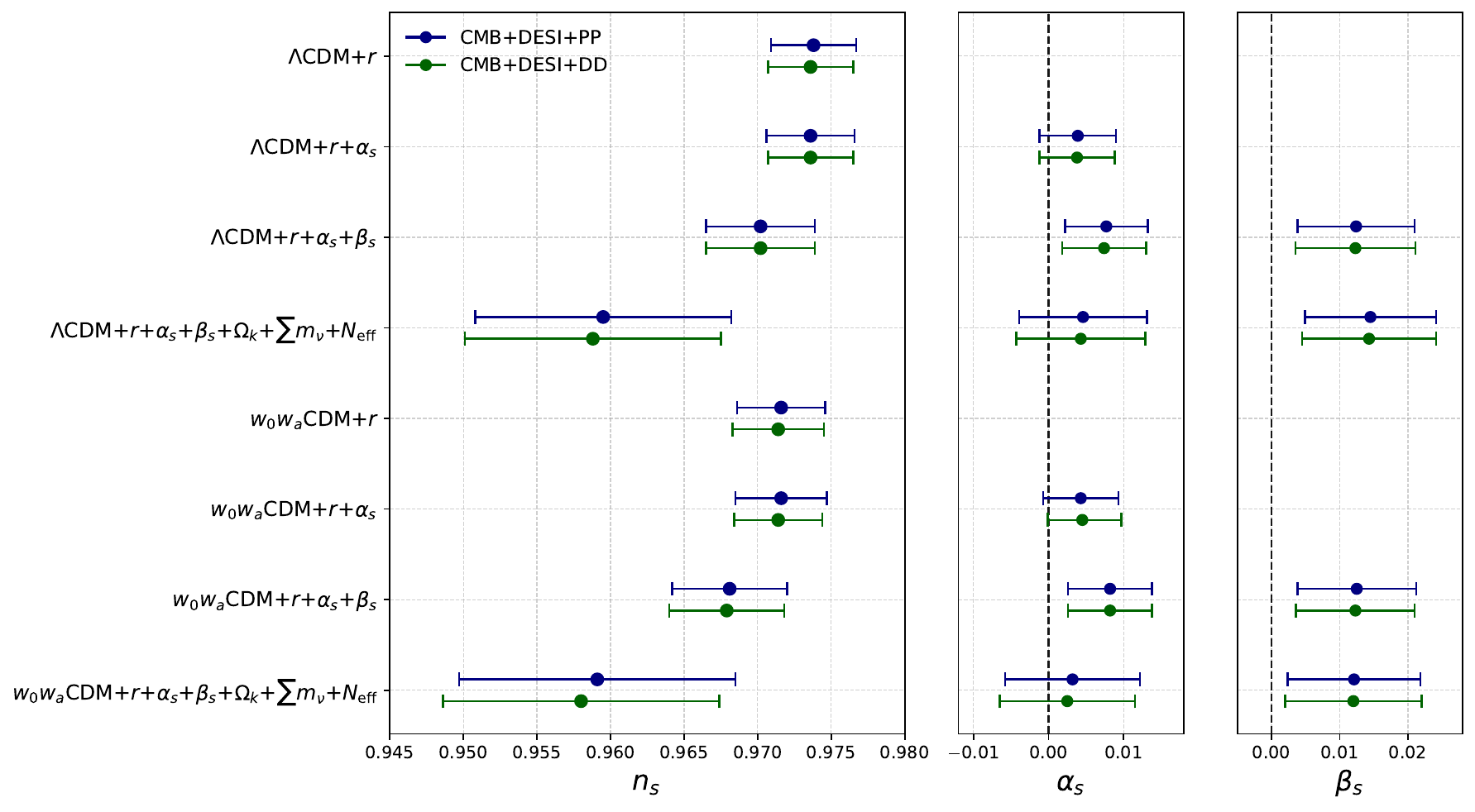}
    \caption{Summary of the marginalized $1\sigma$ constraints on the scalar spectral tilt $n_s$ and, where applicable, on its running $\alpha_s$ and running of the running $\beta_s$, across the extended inflationary cosmologies considered in this work. The main panel shows the constraints on $n_s$, while the two side panels report the corresponding constraints on $\alpha_s$ and $\beta_s$ only for the models in which these parameters are allowed to vary. Results are shown for the two dataset combinations CMB+DESI+PP and CMB+DESI+DD, corresponding to the same model extensions reported in Table~\ref{tab:results.inflation}.}
    \label{fig:whisker_inf}
\end{figure}

The scalar sector carries most of the response to the enlarged parametrizations, with the dominant effect appearing in the inferred value of $n_s$ and, where allowed, in the running parameters $\alpha_s$ and $\beta_s$, as shown in Fig.~\ref{fig:whisker_inf}. For CMB+DESI+PP, the minimal $\Lambda$CDM+$r$ model gives a tightly constrained scalar tilt, $n_s=0.9738\pm0.0029$. Replacing $\Lambda$CDM with $w_0w_a$CDM induces only a mild downward shift, to $n_s=0.9716\pm0.0030$, showing that late-time DE freedom has a visible but limited impact on the scalar tilt. Allowing for $\alpha_s$ alone leaves this picture essentially unchanged: $n_s$ remains stable, and the running is compatible with zero, with $\alpha_s=0.0039\pm0.0051$ in $\Lambda$CDM+$r+\alpha_s$. A more pronounced shift in parameter space appears when $\beta_s$ is also varied. In this case, $n_s$ shifts downward, reaching $n_s=0.9702\pm0.0037$ in $\Lambda$CDM+$r+\alpha_s+\beta_s$, while $\alpha_s$ and $\beta_s$ acquire positive central values, with $\beta_s=0.0124\pm0.0086$. This points to a correlated redistribution of freedom within the scalar sector. The fact that the central values of both $\alpha_s$ and $\beta_s$ are positive across these extended scalar parametrizations suggests a mild preference for additional small-scale power, although the preference never exceeds the $2\sigma$ level across the extensions considered.

The largest displacement of $n_s$ occurs when curvature and neutrino-sector parameters are enlarged simultaneously. In the $\Lambda$CDM extension including $\Omega_k$, $\sum m_\nu$, and $N_{\rm eff}$, together with $r$, $\alpha_s$, and $\beta_s$, the scalar tilt shifts to $n_s=0.9595\pm0.0087$ for CMB+DESI+PP. The analogous $w_0w_a$CDM extension with the same additional parameters gives $n_s=0.9591\pm0.0094$, showing that the additional late-time DE freedom has little impact. The shift is mainly driven by the additional degeneracies introduced by these parameters, in particular by $N_{\rm eff}$, which changes the pre-recombination expansion rate and hence the ratio between the photon-diffusion scale and the acoustic scale, altering the damping-tail power. This can be partially compensated by $n_s$, which controls the relative amount of small-scale primordial power. For all these models, the corresponding CMB+DESI+DD constraints follow the same pattern and lead to indistinguishable conclusions.

\subsection{Slow-Roll and Hubble-Flow Parameters in Extended Cosmological Models}


The slow-roll regime can be characterized in terms of the potential slow-roll parameters
\begin{equation}
\epsilon_V \equiv \frac{M_{\rm pl}^2}{2}\left(\frac{V_\phi}{V}\right)^2,
\qquad
\eta_V \equiv M_{\rm pl}^2\frac{V_{\phi\phi}}{V},
\end{equation}
together with their higher-order extensions
\begin{equation}
\xi_V^2 \equiv M_{\rm pl}^4\frac{V_\phi V_{\phi\phi\phi}}{V^2},
\qquad
\varpi_V^3 \equiv M_{\rm pl}^6\frac{V_\phi^2 V_{\phi\phi\phi\phi}}{V^3}.
\end{equation}
A controlled slow-roll expansion requires these quantities to remain perturbatively small, with $\epsilon_V\ll1$ and $|\eta_V|\ll1$ ensuring accelerated expansion and the higher-order parameters controlling departures from an exactly scale-independent tilt~\cite{Liddle:1994dx,Lidsey:1995np}.

The same dynamics can also be described directly in terms of the evolution of the Hubble rate through the Hubble-flow hierarchy,
\begin{equation}
\epsilon_1 \equiv -\frac{\dot H}{H^2},
\qquad
\epsilon_{i>1}\equiv \frac{d\ln|\epsilon_{i-1}|}{dN},
\end{equation}
where $N\equiv\ln a$ is the number of e-folds. Inflation requires $\epsilon_1<1$, while slow roll corresponds to $\epsilon_1\ll1$ and to a slowly varying hierarchy of higher-order flow parameters. Since the background evolution is driven by the motion of the scalar field along $V(\phi)$, the potential slow-roll and Hubble-flow parameters provide two complementary descriptions of the same inflationary dynamics, with $\epsilon_1\simeq\epsilon_V$ at leading order~\cite{Liddle:1994dx,Lidsey:1995np,Schwarz:2001vv,Leach:2002ar}.

The scalar and tensor spectral parameters can then be expressed in terms of either hierarchy. For the scalar sector, at leading non-trivial order one obtains
\begin{equation}
\begin{aligned}
n_s-1
&\simeq 2\eta_V-6\epsilon_V
\simeq -2\epsilon_1-\epsilon_2 ,
\label{eq:ns_SR}
\end{aligned}
\end{equation}
\begin{equation}
\begin{aligned}
\alpha_s
&\simeq 16\epsilon_V\eta_V-24\epsilon_V^2-2\xi_V^2
\simeq -2\epsilon_1\epsilon_2-\epsilon_2\epsilon_3 ,
\end{aligned}
\end{equation}
\begin{equation}
\begin{aligned}
\beta_s
&\simeq
-192\epsilon_V^3
+192\epsilon_V^2\eta_V
-32\epsilon_V\eta_V^2
-24\epsilon_V\xi_V^2
+2\eta_V\xi_V^2
+2\varpi_V^3 \\
&\simeq
-2\epsilon_1\epsilon_2^2
-2\epsilon_1\epsilon_2\epsilon_3
-\epsilon_2\epsilon_3^2
-\epsilon_2\epsilon_3\epsilon_4 .
\end{aligned}
\end{equation}
Similarly, for the tensor sector,
\begin{equation}
\begin{aligned}
r
&\simeq 16\epsilon_V
\simeq 16\epsilon_1 ,
\end{aligned}
\end{equation}
\begin{equation}
\begin{aligned}
n_T
&\simeq -2\epsilon_V
\simeq -2\epsilon_1 ,
\end{aligned}
\end{equation}
\begin{equation}
\begin{aligned}
\alpha_T
&\simeq 4\epsilon_V\eta_V-8\epsilon_V^2
\simeq -2\epsilon_1\epsilon_2 ,
\end{aligned}
\end{equation}
\begin{equation}
\begin{aligned}
\beta_T
&\simeq
-32\epsilon_V^3
+56\epsilon_V^2\eta_V
-8\epsilon_V\eta_V^2
-4\epsilon_V\xi_V^2 \\
&\simeq
-2\epsilon_1\epsilon_2(\epsilon_2+\epsilon_3).
\end{aligned}
\end{equation}

The relations above define two closed systems of algebraic equations: one connecting $\{n_s,\alpha_s,\beta_s,r\}$ to the potential slow-roll parameters $\{\epsilon_V,\eta_V,\xi_V^2,\varpi_V^3\}$, and one connecting the same spectral quantities to the Hubble-flow hierarchy $\{\epsilon_1,\epsilon_2,\epsilon_3,\epsilon_4\}$. We invert these systems to express the slow-roll variables as derived quantities from the sampled spectral parameters. Within each extended model and for each dataset combination, this allows us to translate the posterior constraints on the primordial spectra into constraints on the underlying inflationary dynamics. The resulting constraints are summarized in Table~\ref{tab:results.slow-roll}.

\begin{table*}[htpb!]
\centering
\renewcommand{\arraystretch}{1}
\resizebox{\textwidth}{!}{
\begin{tabular}{l @{\hspace{0.5 cm}} l @{\hspace{0.5 cm}}c@{\hspace{0.5 cm}}c@{\hspace{0.5 cm}}c@{\hspace{0.5 cm}}cccccc}
\hline\hline
\\
\textbf{Slow-Roll Param.} & \textbf{Model} & \textbf{CMB+DESI+PP} & \textbf{CMB+DESI+DD} \\
\\
\hline
&&&&&\\

$\epsilon_V\simeq \epsilon_1$  
&  $\Lambda\mathrm{CDM}+r+\alpha_s+\beta_s$ &  $<0.00219$ &  $<0.00214$ \\[1ex]
&  $\Lambda\mathrm{CDM}+r+\alpha_s+\beta_s+\Omega_k+\sum m_{\nu}+N_{\rm eff}$   &  $<0.00226$ &  $<0.00219$\\[1ex]
&  $w_0w_a\mathrm{CDM}+r+\alpha_s+\beta_s$  &  $<0.00218$ &  $<0.00220$\\[1ex]
&  $w_0w_a\mathrm{CDM}+r+\alpha_s+\beta_s+\Omega_k+\sum m_{\nu}+N_{\rm eff}$   &  $<0.00222$ &  $<0.00223$\\[1ex]

\hline\\[1ex]

$\eta_V$  
&  $\Lambda\mathrm{CDM}+r+\alpha_s+\beta_s$ &  $-0.0118^{+0.0025}_{-0.0029}$ &  $-0.0118^{+0.0024}_{-0.0028}$ \\[1ex]
&  $\Lambda\mathrm{CDM}+r+\alpha_s+\beta_s+\Omega_k+\sum m_{\nu}+N_{\rm eff}$   &  $-0.0171\pm 0.0047$ &  $-0.0175\pm 0.0047$\\[1ex]
&  $w_0w_a\mathrm{CDM}+r+\alpha_s+\beta_s$  &  $-0.0129^{+0.0025}_{-0.0029}$ &  $-0.0130^{+0.0025}_{-0.0029}$\\[1ex]
&  $w_0w_a\mathrm{CDM}+r+\alpha_s+\beta_s+\Omega_k+\sum m_{\nu}+N_{\rm eff}$   &  $-0.0173 \pm 0.0050$ &  $-0.0179 \pm 0.0050$\\[1ex]

\hline\\[1ex]

$\xi_V^2$  
&  $\Lambda\mathrm{CDM}+r+\alpha_s+\beta_s$ &  $-0.0040 \pm 0.0028$ &  $-0.0038 \pm 0.0028$ \\[1ex]
&  $\Lambda\mathrm{CDM}+r+\alpha_s+\beta_s+\Omega_k+\sum m_{\nu}+N_{\rm eff}$   &  $-0.0024\pm 0.0042$ &  $-0.0023 \pm 0.0043$\\[1ex]
&  $w_0w_a\mathrm{CDM}+r+\alpha_s+\beta_s$  &  $-0.0042 \pm 0.0028$ &  $-0.0042 \pm 0.0028$\\[1ex]
&  $w_0w_a\mathrm{CDM}+r+\alpha_s+\beta_s+\Omega_k+\sum m_{\nu}+N_{\rm eff}$   &  $-0.0018 \pm 0.0045$ &  $-0.0014 \pm 0.0045$\\[1ex]

\hline\\[1ex]

$\varpi_V^3$  
&  $\Lambda\mathrm{CDM}+r+\alpha_s+\beta_s$ &  $0.0061 \pm 0.0042$ &  $0.0061 \pm 0.0043$ \\[1ex]
&  $\Lambda\mathrm{CDM}+r+\alpha_s+\beta_s+\Omega_k+\sum m_{\nu}+N_{\rm eff}$   &  $0.0072\pm 0.0047$ &  $0.0071\pm 0.0048$\\[1ex]
&  $w_0w_a\mathrm{CDM}+r+\alpha_s+\beta_s$  &  $0.0061 \pm 0.0043$ &  $0.0061 \pm 0.0043$\\[1ex]
&  $w_0w_a\mathrm{CDM}+r+\alpha_s+\beta_s+\Omega_k+\sum m_{\nu}+N_{\rm eff}$   &  $0.0060 \pm 0.0048$ &  $0.0059 \pm 0.0050$\\[1ex]

\hline\\[1ex]

$\epsilon_2$  
&  $\Lambda\mathrm{CDM}+r+\alpha_s+\beta_s$ &  $0.0277 \pm 0.0039$ &  $0.0277 \pm 0.0039$ \\[1ex]
&  $\Lambda\mathrm{CDM}+r+\alpha_s+\beta_s+\Omega_k+\sum m_{\nu}+N_{\rm eff}$   &  $0.0384 \pm 0.0088$ &  $0.0392\pm 0.0088$\\[1ex]
&  $w_0w_a\mathrm{CDM}+r+\alpha_s+\beta_s$  &  $0.0299 \pm 0.0041$ &  $0.0300 \pm 0.0041$\\[1ex]
& $w_0w_a\mathrm{CDM}+r+\alpha_s+\beta_s+\Omega_k+\sum m_{\nu}+N_{\rm eff}$   &  $0.0388 \pm 0.0094$ &  $0.0399 \pm 0.0094$\\[1ex]

\hline\\[1ex]

$\epsilon_3$  
&  $\Lambda\mathrm{CDM}+r+\alpha_s+\beta_s$ &  $-0.28 \pm 0.20$ &  $-0.27 \pm 0.20$ \\[1ex]
&  $\Lambda\mathrm{CDM}+r+\alpha_s+\beta_s+\Omega_k+\sum m_{\nu}+N_{\rm eff}$   &  $-0.16^{+0.30}_{-0.17}$ &  $-0.15^{+0.29}_{-0.17}$\\[1ex]
&  $w_0w_a\mathrm{CDM}+r+\alpha_s+\beta_s$  &  $-0.27 \pm 0.19$ &  $-0.27 \pm 0.19$\\[1ex]
&  $w_0w_a\mathrm{CDM}+r+\alpha_s+\beta_s+\Omega_k+\sum m_{\nu}+N_{\rm eff}$   &  $-0.13^{+0.32}_{-0.17}$ &  $-0.11^{+0.30}_{-0.16}$\\[1ex]

\hline\\[1ex]

$V_{\rm inf}^{1/4}$  
& $\Lambda\mathrm{CDM}+r+\alpha_s+\beta_s$ & $<1.49\cdot 10^{16}$ GeV & $<1.48\cdot 10^{16}$ GeV \\[1ex]
&  $\Lambda\mathrm{CDM}+r+\alpha_s+\beta_s+\Omega_k+\sum m_{\nu}+N_{\rm eff}$   &  $< 1.50\cdot 10^{16}$ GeV &  $< 1.49\cdot 10^{16}$ GeV \\[1ex]
&  $w_0w_a\mathrm{CDM}+r+\alpha_s+\beta_s$  &  $< 1.48\cdot 10^{16}$ GeV &  $< 1.48\cdot 10^{16}$ GeV \\[1ex]
&  $w_0w_a\mathrm{CDM}+r+\alpha_s+\beta_s+\Omega_k+\sum m_{\nu}+N_{\rm eff}$   &  $< 1.48\cdot 10^{16}$ GeV &  $< 1.49\cdot 10^{16}$ GeV \\[1ex]
\hline
\bottomrule
\end{tabular}
}
\caption{Constraints on the potential slow-roll parameters, Hubble-flow parameters, and inflationary energy scale derived within different extended cosmological models, using the CMB+DESI+PP and CMB+DESI+DD dataset combinations. Upper limits on $\epsilon_V\simeq\epsilon_1$ and $V_{\rm inf}^{1/4}$ are quoted at 95\% CL, while all other constraints are given at 68\% CL. The constraints on $\epsilon_4$ are poorly determined in all cases and are therefore not reported.}
\label{tab:results.slow-roll}
\end{table*}

The behaviour of the slow-roll parameters mirrors that of the spectral parameters. The quantities controlled directly by the tensor amplitude remain stable across the full set of models. In particular, the bound on the first slow-roll parameter remains at the level $\epsilon_V\simeq\epsilon_1\lesssim 2.3\times10^{-3}$, with no significant dependence on the overall cosmological model. The same stability propagates to the inflationary energy scale, which we estimate from~\cite{Baumann:2009ds}
\begin{equation}
V_{\rm inf}
\simeq
\frac{3M_{\rm pl}^4\pi^2}{2}\,
A_s\,r.
\end{equation}
We find $V_{\rm inf}^{1/4}\lesssim 1.48-1.50\times10^{16}\,{\rm GeV}$ across all the models considered. This stability is expected, since the estimate depends primarily on the upper bound on $r$, with only a mild residual dependence on the scalar amplitude $A_s$. 

The main model dependence appears instead in the parameters that control the scalar tilt. In the potential slow-roll hierarchy, $\eta_V$ moves from $\eta_V\simeq -0.012$ in the models with $r+\alpha_s+\beta_s$ to $\eta_V\simeq -0.017$ once $\Omega_k$, $\sum m_\nu$, and $N_{\rm eff}$ are also varied. This follows directly from Eq.~\eqref{eq:ns_SR}. Since $\epsilon_V$ is small and stable, the downward shift of $n_s$ in the extended models is mainly translated into a more negative $\eta_V$. The same effect is visible in the Hubble-flow hierarchy, where $\epsilon_2$ increases from $\epsilon_2\simeq 0.028$ to $\epsilon_2\simeq 0.040$.

The higher-order slow-roll parameters are less tightly constrained and reflect the freedom associated with the running parameters. The quantity $\xi_V^2$ is mildly negative in the baseline extended-primordial models, around $\xi_V^2\simeq -0.004$, but moves closer to zero and becomes less constrained once $\Omega_k$, $\sum m_\nu$, and $N_{\rm eff}$ are included. This is consistent with the absence of a significant detection of $\alpha_s$. By contrast, $\varpi_V^3$ has positive central values, around $\varpi_V^3\simeq 0.006$, tracking the positive central values of $\beta_s$, although the uncertainties remain large. Similarly, $\epsilon_3$ has negative central values in the baseline cases, while it shifts closer to zero in the most extended models, with substantially enlarged uncertainties.

Overall, extended cosmologies do not significantly affect the first slow-roll parameter or the inflationary energy scale. Their main impact is instead to shift the inferred curvature of the inflationary potential, encoded in $\eta_V$, or equivalently the second Hubble-flow parameter $\epsilon_2$, as well as to broaden the constraints on the higher-order derivatives of the slow-roll hierarchy.

\subsection{Inflationary Model Consistency in Extended Cosmologies}

\begin{figure}[t]
    \centering    \includegraphics[width=0.8\linewidth]{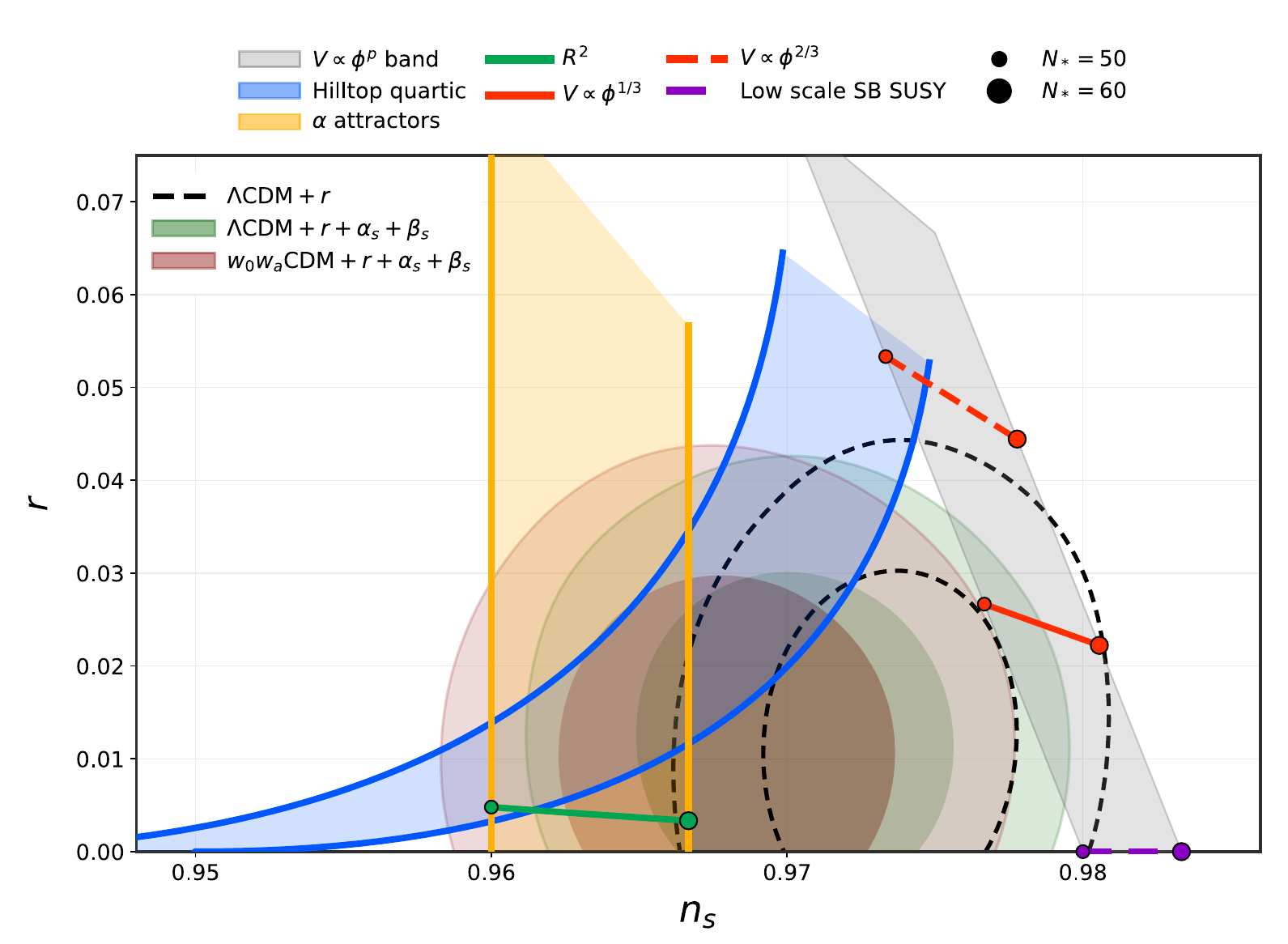}
    \caption{
    Marginalized constraints in the $n_s$--$r$ plane obtained from CMB+DESI+DD, compared with representative predictions from benchmark single-field inflationary models. The black dashed contours correspond to $\Lambda$CDM+$r$, while the green and red contours show $\Lambda$CDM+$r+\alpha_s+\beta_s$ and $w_0w_a$CDM+$r+\alpha_s+\beta_s$, respectively. Contours denote the 68\% and 95\% CL regions. Theoretical predictions are shown for $N_\star=50$--$60$ e-folds before the end of inflation, including plateau-like models such as $R^2$ inflation~\cite{Starobinsky:1980te} and $\alpha$-attractors~\cite{Kallosh:2013yoa}, hilltop quartic models~\cite{Boubekeur:2005zm}, shallow power-law potentials, the broader $V\propto\phi^p$ band~\cite{Linde:1983gd}, and the low-scale SB SUSY prediction~\cite{Dvali:1994ms}.
    }
    \label{fig:ns_r}
\end{figure}

As a final diagnostic, we show in Fig.~\ref{fig:ns_r} the marginalized constraints in the $n_s$--$r$ plane together with representative predictions from standard single-field inflationary models.

The figure makes explicit that the upper limits on $r$ are rather stable, while the inferred position of the allowed region along the $n_s$ axis depends appreciably on the parametrization adopted for the primordial spectra, as well as on the overall cosmological framework.

In the minimal $\Lambda$CDM+$r$ case, shown by the black dashed probability regions, the posterior favours a relatively large value, $n_s\simeq 0.975$. Assuming $N_\star\in[50,60]$ e-folds before the end of inflation~\cite{Liddle:2003as,Dodelson:2003vq,Martin:2010kz,Dai:2014jja}, plateau-like models such as $R^2$ inflation~\cite{Starobinsky:1980te,Mukhanov:1981xt,Whitt:1984pd} and $\alpha$-attractors~\cite{Kallosh:2013yoa,Kallosh:2013hoa,Kallosh:2013tua,Galante:2014ifa} typically predict lower scalar tilts, $n_s\sim 1-2/N_\star\sim 0.96$--$0.966$, and therefore lie close to the edge of, or outside, the allowed 95\% CL region. The lower-$n_s$ part of the hilltop quartic region~\cite{Boubekeur:2005zm} is affected in the same direction, although a sizable overlap remains because the additional scale $\mu$ broadens the prediction, allowing the model to interpolate between very small-$r$, redder hilltop solutions and larger-$n_s$ trajectories as the horizon-exit point moves away from the maximum. By contrast, shallower power-law potentials, such as $V\propto\phi^{1/3}$, shown in red, and, more broadly, $V\propto\phi^p$~\cite{Linde:1983gd,Silverstein:2008sg,McAllister:2008hb}, shown by the grey band, are more naturally aligned with the high-$n_s$ region preferred in the minimal $\Lambda$CDM+$r$ fit. The latter is also marginally consistent with the low-scale SB SUSY prediction~\cite{Dvali:1994ms,Copeland:1994vg}.

In $\Lambda$CDM+$r+\alpha_s+\beta_s$, shown by the green probability contours, part of the scale dependence that in the minimal parametrization is effectively projected onto a larger value of $n_s$ is instead absorbed by $\alpha_s$ and $\beta_s$. The marginalized contours therefore move towards lower $n_s$, improving the consistency with plateau-like models of inflation. This brings $R^2$ inflation and $\alpha$-attractors back within the allowed contours while maintaining overlap with the hilltop quartic region. The contours still extend marginally towards the region occupied by shallow power-law models such as $V\propto\phi^{1/3}$, mainly because of the broader uncertainties. However, the agreement with these models becomes less pronounced once running and running of the running are included in the primordial parametrization.

The same conclusion holds for the $w_0w_a$CDM+$r+\alpha_s+\beta_s$ extension. This case, shown by the red probability contours, produces a similar allowed region, but with $n_s$ shifted slightly further towards lower values. This further strengthens the consistency with plateau-like models, while reducing the overlap with monomial and shallow power-law potentials, which now lie closer to the edge of, or outside, the probability contours.

Although the enlarged primordial parametrization allows for smaller values of $n_s$, the posterior volume in the running sector is mostly displaced towards $\alpha_s>0$ and $\beta_s>0$. This displacement is mild, with both parameters remaining consistent with zero within two standard deviations. Therefore, in the $\alpha_s$--$\beta_s$ plane, the inflationary models considered here would remain within the allowed regions for the different cosmological cases.

Overall, the preference for larger values of $n_s$ found when combining the CMB experiments, compared to Planck-only results, should not be over-interpreted as a model-independent statement. The same caution applies to the discussion on the apparent consistency, or lack thereof, of benchmark inflationary potentials in the $n_s$--$r$ plane. This comparison is subject to important caveats related to the parametrization adopted for the primordial spectra and, more broadly, to the cosmological model.

\clearpage
\section{Cosmological Tensions}
\label{sec.tensions}

As the precision of cosmological observations has continued to improve, several discrepancies and tensions have become increasingly apparent. The best-known example is the Hubble tension~\cite{Verde:2019ivm,DiValentino:2020zio,DiValentino:2021izs,Perivolaropoulos:2021jda,Schoneberg:2021qvd,Shah:2021onj,Abdalla:2022yfr,DiValentino:2022fjm,Kamionkowski:2022pkx,Giare:2023xoc,Hu:2023jqc,Verde:2023lmm,DiValentino:2024yew,Efstathiou:2024dvn,CosmoVerseNetwork:2025alb}. Its statistical significance has now reached at least the conservative $5\sigma$ level when comparing the leading early- and late-time determinations from Planck~\cite{Planck:2018vyg} and SH0ES~\cite{Riess:2021jrx}. However, it can rise to about $7\sigma$ when the combined CMB-based inference is compared with the covariance-weighted local determination from the H0DN collaboration~\cite{H0DN:2025lyy}.

Other long-standing discussions concern the parameter $\sigma_8$, which quantifies the amplitude of linear matter fluctuations smoothed on scales of $8\,h^{-1}\,\mathrm{Mpc}$, and the derived quantity $S_8 \equiv \sigma_8 (\Omega_m/0.3)^{0.5}$. Historically, weak-lensing surveys have often preferred values of $S_8$ lower than those inferred from CMB data within $\Lambda$CDM, giving rise to the so-called $S_8$ tension~\cite{DiValentino:2020vvd,DiValentino:2018gcu,Nunes:2021ipq,DES:2021bvc,DES:2021vln,KiDS:2020suj,Asgari:2019fkq,Joudaki:2019pmv,DAmico:2019fhj,Kilo-DegreeSurvey:2023gfr,Troster:2019ean,Heymans:2020gsg,Dalal:2023olq,Chen:2024vvk,ACT:2024okh,DES:2024oud,Harnois-Deraps:2024ucb,Dvornik:2022xap,DES:2021wwk,Wright:2025xka,DES:2026fyc}. One of the most discussed cases was the KiDS-1000 cosmic-shear analysis, which reported a discrepancy with Planck at the level of about $3.4\sigma$~\cite{KiDS:2020suj}. More recently, the KiDS-Legacy reanalysis found substantially higher values of $S_8$, in good agreement with Planck~\cite{Wright:2025xka}. The shift was mainly driven by improved redshift-distribution estimation and calibration, together with updated treatments of astrophysical systematics such as intrinsic alignments and baryon feedback~\cite{Wright:2025xka,Stolzner:2025htz}. At the same time, however, the recent Dark Energy Survey Year 6 (DES-Y6) analysis still finds a residual mismatch with CMB constraints, corresponding to about $2.6\sigma$ when projected onto $S_8$, and up to $2.8\sigma$ in the full parameter space~\cite{DES:2026fyc}.

Last but not least, as already discussed at several points in the previous sections, following the DESI results, other discrepancies have also come to light. In particular, within a spatially flat $\Lambda$CDM cosmology, the transverse and radial BAO distances measured by DESI depend exclusively on the combination $r_d H_0$ and on $\Omega_m$. DESI tends to favour larger values of the former and lower values of the latter, thereby pointing to a region of parameter space that is shifted with respect to that preferred by CMB data at $\sim 2.3\sigma$~\cite{DESI:2025zgx,DESI:2025fii}.

In what follows, we do not attempt a comprehensive review or resolution of these anomalies, but rather investigate how the parameters characterizing them behave when inferred in extended cosmological parameter spaces.

\begin{table*}[htp!]
\centering
\renewcommand{\arraystretch}{1}
\resizebox{\textwidth}{!}{
\begin{tabular}{l @{\hspace{0.5 cm}} l @{\hspace{0.5 cm}} c @{\hspace{0.5 cm}} c @{\hspace{0.5 cm}} c @{\hspace{0.5 cm}} c @{\hspace{0.5 cm}} c}
\hline\hline
\\[-1.5ex]
\textbf{Model} & \textbf{Dataset} & \boldmath{$H_0$} \textbf{[km/s/Mpc]}& \boldmath{$\Omega_m$} & \boldmath{$\sigma_8$} & \boldmath{$S_8$} & \boldmath{$r_\mathrm{d}$} \textbf{[Mpc]} \\
\\[-0.5ex]
\hline
& & & & & & \\

$\Lambda\mathrm{CDM}$
& CMB+DESI+PP & $68.17\pm 0.25$ & $0.3040\pm 0.0033$ & $0.8182\pm 0.0042$ & $0.8236\pm 0.0062$ & $147.48\pm 0.17$ \\[1ex]
& CMB+DESI+DD & $68.13\pm 0.25$ & $0.3044\pm 0.0034$ & $0.8183\pm 0.0042$ & $0.8243\pm 0.0061$ & $147.47\pm 0.17$ \\[1.5ex]

$\Lambda\mathrm{CDM}+\sum m_{\nu}$
& CMB+DESI+PP & $68.35^{+0.28}_{-0.24}$ & $0.3021^{+0.0032}_{-0.0036}$ & $0.8241^{+0.0053}_{-0.0046}$ & $0.8270\pm 0.0064$ & $147.41\pm 0.18$ \\[1ex]
& CMB+DESI+DD & $68.33\pm 0.26$ & $0.3024\pm 0.0034$ & $0.8241^{+0.0053}_{-0.0046}$ & $0.8274\pm 0.0064$ & $147.41\pm 0.18$ \\[1.5ex]

$\Lambda\mathrm{CDM}+N_{\rm eff}$
& CMB+DESI+PP & $67.91\pm 0.68$ & $0.3048\pm 0.0037$ & $0.8165\pm 0.0063$ & $0.8230\pm 0.0066$ & $147.9\pm 1.1$ \\[1ex]
& CMB+DESI+DD & $67.85\pm 0.68$ & $0.3052\pm 0.0038$ & $0.8161\pm 0.0063$ & $0.8231\pm 0.0067$ & $148.0\pm 1.1$ \\[1.5ex]

$\Lambda\mathrm{CDM}+\sum m_{\nu}+N_{\rm eff}$
& CMB+DESI+PP & $67.89\pm 0.68$ & $0.3033\pm 0.0038$ & $0.8209\pm 0.0065$ & $0.8254\pm 0.0067$ & $148.2\pm 1.1$ \\[1ex]
& CMB+DESI+DD & $67.86\pm 0.68$ & $0.3037\pm 0.0038$ & $0.8207\pm 0.0067$ & $0.8257\pm 0.0067$ & $148.2\pm 1.1$ \\[1.5ex]

$\Lambda\mathrm{CDM}+\Omega_k$
& CMB+DESI+PP & $68.57\pm 0.32$ & $0.3043\pm 0.0033$ & $0.8230\pm 0.0047$ & $0.8288\pm 0.0066$ & $147.05\pm 0.26$ \\[1ex]
& CMB+DESI+DD & $68.56\pm 0.31$ & $0.3045\pm 0.0033$ & $0.8231\pm 0.0047$ & $0.8293\pm 0.0066$ & $147.04\pm 0.26$ \\[1.5ex]

$\Lambda\mathrm{CDM}+r$
& CMB+DESI+PP & $68.17\pm 0.25$ & $0.3039\pm 0.0033$ & $0.8179\pm 0.0041$ & $0.8231\pm 0.0061$ & $147.49\pm 0.17$ \\[1ex]
& CMB+DESI+DD & $68.14\pm 0.24$ & $0.3043\pm 0.0033$ & $0.8178\pm 0.0042$ & $0.8237\pm 0.0063$ & $147.47\pm 0.18$ \\[1.5ex]

$\Lambda\mathrm{CDM}+r+\alpha_s$
& CMB+DESI+PP & $68.14\pm 0.25$ & $0.3041\pm 0.0033$ & $0.8176\pm 0.0041$ & $0.8232\pm 0.0062$ & $147.52\pm 0.18$ \\[1ex]
& CMB+DESI+DD & $68.12\pm 0.25$ & $0.3044\pm 0.0034$ & $0.8178\pm 0.0040$ & $0.8237\pm 0.0062$ & $147.51\pm 0.18$ \\[1.5ex]

$\Lambda\mathrm{CDM}+r+\alpha_s+\beta_s$
& CMB+DESI+PP & $68.10\pm 0.25$ & $0.3046\pm 0.0034$ & $0.8201^{+0.0043}_{-0.0048}$ & $0.8264\pm 0.0066$ & $147.53\pm 0.18$ \\[1ex]
& CMB+DESI+DD & $68.09\pm 0.25$ & $0.3047\pm 0.0033$ & $0.8201\pm 0.0046$ & $0.8265\pm 0.0067$ & $147.52\pm 0.18$ \\[1.5ex]

$\Lambda\mathrm{CDM}+r+\alpha_s+\beta_s+\Omega_k+\sum m_{\nu}+N_{\rm eff}$
& CMB+DESI+PP & $68.01\pm 0.96$ & $0.3059\pm 0.0047$ & $0.8238\pm 0.0093$ & $0.8318\pm 0.0080$ & $148.2\pm 1.7$ \\[1ex]
& CMB+DESI+DD & $67.94\pm 0.94$ & $0.3065^{+0.0043}_{-0.0048}$ & $0.8236\pm 0.0092$ & $0.8324\pm 0.0080$ & $148.3\pm 1.7$ \\[1.5ex]

$w_0w_a\mathrm{CDM}$
& CMB+DESI+PP & $67.63\pm 0.61$ & $0.3109\pm 0.0058$ & $0.8187\pm 0.0073$ & $0.8334\pm 0.0069$ & $147.26\pm 0.20$ \\[1ex]
& CMB+DESI+DD & $67.44\pm 0.54$ & $0.3129\pm 0.0053$ & $0.8176\pm 0.0070$ & $0.8350\pm 0.0069$ & $147.23\pm 0.20$ \\[1.5ex]

$w_0w_a\mathrm{CDM}+\sum m_{\nu}$
& CMB+DESI+PP & $67.65\pm 0.60$ & $0.3104\pm 0.0057$ & $0.8200\pm 0.0081$ & $0.8340\pm 0.0071$ & $147.26\pm 0.20$ \\[1ex]
& CMB+DESI+DD & $67.44\pm 0.54$ & $0.3127\pm 0.0054$ & $0.8183\pm 0.0078$ & $0.8353\pm 0.0072$ & $147.24\pm 0.20$ \\[1.5ex]

$w_0w_a\mathrm{CDM}+N_{\rm eff}$
& CMB+DESI+PP & $66.94\pm 0.81$ & $0.3122\pm 0.0058$ & $0.8150\pm 0.0078$ & $0.8313\pm 0.0070$ & $148.8\pm 1.2$ \\[1ex]
& CMB+DESI+DD & $66.65\pm 0.78$ & $0.3146\pm 0.0054$ & $0.8133\pm 0.0075$ & $0.8327\pm 0.0070$ & $148.9\pm 1.2$ \\[1.5ex]

$w_0w_a\mathrm{CDM}+\sum m_{\nu}+N_{\rm eff}$
& CMB+DESI+PP & $66.92\pm 0.81$ & $0.3117\pm 0.0059$ & $0.8166\pm 0.0083$ & $0.8323\pm 0.0071$ & $148.8\pm 1.2$ \\[1ex]
& CMB+DESI+DD & $66.66\pm 0.77$ & $0.3142\pm 0.0056$ & $0.8147\pm 0.0083$ & $0.8337\pm 0.0072$ & $148.9\pm 1.2$ \\[1.5ex]

$w_0w_a\mathrm{CDM}+\Omega_k$
& CMB+DESI+PP & $67.76\pm 0.61$ & $0.3111\pm 0.0056$ & $0.8194\pm 0.0074$ & $0.8343\pm 0.0070$ & $147.11\pm 0.26$ \\[1ex]
& CMB+DESI+DD & $67.55\pm 0.58$ & $0.3129\pm 0.0054$ & $0.8180\pm 0.0071$ & $0.8353\pm 0.0070$ & $147.12\pm 0.26$ \\[1.5ex]

$w_0w_a\mathrm{CDM}+r$
& CMB+DESI+PP & $67.65\pm 0.61$ & $0.3107\pm 0.0057$ & $0.8184\pm 0.0074$ & $0.8328\pm 0.0068$ & $147.27\pm 0.20$ \\[1ex]
& CMB+DESI+DD & $67.45\pm 0.54$ & $0.3127\pm 0.0052$ & $0.8172\pm 0.0070$ & $0.8342\pm 0.0070$ & $147.25\pm 0.21$ \\[1.5ex]

$w_0w_a\mathrm{CDM}+r+\alpha_s$
& CMB+DESI+PP & $67.63\pm 0.59$ & $0.3108\pm 0.0056$ & $0.8183\pm 0.0073$ & $0.8328\pm 0.0068$ & $147.31\pm 0.20$ \\[1ex]
& CMB+DESI+DD & $67.44\pm 0.55$ & $0.3127\pm 0.0053$ & $0.8171\pm 0.0070$ & $0.8342\pm 0.0068$ & $147.29\pm 0.20$ \\[1.5ex]

$w_0w_a\mathrm{CDM}+r+\alpha_s+\beta_s$
& CMB+DESI+PP & $67.59\pm 0.60$ & $0.3113\pm 0.0057$ & $0.8208\pm 0.0075$ & $0.8360\pm 0.0072$ & $147.32\pm 0.21$ \\[1ex]
& CMB+DESI+DD & $67.42\pm 0.54$ & $0.3131\pm 0.0053$ & $0.8201\pm 0.0070$ & $0.8377\pm 0.0072$ & $147.29\pm 0.21$ \\[1.5ex]

$w_0w_a\mathrm{CDM}+r+\alpha_s+\beta_s+\Omega_k+\sum m_{\nu}+N_{\rm eff}$
& CMB+DESI+PP & $67.1\pm 1.0$ & $0.3135\pm 0.0061$ & $0.817\pm 0.010$ & $0.8351\pm 0.0082$ & $148.5\pm 1.9$ \\[1ex]
& CMB+DESI+DD & $66.8\pm 1.0$ & $0.3155\pm 0.0059$ & $0.8148\pm 0.0099$ & $0.8355\pm 0.0084$ & $148.7\pm 1.9$ \\[1.5ex]
\hline
\bottomrule
\end{tabular}
}
\caption{Summary of the marginalized constraints on $H_0$, $\Omega_m$, $\sigma_8$, $S_8$, and $r_\mathrm{d}$ for the cosmological models considered in the tension analysis, using the CMB+DESI+PP and CMB+DESI+DD dataset combinations. Quoted constraints correspond to 68\% CL.}
\label{tab:results.tension.corrected}
\end{table*}

\subsection{$\Omega_m$ in Extended Cosmological Models} 

\begin{figure}[ht!]
    \centering \includegraphics[width=\linewidth]{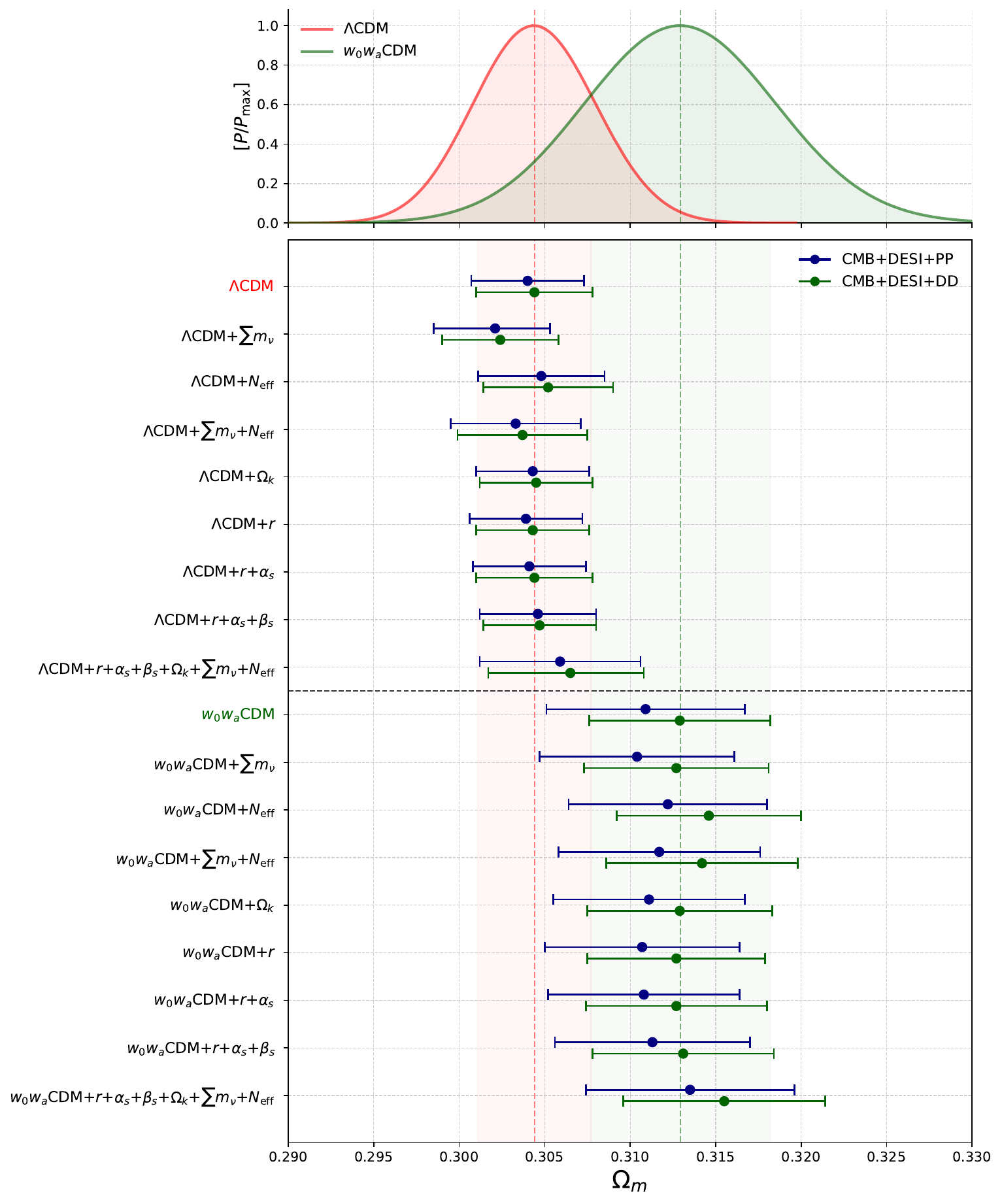}
    \caption{Marginalized constraints on $\Omega_m$ for the cosmological models considered in the analysis. The lower panel reports the 68\% CL constraints from the CMB+DESI+PP and CMB+DESI+DD dataset combinations for all model extensions listed in Table~\ref{tab:results.tension.corrected}. The upper panel shows the corresponding one-dimensional posteriors for the baseline $\Lambda$CDM and $w_0w_a$CDM cases.}
    \label{fig:whisker_Omegam}
\end{figure}

The constraints on $\Omega_m$ are summarized in Table~\ref{tab:results.tension.corrected} and in Fig.~\ref{fig:whisker_Omegam}. The inferred values of $\Omega_m$ remain remarkably stable within each of the two broad late-time model classes considered in this work, namely $\Lambda$CDM-like and $w_0w_a$CDM-like cosmologies. The dominant shift appears when moving from one class to the other, while relaxing additional sectors such as curvature, neutrinos, and inflationary parameters induces only comparatively minor variations within each macro-class. Similarly, the choice of SN sample (namely PP or DD) has only a comparatively minor impact on the inferred matter density.

All $\Lambda$CDM-based extensions cluster in a relatively narrow region around $\Omega_m \simeq 0.303$--$0.306$, whereas the $w_0w_a$CDM-based extensions are systematically shifted toward higher values, around $\Omega_m \simeq 0.311$--$0.316$. Fig.~\ref{fig:whisker_Omegam} makes this separation particularly clear.

A similar conclusion follows from the behaviour of the uncertainties. Within the $\Lambda$CDM branch, $\Omega_m$ is typically measured with an absolute uncertainty of order $\Delta\Omega_m \sim 0.0033$--$0.0038$, corresponding to a precision at the level of roughly $1.1\%$--$1.3\%$ and degrading only to about $1.5\%$ in the most extended case. Within the $w_0w_a$CDM branch, the uncertainty broadens to $\Delta\Omega_m \sim 0.0052$--$0.0059$, corresponding to a typical precision of about $1.7\%$--$1.9\%$, and reaches about $0.006$ only in the fully extended model. Thus, the shift from $\Lambda$CDM to $w_0w_a$CDM is accompanied by a moderate loss of constraining power. This widening, however, is not the dominant effect underlying the coherent displacement in the central value, which identifies the late-time DE parametrization as the primary source of the inferred shift in $\Omega_m$.


\subsection{$H_0$ in Extended Cosmological Models}

The constraints on $H_0$ reported in Table~\ref{tab:results.tension.corrected} show that, across all model extensions considered in this work, the inferred Hubble constant never shifts significantly toward values of order $H_0\sim 73$ km/s/Mpc, as suggested by SH0ES and other local determinations. Instead, the posterior remains centered close to the standard CMB $\Lambda$CDM value, $H_0\sim 68$ km/s/Mpc.

Within the minimally extended $\Lambda$CDM-like cases, $H_0$ is measured with an uncertainty of $\sim 0.24$--$0.32$ km/s/Mpc, corresponding to a precision of about $0.4\%$. The uncertainties degrade to $\sim 0.7$ km/s/Mpc ($\sim 1\%$) when $N_{\rm eff}$ is allowed to vary and to $\sim 0.95$ km/s/Mpc ($1.4\%$) in the fully extended $\Lambda$CDM-based model. A similar pattern is found in the $w_0w_a$CDM-like branch. The typical uncertainty is already larger ($\sim 0.6$ km/s/Mpc, i.e., $0.85\%$) and reaches $\sim 1$ km/s/Mpc ($1.5\%$) in the fully extended case.

While the error bar on $H_0$ does broaden when the parameter space is enlarged, no extension shifts the central value in the direction required to reconcile early- and late-time determinations. On the contrary, in the most extended scenarios the broadening is accompanied by a displacement of the central value toward slightly lower values. This indicates that none of the extensions explored here provides a genuine route toward resolving the Hubble tension.

\begin{figure}[t!]
    \centering \includegraphics[width=\linewidth]{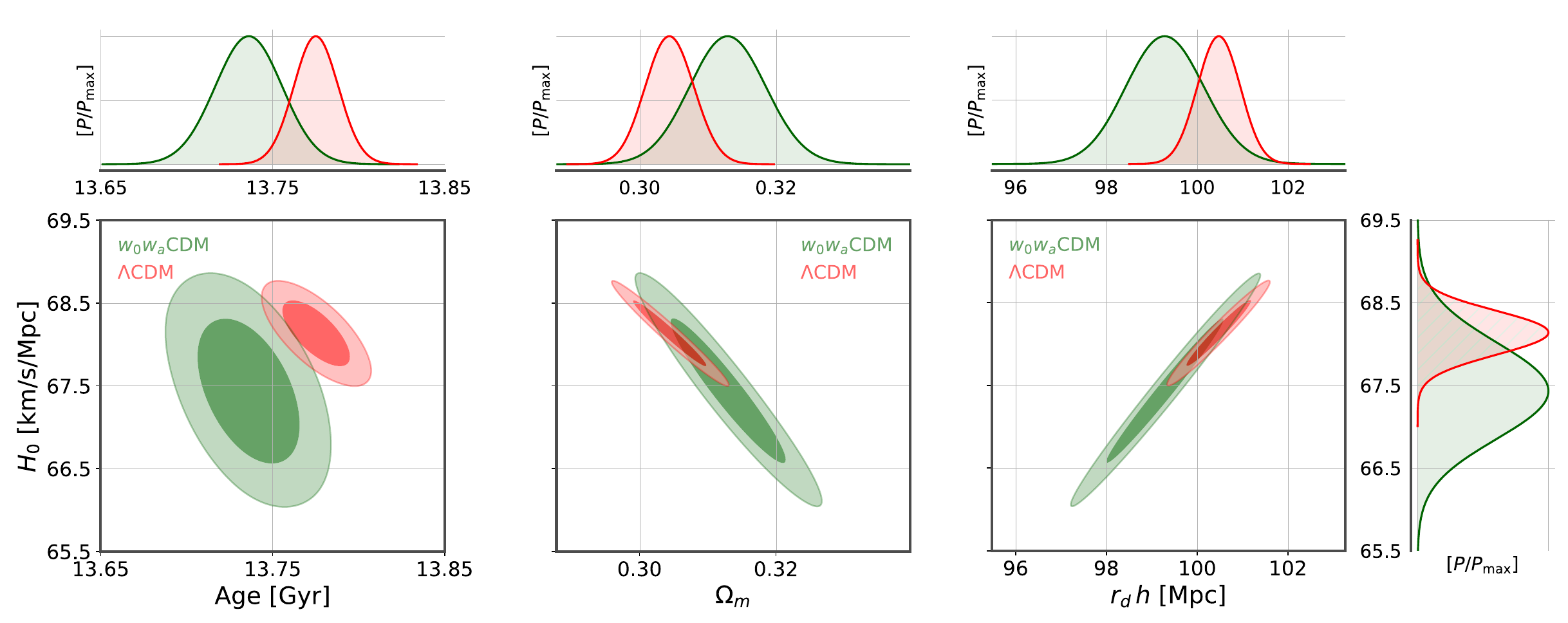}
    \caption{Marginalized constraints in the $(H_0,{\rm age})$, $(H_0,\Omega_m)$, and $(H_0,r_dh)$ planes for $\Lambda$CDM and $w_0w_a$CDM, using CMB+DESI+DD.}
    \label{fig:H0_tension}
\end{figure}

This conclusion is particularly transparent in the comparison between $\Lambda$CDM and $w_0w_a$CDM shown in Fig.~\ref{fig:H0_tension} for CMB+DESI+DD. The three panels illustrate that dynamical DE not only fails to move the inferred Hubble constant toward the local-universe determination, but actually shifts the preferred cosmology in the opposite direction. In the $H_0$-age plane, the $w_0w_a$CDM solution is displaced toward lower $H_0$ and older cosmic ages, showing that the extra late-time freedom favors a cosmology with a longer expansion time. In the $H_0$-$\Omega_m$ plane, the same branch is shifted toward higher matter density and a lower Hubble constant, a direction that worsens the discrepancy with local measurements. Finally, in the $H_0$-$r_d h$ plane (where $h=H_0/[100\,\mathrm{km\,s^{-1}\,Mpc^{-1}}]$), the two branches follow distinct but tightly constrained degeneracy directions, with the $w_0w_a$CDM branch shifted toward smaller $H_0$ and slightly lower $r_d h$, again worsening the tension with local estimates. Altogether, Fig.~\ref{fig:H0_tension} shows that dynamical DE favors a correlated shift toward \textit{lower} $H_0$, higher $\Omega_m$, and older ages, thereby making the Hubble tension, if anything, more severe.


\subsection{$S_8$ and $\sigma_8$ in Extended Cosmological Models}

Unlike the Hubble tension, which compares a CMB-calibrated cosmological inference with late-time determinations that are essentially model independent, the $S_8$ tension involves quantities that are themselves extracted within an assumed cosmological framework~\cite{DES:2018ufa,KiDS:2020ghu,DES:2022ccp}. Therefore, a direct comparison between the values of $S_8$ inferred across the different extended models and those reported by weak-lensing surveys should be treated with some caution~\cite{DiValentino:2018gcu}. A fully consistent assessment would require reanalyzing those datasets within the same extended cosmologies considered here, but this is beyond the scope of the present work. For this reason, here we do not attempt a dedicated reassessment of the $S_8$ tension in each enlarged parameter space, but rather focus on the behaviour of the underlying parameters $\sigma_8$ and $\Omega_m$, from which $S_8$ is constructed.

The results reported in Table~\ref{tab:results.tension.corrected} show that $\sigma_8$ remains comparatively stable across the full set of cosmological extensions considered in this work. Within the $\Lambda$CDM-like branch, $\sigma_8$ typically lies in the range $\simeq 0.816$--$0.824$, with absolute uncertainties of order $\Delta\sigma_8 \sim 0.004$--$0.007$ in the minimally and moderately extended cases, corresponding to a precision of roughly $0.5\%$--$0.8\%$, and broadening to about $\Delta\sigma_8 \sim 0.009$ ($\sim 1.1\%$) only in the fully extended scenario. A similar pattern is found in the $w_0w_a$CDM-like branch, where $\sigma_8$ remains in the range $\simeq 0.813$--$0.821$, with typical uncertainties of order $\Delta\sigma_8 \sim 0.008$ ($\sim 1.0\%$), reaching about $\Delta\sigma_8 \sim 0.010$ only in the most extended model.

\begin{figure}[t!]
    \centering \includegraphics[width=0.73\linewidth]{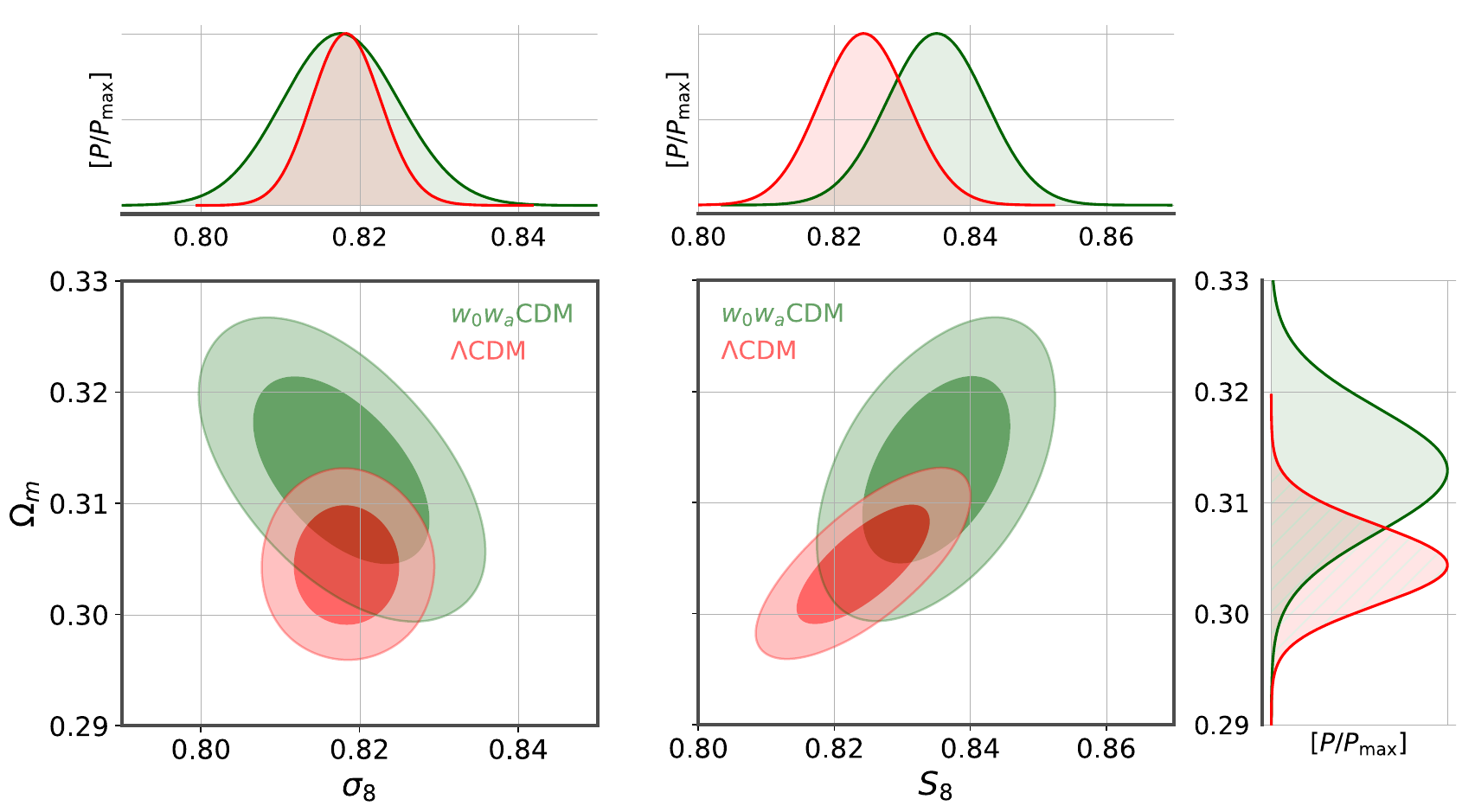}
    \caption{Marginalized constraints in the $(\sigma_8,\Omega_m)$ and $(S_8,\Omega_m)$ planes for $\Lambda$CDM and $w_0w_a$CDM, obtained from the CMB+DESI+DD dataset combination. }
    \label{fig:S8_tension}
\end{figure}

Overall, while $\sigma_8$ is not strictly unchanged, its variations remain comparatively modest and do not exhibit the same coherent branch-to-branch displacement seen in $\Omega_m$. As a result, the main differences in the inferred values of $S_8$ are driven by the latter. This is illustrated in Fig.~\ref{fig:S8_tension}, where the left panel shows the constraints in the $(\sigma_8,\Omega_m)$ plane and the right panel the corresponding constraints in the $(S_8,\Omega_m)$ plane, both obtained within the baseline $\Lambda$CDM and $w_0w_a$CDM models from the CMB+DESI+DD dataset combination. In the $(\sigma_8,\Omega_m)$ plane, the two branches are separated primarily along the $\Omega_m$ direction. The preferred value of $\sigma_8$ remains broadly similar, while $\Omega_m$ is coherently shifted upward in the $w_0w_a$CDM case. By construction, this displacement propagates into the derived parameter $S_8$, as shown in the right panel, where the $w_0w_a$CDM branch is shifted toward higher $S_8$ because of the higher matter density.

\section{Model Comparison}

\label{sec.model_comparison}
\begin{table*}[t!]
\centering
\renewcommand{\arraystretch}{1.45}
\resizebox{\textwidth}{!}{
\begin{tabular}{lc|cc|cc|}
\hline\hline
& & \multicolumn{2}{c|}{\textbf{CMB+DESI+PP}} & \multicolumn{2}{c|}{\textbf{CMB+DESI+DD}} \\
\cline{3-4}\cline{5-6}
\textbf{Model} & $\Delta k$ & $\Delta\chi^2$ & $\Delta$AIC & $\Delta\chi^2$ & $\Delta$AIC \\
\hline
$\Lambda$CDM & 0 & 0.00 & 0.00 & 0.00 & 0.00 \\
$\Lambda$CDM+$\sum m_\nu$ & 1 & -4.53 & -2.53 & -2.65 & -0.65 \\
$\Lambda$CDM+$N_{\rm eff}$ & 1 & -0.49 & 1.51 & 0.29 & 2.29 \\
$\Lambda$CDM+$\sum m_\nu+N_{\rm eff}$ & 2 & -5.80 & -1.80 & -4.68 & -0.68 \\
$\Lambda$CDM+$\Omega_k$ & 1 & -4.39 & -2.39 & -4.87 & -2.87 \\
$\Lambda$CDM+$r$ & 1 & -0.55 & 1.45 & 0.43 & 2.43 \\
$\Lambda$CDM+$r+\alpha_s$ & 2 & -0.54 & 3.46 & 0.20 & 4.20 \\
$\Lambda$CDM+$r+\alpha_s+\beta_s$ & 3 & -2.15 & 3.85 & -2.01 & 3.99 \\
$\Lambda$CDM+$r+\alpha_s+\beta_s+\Omega_k+\sum m_\nu+N_{\rm eff}$ & 6 & -9.53 & 2.47 & -9.19 & 2.81 \\
$w_0w_a$CDM & 2 & -10.19 & -6.19 & -13.03 & -9.03 \\
$w_0w_a$CDM+$\sum m_\nu$ & 3 & -11.13 & -5.13 & -13.61 & -7.61 \\
$w_0w_a$CDM+$N_{\rm eff}$ & 3 & -10.94 & -4.94 & -15.11 & -9.11 \\
$w_0w_a$CDM+$\sum m_\nu+N_{\rm eff}$ & 4 & -12.29 & -4.29 & -15.48 & -7.48 \\
$w_0w_a$CDM+$\Omega_k$ & 3 & -9.41 & -3.41 & -13.78 & -7.78 \\
$w_0w_a$CDM+$r$ & 3 & -10.17 & -4.17 & -12.73 & -6.73 \\
$w_0w_a$CDM+$r+\alpha_s$ & 4 & -10.22 & -2.22 & -13.14 & -5.14 \\
$w_0w_a$CDM+$r+\alpha_s+\beta_s$ & 5 & -12.58 & -2.58 & -14.73 & -4.73 \\
$w_0w_a$CDM+$r+\alpha_s+\beta_s+\Omega_k+\sum m_\nu+N_{\rm eff}$ & 8 & -15.41 & 0.59 & -17.08 & -1.08 \\
\hline\hline
\end{tabular}
}
\caption{Relative model-comparison statistics for the cosmological models considered in this work. For each dataset combination, we report $\Delta\chi^2$ with respect to baseline flat $\Lambda$CDM and the corresponding $\Delta$AIC values. The comparison is performed separately for the CMB+DESI+PP and CMB+DESI+DD dataset combinations. Here $\Delta k$ denotes the number of additional free parameters relative to flat $\Lambda$CDM. Negative values of $\Delta\chi^2$ and $\Delta$AIC indicate an improvement relative to baseline $\Lambda$CDM.}
\label{tab:model_comparison}
\end{table*}

In this section, we briefly assess how the overall fit quality changes across the various extensions considered, and whether any of them yields a meaningful improvement over baseline $\Lambda$CDM.

To this end, Table~\ref{tab:model_comparison} reports, for both CMB+DESI+PP and CMB+DESI+DD, the variation in best-fit $\chi^2$ with respect to flat $\Lambda$CDM, together with the corresponding Akaike Information Criterion (AIC)~\cite{Akaike:1974vps} difference, $\Delta {\rm AIC}=\Delta\chi^2+2\Delta k$, where $\Delta k$ denotes the number of additional free parameters relative to $\Lambda$CDM. By construction, negative values of $\Delta\chi^2$ indicate that the extended model provides a better fit than baseline $\Lambda$CDM, while negative values of $\Delta {\rm AIC}$ indicate that such an improvement remains preferred even after accounting for the increase in parameter volume. As a rough guideline, we interpret $\Delta {\rm AIC}\gtrsim -2$ as indicating little to no preference, $\Delta {\rm AIC}\sim -3$ as corresponding to mild evidence, $\Delta {\rm AIC}\sim -5$ as pointing to a clear preference, while $\Delta {\rm AIC}\lesssim -10$ may be regarded as strong support for the model with lower AIC.

Within the $\Lambda$CDM-like branch, several extensions improve the best-fit $\chi^2$, but almost never enough to imply a robust preference once the larger parameter volume is taken into account. This conclusion is robust across both SN choices. For CMB+DESI+PP, the strongest cases are $\Lambda$CDM+$\sum m_\nu$ and $\Lambda$CDM+$\Omega_k$, both reaching only $\Delta {\rm AIC}\simeq -2.5$, while $\Lambda$CDM+$\sum m_\nu+N_{\rm eff}$ remains weaker ($\Delta {\rm AIC}\simeq -1.8$). For CMB+DESI+DD, $\Lambda$CDM+$\Omega_k$ provides the largest improvement ($\Delta {\rm AIC}\simeq -2.9$), whereas all other cases remain weak or at most marginally preferred. More extended $\Lambda$CDM-like parameter spaces can lower $\chi^2$ more substantially, but this gain is largely offset by the higher dimensionality, so that the corresponding $\Delta {\rm AIC}$ turns positive. Overall, within the cosmological-constant branch, none of the extensions considered here emerges as strongly preferred over minimal $\Lambda$CDM.

Once the late-time DE sector is allowed to become dynamical, the gain in $\Delta\chi^2$ with respect to the corresponding $\Lambda$CDM-like models becomes substantially larger, and remains statistically significant even after the AIC penalty is taken into account. For CMB+DESI+PP, the baseline $w_0w_a$CDM model reaches $\Delta {\rm AIC}\simeq -6.2$, while several of its extensions remain in the range $\Delta {\rm AIC}\sim -4$ to $-5$. The effect becomes even more pronounced for CMB+DESI+DD, where the same baseline $w_0w_a$CDM model reaches $\Delta {\rm AIC}\simeq -9.0$, and several extensions lie between roughly $-7$ and $-9$, pointing to a substantially stronger preference for dynamical DE. The role of the SN choice is therefore much more significant in this branch, with DES-Dovekie systematically strengthening the preference for dynamical DE, consistently with the results discussed in Sec.~\ref{sec.DE}. 

All in all, the only robust and recurrent gain is associated with moving from a cosmological constant to a dynamical late-time DE description.
\section{Conclusions}
\label{sec.conclusions}

In this work, we have carried out a systematic reassessment of cosmological constraints beyond baseline $\Lambda$CDM. We progressively relaxed the main assumptions defining the key sectors of the standard cosmological model, namely late-time DE, spatial curvature, the neutrino sector, and the primordial inflationary sector. In doing so, we derived intertwined constraints on both standard and extended parameters, and clarified how parameter bounds, degeneracy directions, and inferred cosmological conclusions are reshaped across increasingly enlarged cosmological frameworks. We paid particular attention to the major current challenges of $\Lambda$CDM, quantifying the extent to which they are contingent on the assumptions built into the baseline framework and how robust they remain once these assumptions are relaxed.

The clearest and most recurrent indication of physics beyond baseline $\Lambda$CDM emerging from our analysis concerns the late-time DE sector. Across the different dataset combinations and extended cosmologies considered in this work, promoting DE to a dynamical component described by the CPL parametrization yields the most stable departure from $\Lambda$CDM, both in terms of parameter shifts and overall fit improvement; see Table~\ref{tab:model_comparison}. The statistical preference for dynamical DE across all the cosmological models is summarized in Table~\ref{tab:results.DE} and Fig.~\ref{fig:DDE_sigmas}. When additional sectors are opened simultaneously, this preference is only mildly reduced, with the largest degradation arising when curvature and, to a lesser extent, the neutrino sector are allowed to vary. This weakening primarily reflects the broadening of the allowed $(w_0,w_a)$ region, as quantified by the degradation of the Figure of Merit, rather than a qualitative change in the preferred DE evolution. Indeed, the preferred region remains consistently displaced toward the characteristic CPL behaviour with $w_0>-1$ and $w_a<0$, in agreement with the recent DESI results. The inferred evolution of the EoS also remains essentially unchanged when additional sectors are progressively opened, consistently favoring an evolution that is mildly quintessence-like today, crosses the phantom divide near $z\sim0.35$, and becomes phantom-like at higher redshift; see Fig.~\ref{fig:EoS} and Fig.~\ref{fig:summary_plot_DDE}.

The results for extensions involving spatial curvature are summarized in Table~\ref{tab:results.omk} and Fig.~\ref{fig:omk_1D}. When the curvature parameter is allowed to vary within a cosmological-constant description of DE, we find a mild preference for $\Omega_k>0$ that remains at the level of about $2.2\sigma$ even in the fully extended case. However, this indication is strongly degraded once the late-time DE sector is promoted to a dynamical component. In the corresponding $w_0w_a$CDM extensions, the posterior shifts substantially toward the flat limit and the significance drops to values fully consistent with spatial flatness. This suggests that the apparent preference for positive $\Omega_k$ in $\Lambda$CDM-like models is at least partly absorbing the same late-time geometric freedom that, once the DE sector is opened, is more naturally described by dynamical DE. Overall, our results do not support a strong case for a departure from spatial flatness, but indicate that current curvature constraints are strongly dependent on the assumptions made in the late-time DE sector.

The neutrino sector provides yet another clear example of how strongly cosmological inference can depend on the assumed framework. The upper bounds on the total neutrino mass are summarized in Table~\ref{tab:results.neutrinos}. They vary substantially across the cosmological extensions considered here, ranging from $\sum m_\nu\lesssim 0.06$ eV in minimal $\Lambda$CDM-like scenarios (namely values compressed near the NO floor) to much weaker limits of order $\sum m_\nu\lesssim 0.1$–$0.2$ eV once late-time DE, curvature, and additional sectors are allowed to vary, see Fig.~\ref{fig:mnu_neff}. As a consequence, both the cosmological preference for NO over IO and the apparent tension between cosmological mass bounds and oscillation-informed expectations are themselves highly model dependent, see Table~\ref{tab:results.neutrinos} and Fig.~\ref{fig:mnu_summary}. In minimal $\Lambda$CDM-based analyses, the strong compression of the posterior toward very small masses yields a moderate-to-strong preference for NO and a marked mismatch with oscillation data. Both effects are significantly weakened, and in some cases erased, in enlarged cosmological parameter spaces. By contrast, the constraints on $N_{\rm eff}$ remain comparatively stable across all the models considered. While the data show a mild yet persistent preference for slightly smaller values, the results remain broadly consistent with the standard expectation $N_{\rm eff}=3.04$, see Fig.~\ref{fig:Neff}.

The results for the inflationary sector are summarized in Table~\ref{tab:results.inflation} for the scalar and tensor spectral parameters, obtained under the assumption of the slow-roll consistency relations, and in Table~\ref{tab:results.slow-roll} for the corresponding derived slow-roll and Hubble-flow parameters. Overall, the tensor sector remains remarkably stable across all the cosmological extensions considered in this work. The upper bound on $r$ varies only minimally, and no evidence for a non-vanishing primordial tensor signal emerges in any of the models explored. By contrast, the scalar sector is significantly more sensitive to the assumed cosmological framework, and especially to the parametrization adopted for the primordial spectrum. Allowing for non-zero running $\alpha_s$ and running of the running $\beta_s$ reabsorbs part of the preference for larger values of $n_s$ found in simpler parametrizations with $\alpha_s=\beta_s=0$. While both $\alpha_s$ and $\beta_s$ show mildly positive central values, they remain compatible with zero at less than $2\sigma$, while shifting the preferred scalar tilt toward smaller values, see Fig.~\ref{fig:whisker_inf}. These shifts have non-trivial implications for inflationary model building. As shown in Fig.~\ref{fig:ns_r}, the apparent agreement or disagreement of benchmark models in the $n_s$-$r$ plane depends non-negligibly on both the primordial parametrization and the overall cosmological framework adopted in the analysis. In this sense, we stress that constraints on inflationary model space should not be interpreted as model independent once extended or non-standard cosmologies are taken into account.

Finally, the impact of the extended cosmologies considered here on the main cosmological tensions is summarized in Table~\ref{tab:results.tension.corrected}. None of the extensions explored in this work provides a viable route toward resolving the Hubble tension. In particular, promoting DE to a dynamical component does not shift the inferred value of $H_0$ toward local determinations, but, if anything, moves the preferred cosmology toward lower $H_0$ and higher $\Omega_m$, thereby worsening the discrepancy, see Fig.~\ref{fig:H0_tension}. At the same time, the inferred matter density $\Omega_m$ remains very stable within each of the two late-time branches, but is systematically shifted to higher values in $w_0w_a$CDM-like cosmologies relative to $\Lambda$CDM-like ones, see Fig.~\ref{fig:whisker_Omegam}. As for $\sigma_8$, it remains comparatively stable across the different models, so that the main branch-to-branch differences relevant for $S_8$ are driven primarily by the shift in $\Omega_m$, see Fig.~\ref{fig:S8_tension}.

All in all, our results show that, among the extensions considered in this work, late-time dynamical DE stands out as the only one that is consistently favored by the data, see Table~\ref{tab:model_comparison}. At the same time, several of the sharpest inferences drawn within the baseline $\Lambda$CDM framework, from the preference for a given neutrino mass ordering, to the tension between cosmological mass bounds and oscillation data, to the apparent indication for values of $n_s$ larger than those expected in several benchmark inflationary models, reflect the narrowness of the framework, rather than the robustness of the underlying physical conclusions. Once the standard assumptions of $\Lambda$CDM are relaxed, many of these apparent signals lose much of their statistical force, shift, or become strongly model dependent. Against this intertwined landscape, late-time dynamical DE emerges as the only recurrent and coherent departure from the baseline cosmological model across the different datasets and extended parameter spaces considered in this work and has the strongest impact on the inferred conclusions in the other sectors of the model.

\acknowledgments
W.G. acknowledges support from National Aeronautics and Space Administration
(NASA) under Grant No. 80NSSC24K0898. DHL is supported by an EPSRC studentship. EDV is supported by a Royal Society Dorothy Hodgkin Research Fellowship.
This article is based upon work from the COST Action CA21136 ``Addressing observational tensions in cosmology with systematics and fundamental physics'' (CosmoVerse), supported by COST (European Cooperation in Science and Technology). We acknowledge IT Services at The University of Sheffield for the provision of services for High Performance Computing.
\clearpage
\appendix
\section{Summary of Constraints}
\label{appendix.Tables}



\begin{table*}[h]
\begin{center}
\renewcommand{\arraystretch}{1.5}
\resizebox{\textwidth}{!}{
\begin{tabular}{l  c c}
\hline
\textbf{Parameter} & \textbf{ CMB+DESI+PP } & \textbf{ CMB+DESI+DD } \\
\hline\hline

$ \Omega_\mathrm{b} h^2  $ & $  0.022482\pm 0.000090\, ( 0.02248^{+0.00018}_{-0.00017} ) $ & $  0.022480\pm 0.000092\, ( 0.02248^{+0.00018}_{-0.00018} ) $ \\
$ \Omega_\mathrm{c} h^2  $ & $  0.11811\pm 0.00060\, ( 0.1181^{+0.0012}_{-0.0012} ) $ & $  0.11818\pm 0.00060\, ( 0.1182^{+0.0012}_{-0.0012} ) $ \\
$ \tau_\mathrm{reio}  $ & $  0.0661^{+0.0056}_{-0.0063}\, ( 0.066^{+0.012}_{-0.011} ) $ & $  0.0661^{+0.0057}_{-0.0064}\, ( 0.066^{+0.012}_{-0.011} ) $ \\
$ 100\theta_\mathrm{MC}  $ & $  1.04092\pm 0.00022\, ( 1.04092^{+0.00043}_{-0.00043} ) $ & $  1.04090\pm 0.00022\, ( 1.04090^{+0.00045}_{-0.00044} ) $ \\
$ n_\mathrm{s}  $ & $  0.9735\pm 0.0029\, ( 0.9735^{+0.0058}_{-0.0059} ) $ & $  0.9733\pm 0.0029\, ( 0.9733^{+0.0058}_{-0.0057} ) $ \\
$ \log(10^{10} A_\mathrm{s})  $ & $  3.067\pm 0.011\, ( 3.067^{+0.021}_{-0.020} ) $ & $  3.067\pm 0.011\, ( 3.067^{+0.022}_{-0.020} ) $ \\
$ H_0  $ & $  68.17\pm 0.25\, ( 68.17^{+0.48}_{-0.48} ) $ & $  68.13\pm 0.25\, ( 68.13^{+0.49}_{-0.48} ) $ \\
\hline \hline
\end{tabular} }
\end{center}
\caption{ Results for $\Lambda\mathrm{CDM}$ at $68\% (95\%)$ CL. }
\label{ tab:lcdm_results }
\end{table*}




\begin{table*}
\begin{center}
\renewcommand{\arraystretch}{1.5}
\resizebox{\textwidth}{!}{
\begin{tabular}{l  c c}
\hline
\textbf{Parameter} & \textbf{ CMB+DESI+PP } & \textbf{ CMB+DESI+DD } \\
\hline\hline
$ \Omega_\mathrm{b} h^2  $ & $  0.022477\pm 0.000092\, ( 0.02248^{+0.00018}_{-0.00018} ) $ & $  0.022471\pm 0.000091\, ( 0.02247^{+0.00018}_{-0.00017} ) $ \\
$ \Omega_\mathrm{c} h^2  $ & $  0.11840\pm 0.00061\, ( 0.1184^{+0.0012}_{-0.0012} ) $ & $  0.11844\pm 0.00061\, ( 0.1184^{+0.0012}_{-0.0012} ) $ \\
$ \tau_\mathrm{reio}  $ & $  0.0635^{+0.0054}_{-0.0061}\, ( 0.064^{+0.012}_{-0.011} ) $ & $  0.0636^{+0.0053}_{-0.0063}\, ( 0.064^{+0.012}_{-0.011} ) $ \\
$ 100\theta_\mathrm{MC}  $ & $  1.04090\pm 0.00023\, ( 1.04090^{+0.00045}_{-0.00045} ) $ & $  1.04090\pm 0.00023\, ( 1.04090^{+0.00045}_{-0.00044} ) $ \\
$ n_\mathrm{s}  $ & $  0.9728\pm 0.0029\, ( 0.9728^{+0.0057}_{-0.0059} ) $ & $  0.9728\pm 0.0030\, ( 0.9728^{+0.0059}_{-0.0058} ) $ \\
$ \log(10^{10} A_\mathrm{s})  $ & $  3.0613^{+0.0098}_{-0.011}\, ( 3.061^{+0.022}_{-0.020} ) $ & $  3.0612^{+0.0099}_{-0.011}\, ( 3.061^{+0.022}_{-0.020} ) $ \\
$ H_0  $ & $  68.35^{+0.28}_{-0.24}\, ( 68.35^{+0.50}_{-0.54} ) $ & $  68.33\pm 0.26\, ( 68.33^{+0.51}_{-0.53} ) $ \\
\hline \hline
\end{tabular} }
\end{center}
\caption{ Results for $\Lambda\mathrm{CDM}+\sum m_\nu$ at $68\% (95\%)$ CL. }
\label{ tab:lcdm_mnu_results }
\end{table*}




\begin{table*}
\begin{center}
\renewcommand{\arraystretch}{1.5}
\resizebox{\textwidth}{!}{
\begin{tabular}{l  c c}
\hline
\textbf{Parameter} & \textbf{ CMB+DESI+PP } & \textbf{ CMB+DESI+DD } \\
\hline\hline
$ \Omega_\mathrm{b} h^2  $ & $  0.02245\pm 0.00013\, ( 0.02245^{+0.00024}_{-0.00024} ) $ & $  0.02244\pm 0.00013\, ( 0.02244^{+0.00024}_{-0.00025} ) $ \\
$ \Omega_\mathrm{c} h^2  $ & $  0.1175\pm 0.0018\, ( 0.1175^{+0.0036}_{-0.0035} ) $ & $  0.1174\pm 0.0018\, ( 0.1174^{+0.0037}_{-0.0035} ) $ \\
$ \tau_\mathrm{reio}  $ & $  0.0663^{+0.0055}_{-0.0063}\, ( 0.066^{+0.012}_{-0.011} ) $ & $  0.0661^{+0.0056}_{-0.0063}\, ( 0.066^{+0.012}_{-0.011} ) $ \\
$ 100\theta_\mathrm{MC}  $ & $  1.04097\pm 0.00028\, ( 1.04097^{+0.00054}_{-0.00054} ) $ & $  1.04097\pm 0.00027\, ( 1.04097^{+0.00053}_{-0.00054} ) $ \\
$ n_\mathrm{s}  $ & $  0.9717\pm 0.0051\, ( 0.972^{+0.010}_{-0.010} ) $ & $  0.9713\pm 0.0052\, ( 0.971^{+0.010}_{-0.010} ) $ \\
$ \log(10^{10} A_\mathrm{s})  $ & $  3.065\pm 0.011\, ( 3.065^{+0.023}_{-0.021} ) $ & $  3.065\pm 0.011\, ( 3.065^{+0.023}_{-0.021} ) $ \\
$ H_0  $ & $  67.91\pm 0.68\, ( 67.9^{+1.3}_{-1.3} ) $ & $  67.85\pm 0.68\, ( 67.8^{+1.3}_{-1.4} ) $ \\
\hline \hline
\end{tabular} }
\end{center}
\caption{ Results for $\Lambda\mathrm{CDM}+N_{\mathrm{eff}}$ at $68\% (95\%)$ CL. }
\label{ tab:lcdm_nnu_results }
\end{table*}




\begin{table*}
\begin{center}
\renewcommand{\arraystretch}{1.5}
\resizebox{\textwidth}{!}{
\begin{tabular}{l  c c}
\hline
\textbf{Parameter} & \textbf{ CMB+DESI+PP } & \textbf{ CMB+DESI+DD } \\
\hline\hline
$ \Omega_\mathrm{b} h^2  $ & $  0.02241\pm 0.00013\, ( 0.02241^{+0.00024}_{-0.00025} ) $ & $  0.02241\pm 0.00012\, ( 0.02241^{+0.00024}_{-0.00024} ) $ \\
$ \Omega_\mathrm{c} h^2  $ & $  0.1171\pm 0.0018\, ( 0.1171^{+0.0036}_{-0.0035} ) $ & $  0.1172\pm 0.0018\, ( 0.1172^{+0.0035}_{-0.0035} ) $ \\
$ \tau_\mathrm{reio}  $ & $  0.0638^{+0.0053}_{-0.0061}\, ( 0.064^{+0.012}_{-0.011} ) $ & $  0.0636^{+0.0054}_{-0.0063}\, ( 0.064^{+0.012}_{-0.011} ) $ \\
$ 100\theta_\mathrm{MC}  $ & $  1.04101\pm 0.00028\, ( 1.04101^{+0.00054}_{-0.00053} ) $ & $  1.04101\pm 0.00027\, ( 1.04101^{+0.00054}_{-0.00053} ) $ \\
$ n_\mathrm{s}  $ & $  0.9697\pm 0.0052\, ( 0.970^{+0.010}_{-0.010} ) $ & $  0.9695\pm 0.0052\, ( 0.970^{+0.010}_{-0.010} ) $ \\
$ \log(10^{10} A_\mathrm{s})  $ & $  3.059^{+0.010}_{-0.012}\, ( 3.059^{+0.023}_{-0.021} ) $ & $  3.058\pm 0.011\, ( 3.058^{+0.023}_{-0.021} ) $ \\
$ H_0  $ & $  67.89\pm 0.68\, ( 67.9^{+1.3}_{-1.3} ) $ & $  67.86\pm 0.68\, ( 67.9^{+1.3}_{-1.3} ) $ \\
\hline \hline
\end{tabular} }
\end{center}
\caption{ Results for $\Lambda\mathrm{CDM}+\sum m_\nu + N_{\mathrm{eff}}$ at $68\% (95\%)$ CL. }
\label{ tab:lcdm_mnu_nnu_results }
\end{table*}




\begin{table*}
\begin{center}
\renewcommand{\arraystretch}{1.5}
\resizebox{\textwidth}{!}{
\begin{tabular}{l  c c}
\hline
\textbf{Parameter} & \textbf{ CMB+DESI+PP } & \textbf{ CMB+DESI+DD } \\
\hline\hline
$ \Omega_\mathrm{b} h^2  $ & $  0.022419\pm 0.000096\, ( 0.02242^{+0.00019}_{-0.00019} ) $ & $  0.022413\pm 0.000096\, ( 0.02241^{+0.00019}_{-0.00019} ) $ \\
$ \Omega_\mathrm{c} h^2  $ & $  0.1200\pm 0.0010\, ( 0.1200^{+0.0020}_{-0.0020} ) $ & $  0.1201\pm 0.0010\, ( 0.1201^{+0.0020}_{-0.0020} ) $ \\
$ \tau_\mathrm{reio}  $ & $  0.0622^{+0.0055}_{-0.0063}\, ( 0.062^{+0.012}_{-0.011} ) $ & $  0.0619^{+0.0053}_{-0.0062}\, ( 0.062^{+0.012}_{-0.011} ) $ \\
$ 100\theta_\mathrm{MC}  $ & $  1.04072\pm 0.00024\, ( 1.04072^{+0.00049}_{-0.00047} ) $ & $  1.04072\pm 0.00024\, ( 1.04072^{+0.00046}_{-0.00046} ) $ \\
$ n_\mathrm{s}  $ & $  0.9693\pm 0.0034\, ( 0.9693^{+0.0066}_{-0.0066} ) $ & $  0.9692\pm 0.0034\, ( 0.9692^{+0.0065}_{-0.0067} ) $ \\
$ \log(10^{10} A_\mathrm{s})  $ & $  3.062\pm 0.010\, ( 3.062^{+0.021}_{-0.019} ) $ & $  3.062\pm 0.010\, ( 3.062^{+0.021}_{-0.019} ) $ \\
$ H_0  $ & $  68.57\pm 0.32\, ( 68.57^{+0.62}_{-0.61} ) $ & $  68.56\pm 0.31\, ( 68.56^{+0.63}_{-0.60} ) $ \\
\hline \hline
\end{tabular} }
\end{center}
\caption{ Results for $\Lambda\mathrm{CDM}+\Omega_k$ at $68\% (95\%)$ CL. }
\label{ tab:lcdm_omk_results }
\end{table*}




\begin{table*}
\begin{center}
\renewcommand{\arraystretch}{1.5}
\resizebox{\textwidth}{!}{
\begin{tabular}{l  c c}
\hline
\textbf{Parameter} & \textbf{ CMB+DESI+PP } & \textbf{ CMB+DESI+DD } \\
\hline\hline
$ \Omega_\mathrm{b} h^2  $ & $  0.022484\pm 0.000090\, ( 0.02248^{+0.00018}_{-0.00018} ) $ & $  0.022478\pm 0.000091\, ( 0.02248^{+0.00018}_{-0.00018} ) $ \\
$ \Omega_\mathrm{c} h^2  $ & $  0.11809\pm 0.00060\, ( 0.1181^{+0.0012}_{-0.0012} ) $ & $  0.11817\pm 0.00060\, ( 0.1182^{+0.0012}_{-0.0012} ) $ \\
$ \tau_\mathrm{reio}  $ & $  0.0659^{+0.0054}_{-0.0064}\, ( 0.066^{+0.013}_{-0.011} ) $ & $  0.0655^{+0.0056}_{-0.0063}\, ( 0.065^{+0.012}_{-0.011} ) $ \\
$ 100\theta_\mathrm{MC}  $ & $  1.04091\pm 0.00022\, ( 1.04091^{+0.00044}_{-0.00044} ) $ & $  1.04090\pm 0.00022\, ( 1.04090^{+0.00045}_{-0.00043} ) $ \\
$ n_\mathrm{s}  $ & $  0.9738\pm 0.0029\, ( 0.9738^{+0.0056}_{-0.0056} ) $ & $  0.9736\pm 0.0029\, ( 0.9736^{+0.0059}_{-0.0058} ) $ \\
$ \log(10^{10} A_\mathrm{s})  $ & $  3.0662^{+0.0099}_{-0.011}\, ( 3.066^{+0.022}_{-0.020} ) $ & $  3.065\pm 0.011\, ( 3.065^{+0.022}_{-0.020} ) $ \\
$ H_0  $ & $  68.17\pm 0.25\, ( 68.17^{+0.48}_{-0.48} ) $ & $  68.14\pm 0.24\, ( 68.14^{+0.47}_{-0.48} ) $ \\
\hline \hline
\end{tabular} }
\end{center}
\caption{ Results for $\Lambda\mathrm{CDM}+r$ at $68\% (95\%)$ CL. }
\label{ tab:lcdm_r_results }
\end{table*}




\begin{table*}
\begin{center}
\renewcommand{\arraystretch}{1.5}
\resizebox{\textwidth}{!}{
\begin{tabular}{l  c c}
\hline
\textbf{Parameter} & \textbf{ CMB+DESI+PP } & \textbf{ CMB+DESI+DD } \\
\hline\hline
$ \Omega_\mathrm{b} h^2  $ & $  0.022450\pm 0.000099\, ( 0.02245^{+0.00019}_{-0.00019} ) $ & $  0.022450\pm 0.000099\, ( 0.02245^{+0.00020}_{-0.00019} ) $ \\
$ \Omega_\mathrm{c} h^2  $ & $  0.11811\pm 0.00060\, ( 0.1181^{+0.0012}_{-0.0012} ) $ & $  0.11815\pm 0.00061\, ( 0.1182^{+0.0012}_{-0.0012} ) $ \\
$ \tau_\mathrm{reio}  $ & $  0.0647^{+0.0055}_{-0.0064}\, ( 0.065^{+0.012}_{-0.011} ) $ & $  0.0647^{+0.0054}_{-0.0064}\, ( 0.065^{+0.012}_{-0.011} ) $ \\
$ 100\theta_\mathrm{MC}  $ & $  1.04091\pm 0.00022\, ( 1.04091^{+0.00044}_{-0.00043} ) $ & $  1.04090\pm 0.00023\, ( 1.04090^{+0.00044}_{-0.00045} ) $ \\
$ n_\mathrm{s}  $ & $  0.9736\pm 0.0030\, ( 0.9736^{+0.0059}_{-0.0058} ) $ & $  0.9736\pm 0.0029\, ( 0.9736^{+0.0059}_{-0.0056} ) $ \\
$ \log(10^{10} A_\mathrm{s})  $ & $  3.063^{+0.011}_{-0.012}\, ( 3.063^{+0.023}_{-0.021} ) $ & $  3.063^{+0.010}_{-0.012}\, ( 3.063^{+0.022}_{-0.020} ) $ \\
$ H_0  $ & $  68.14\pm 0.25\, ( 68.14^{+0.49}_{-0.49} ) $ & $  68.12\pm 0.25\, ( 68.12^{+0.50}_{-0.49} ) $ \\
\hline \hline
\end{tabular} }
\end{center}
\caption{ Results for $\Lambda\mathrm{CDM}+r+\alpha_s$ at $68\%\,(95\%)$ CL. }
\label{ tab:lcdm_r_nrun_results }
\end{table*}




\begin{table*}
\begin{center}
\renewcommand{\arraystretch}{1.5}
\resizebox{\textwidth}{!}{
\begin{tabular}{l  c c}
\hline
\textbf{Parameter} & \textbf{ CMB+DESI+PP } & \textbf{ CMB+DESI+DD } \\
\hline\hline

$ \Omega_\mathrm{b} h^2  $ & $  0.02243\pm 0.00010\, ( 0.02243^{+0.00020}_{-0.00020} ) $ & $  0.022428\pm 0.000099\, ( 0.02243^{+0.00019}_{-0.00019} ) $ \\
$ \Omega_\mathrm{c} h^2  $ & $  0.11817\pm 0.00060\, ( 0.1182^{+0.0012}_{-0.0012} ) $ & $  0.11819\pm 0.00059\, ( 0.1182^{+0.0011}_{-0.0012} ) $ \\
$ \tau_\mathrm{reio}  $ & $  0.0669^{+0.0057}_{-0.0066}\, ( 0.067^{+0.012}_{-0.012} ) $ & $  0.0667^{+0.0058}_{-0.0068}\, ( 0.067^{+0.013}_{-0.012} ) $ \\
$ 100\theta_\mathrm{MC}  $ & $  1.04091\pm 0.00022\, ( 1.04091^{+0.00042}_{-0.00044} ) $ & $  1.04090\pm 0.00023\, ( 1.04090^{+0.00045}_{-0.00045} ) $ \\
$ n_\mathrm{s}  $ & $  0.9702\pm 0.0037\, ( 0.9702^{+0.0072}_{-0.0073} ) $ & $  0.9702\pm 0.0037\, ( 0.9702^{+0.0074}_{-0.0071} ) $ \\
$ \log(10^{10} A_\mathrm{s})  $ & $  3.067^{+0.011}_{-0.012}\, ( 3.067^{+0.024}_{-0.022} ) $ & $  3.066^{+0.011}_{-0.012}\, ( 3.066^{+0.024}_{-0.022} ) $ \\
$ H_0  $ & $  68.10\pm 0.25\, ( 68.10^{+0.48}_{-0.49} ) $ & $  68.09\pm 0.25\, ( 68.09^{+0.50}_{-0.47} ) $ \\
\hline \hline
\end{tabular} }
\end{center}
\caption{ Results for $\Lambda\mathrm{CDM}+r+\alpha_s+\beta_s$ at $68\%\,(95\%)$ CL. }
\label{ tab:lcdm_r_nrun_nrunrun_results }
\end{table*}




\begin{table*}
\begin{center}
\renewcommand{\arraystretch}{1.5}
\resizebox{\textwidth}{!}{
\begin{tabular}{l  c c c}
\hline
\textbf{Parameter} & \textbf{ CMB+DESI+PP } & \textbf{ CMB+DESI+DD } \\
\hline\hline

$ \Omega_\mathrm{b} h^2  $ & $  0.02227\pm 0.00015\, ( 0.02227^{+0.00030}_{-0.00030} ) $ & $  0.02226\pm 0.00015\, ( 0.02226^{+0.00030}_{-0.00031} ) $ \\
$ \Omega_\mathrm{c} h^2  $ & $  0.1187\pm 0.0027\, ( 0.1187^{+0.0054}_{-0.0052} ) $ & $  0.1187\pm 0.0027\, ( 0.1187^{+0.0053}_{-0.0050} ) $ \\
$ \tau_\mathrm{reio}  $ & $  0.0632^{+0.0057}_{-0.0069}\, ( 0.063^{+0.013}_{-0.012} ) $ & $  0.0630^{+0.0057}_{-0.0067}\, ( 0.063^{+0.013}_{-0.012} ) $ \\
$ 100\theta_\mathrm{MC}  $ & $  1.04086\pm 0.00033\, ( 1.04086^{+0.00066}_{-0.00063} ) $ & $  1.04087\pm 0.00032\, ( 1.04087^{+0.00064}_{-0.00063} ) $ \\
$ n_\mathrm{s}  $ & $  0.9595\pm 0.0087\, ( 0.959^{+0.017}_{-0.017} ) $ & $  0.9588\pm 0.0087\, ( 0.959^{+0.017}_{-0.017} ) $ \\
$ \log(10^{10} A_\mathrm{s})  $ & $  3.060^{+0.012}_{-0.014}\, ( 3.060^{+0.028}_{-0.025} ) $ & $  3.059^{+0.012}_{-0.014}\, ( 3.059^{+0.027}_{-0.024} ) $ \\
$ H_0  $ & $  68.01\pm 0.96\, ( 68.0^{+1.9}_{-1.9} ) $ & $  67.94\pm 0.94\, ( 67.9^{+1.9}_{-1.8} ) $ \\
\hline \hline
\end{tabular} }
\end{center}
\caption{ Results for $\Lambda\mathrm{CDM}+r+\alpha_s+\beta_s+\Omega_k+\sum m_\nu+N_\mathrm{eff}$ at $68\%\,(95\%)$ CL. }
\label{ tab:lcdm_all_results }
\end{table*}




\begin{table*}
\begin{center}
\renewcommand{\arraystretch}{1.5}
\resizebox{\textwidth}{!}{
\begin{tabular}{l  c c}
\hline
\textbf{Parameter} & \textbf{ CMB+DESI+PP } & \textbf{ CMB+DESI+DD } \\
\hline\hline

$ \Omega_\mathrm{b} h^2  $ & $  0.022451\pm 0.000095\, ( 0.02245^{+0.00018}_{-0.00019} ) $ & $  0.022447\pm 0.000093\, ( 0.02245^{+0.00018}_{-0.00018} ) $ \\
$ \Omega_\mathrm{c} h^2  $ & $  0.11908\pm 0.00075\, ( 0.1191^{+0.0015}_{-0.0015} ) $ & $  0.11920\pm 0.00075\, ( 0.1192^{+0.0014}_{-0.0015} ) $ \\
$ \tau_\mathrm{reio}  $ & $  0.0608^{+0.0055}_{-0.0064}\, ( 0.061^{+0.012}_{-0.011} ) $ & $  0.0600^{+0.0052}_{-0.0060}\, ( 0.060^{+0.012}_{-0.010} ) $ \\
$ 100\theta_\mathrm{MC}  $ & $  1.04081\pm 0.00023\, ( 1.04081^{+0.00045}_{-0.00044} ) $ & $  1.04080\pm 0.00023\, ( 1.04080^{+0.00046}_{-0.00045} ) $ \\
$ n_\mathrm{s}  $ & $  0.9713\pm 0.0030\, ( 0.9713^{+0.0060}_{-0.0060} ) $ & $  0.9710\pm 0.0030\, ( 0.9710^{+0.0059}_{-0.0059} ) $ \\
$ \log(10^{10} A_\mathrm{s})  $ & $  3.056^{+0.010}_{-0.012}\, ( 3.056^{+0.022}_{-0.021} ) $ & $  3.0544^{+0.0095}_{-0.011}\, ( 3.054^{+0.021}_{-0.019} ) $ \\
$ H_0  $ & $  67.63\pm 0.61\, ( 67.6^{+1.2}_{-1.2} ) $ & $  67.44\pm 0.54\, ( 67.4^{+1.1}_{-1.1} ) $ \\
\hline \hline
\end{tabular} }
\end{center}
\caption{ Results for $w_0w_a\mathrm{CDM}$ at $68\%\,(95\%)$ CL. }
\label{ tab:cpl_results }
\end{table*}




\begin{table*}
\begin{center}
\renewcommand{\arraystretch}{1.5}
\resizebox{\textwidth}{!}{
\begin{tabular}{l  c c}
\hline
\textbf{Parameter} & \textbf{ CMB+DESI+PP } & \textbf{ CMB+DESI+DD } \\
\hline\hline

$ \Omega_\mathrm{b} h^2  $ & $  0.022452\pm 0.000092\, ( 0.02245^{+0.00018}_{-0.00018} ) $ & $  0.022447\pm 0.000094\, ( 0.02245^{+0.00018}_{-0.00018} ) $ \\
$ \Omega_\mathrm{c} h^2  $ & $  0.11908\pm 0.00074\, ( 0.1191^{+0.0014}_{-0.0015} ) $ & $  0.11920\pm 0.00075\, ( 0.1192^{+0.0014}_{-0.0015} ) $ \\
$ \tau_\mathrm{reio}  $ & $  0.0604^{+0.0054}_{-0.0063}\, ( 0.060^{+0.012}_{-0.011} ) $ & $  0.0600^{+0.0052}_{-0.0063}\, ( 0.060^{+0.012}_{-0.011} ) $ \\
$ 100\theta_\mathrm{MC}  $ & $  1.04082\pm 0.00023\, ( 1.04082^{+0.00045}_{-0.00045} ) $ & $  1.04081\pm 0.00023\, ( 1.04081^{+0.00045}_{-0.00045} ) $ \\
$ n_\mathrm{s}  $ & $  0.9713\pm 0.0031\, ( 0.9713^{+0.0060}_{-0.0060} ) $ & $  0.9711\pm 0.0031\, ( 0.9711^{+0.0060}_{-0.0060} ) $ \\
$ \log(10^{10} A_\mathrm{s})  $ & $  3.055\pm 0.011\, ( 3.055^{+0.023}_{-0.021} ) $ & $  3.0541^{+0.0099}_{-0.012}\, ( 3.054^{+0.023}_{-0.020} ) $ \\
$ H_0  $ & $  67.65\pm 0.60\, ( 67.6^{+1.2}_{-1.2} ) $ & $  67.44\pm 0.54\, ( 67.4^{+1.1}_{-1.0} ) $ \\
\hline \hline
\end{tabular} }
\end{center}
\caption{ Results for $w_0w_a\mathrm{CDM}+\sum m_\nu$ at $68\%\,(95\%)$ CL. }
\label{ tab:cpl_mnu_results }
\end{table*}




\begin{table*}
\begin{center}
\renewcommand{\arraystretch}{1.5}
\resizebox{\textwidth}{!}{
\begin{tabular}{l  c c}
\hline
\textbf{Parameter} & \textbf{ CMB+DESI+PP } & \textbf{ CMB+DESI+DD } \\
\hline\hline

$ \Omega_\mathrm{b} h^2  $ & $  0.02232\pm 0.00014\, ( 0.02232^{+0.00027}_{-0.00027} ) $ & $  0.02230\pm 0.00014\, ( 0.02230^{+0.00027}_{-0.00027} ) $ \\
$ \Omega_\mathrm{c} h^2  $ & $  0.1169\pm 0.0018\, ( 0.1169^{+0.0037}_{-0.0036} ) $ & $  0.1168\pm 0.0018\, ( 0.1168^{+0.0036}_{-0.0036} ) $ \\
$ \tau_\mathrm{reio}  $ & $  0.0600^{+0.0052}_{-0.0060}\, ( 0.060^{+0.012}_{-0.011} ) $ & $  0.0594^{+0.0052}_{-0.0061}\, ( 0.059^{+0.012}_{-0.011} ) $ \\
$ 100\theta_\mathrm{MC}  $ & $  1.04102\pm 0.00028\, ( 1.04102^{+0.00056}_{-0.00054} ) $ & $  1.04102\pm 0.00028\, ( 1.04102^{+0.00056}_{-0.00054} ) $ \\
$ n_\mathrm{s}  $ & $  0.9647\pm 0.0059\, ( 0.965^{+0.012}_{-0.011} ) $ & $  0.9637\pm 0.0059\, ( 0.964^{+0.012}_{-0.011} ) $ \\
$ \log(10^{10} A_\mathrm{s})  $ & $  3.048\pm 0.012\, ( 3.048^{+0.024}_{-0.023} ) $ & $  3.046\pm 0.012\, ( 3.046^{+0.024}_{-0.023} ) $ \\
$ H_0  $ & $  66.94\pm 0.81\, ( 66.9^{+1.6}_{-1.6} ) $ & $  66.65\pm 0.78\, ( 66.6^{+1.5}_{-1.5} ) $ \\
\hline \hline
\end{tabular} }
\end{center}
\caption{ Results for $w_0w_a\mathrm{CDM}+N_\mathrm{eff}$ at $68\%\,(95\%)$ CL. }
\label{ tab:cpl_nnu_results }
\end{table*}




\begin{table*}
\begin{center}
\renewcommand{\arraystretch}{1.5}
\resizebox{\textwidth}{!}{
\begin{tabular}{l  c c}
\hline
\textbf{Parameter} & \textbf{ CMB+DESI+PP } & \textbf{ CMB+DESI+DD } \\
\hline\hline

$ \Omega_\mathrm{b} h^2  $ & $  0.02232\pm 0.00014\, ( 0.02232^{+0.00027}_{-0.00027} ) $ & $  0.02230\pm 0.00014\, ( 0.02230^{+0.00026}_{-0.00026} ) $ \\
$ \Omega_\mathrm{c} h^2  $ & $  0.1168\pm 0.0018\, ( 0.1168^{+0.0037}_{-0.0035} ) $ & $  0.1168\pm 0.0018\, ( 0.1168^{+0.0036}_{-0.0035} ) $ \\
$ \tau_\mathrm{reio}  $ & $  0.0595^{+0.0052}_{-0.0061}\, ( 0.059^{+0.012}_{-0.011} ) $ & $  0.0590^{+0.0051}_{-0.0064}\, ( 0.059^{+0.012}_{-0.011} ) $ \\
$ 100\theta_\mathrm{MC}  $ & $  1.04102\pm 0.00028\, ( 1.04102^{+0.00055}_{-0.00056} ) $ & $  1.04103\pm 0.00028\, ( 1.04103^{+0.00056}_{-0.00054} ) $ \\
$ n_\mathrm{s}  $ & $  0.9645\pm 0.0059\, ( 0.964^{+0.012}_{-0.012} ) $ & $  0.9637\pm 0.0059\, ( 0.964^{+0.011}_{-0.011} ) $ \\
$ \log(10^{10} A_\mathrm{s})  $ & $  3.047^{+0.011}_{-0.013}\, ( 3.047^{+0.025}_{-0.023} ) $ & $  3.046^{+0.011}_{-0.013}\, ( 3.046^{+0.026}_{-0.023} ) $ \\
$ H_0  $ & $  66.92\pm 0.81\, ( 66.9^{+1.6}_{-1.6} ) $ & $  66.66\pm 0.77\, ( 66.7^{+1.5}_{-1.5} ) $ \\
\hline \hline
\end{tabular} }
\end{center}
\caption{ Results for $w_0w_a\mathrm{CDM}+\sum m_\nu+N_\mathrm{eff}$ at $68\%\,(95\%)$ CL. }
\label{ tab:cpl_mnu_nnu_results }
\end{table*}




\begin{table*}
\begin{center}
\renewcommand{\arraystretch}{1.5}
\resizebox{\textwidth}{!}{
\begin{tabular}{l  c c}
\hline
\textbf{Parameter} & \textbf{ CMB+DESI+PP } & \textbf{ CMB+DESI+DD } \\
\hline\hline

$ \Omega_\mathrm{b} h^2  $ & $  0.022433\pm 0.000097\, ( 0.02243^{+0.00019}_{-0.00019} ) $ & $  0.022434\pm 0.000096\, ( 0.02243^{+0.00019}_{-0.00019} ) $ \\
$ \Omega_\mathrm{c} h^2  $ & $  0.1197\pm 0.0011\, ( 0.1197^{+0.0021}_{-0.0020} ) $ & $  0.1197\pm 0.0011\, ( 0.1197^{+0.0021}_{-0.0021} ) $ \\
$ \tau_\mathrm{reio}  $ & $  0.0602^{+0.0054}_{-0.0061}\, ( 0.060^{+0.012}_{-0.011} ) $ & $  0.0595^{+0.0051}_{-0.0063}\, ( 0.060^{+0.012}_{-0.011} ) $ \\
$ 100\theta_\mathrm{MC}  $ & $  1.04074\pm 0.00023\, ( 1.04074^{+0.00045}_{-0.00046} ) $ & $  1.04074\pm 0.00024\, ( 1.04074^{+0.00046}_{-0.00046} ) $ \\
$ n_\mathrm{s}  $ & $  0.9699\pm 0.0034\, ( 0.9699^{+0.0067}_{-0.0068} ) $ & $  0.9700\pm 0.0035\, ( 0.9700^{+0.0070}_{-0.0068} ) $ \\
$ \log(10^{10} A_\mathrm{s})  $ & $  3.0563^{+0.0098}_{-0.011}\, ( 3.056^{+0.022}_{-0.020} ) $ & $  3.0546^{+0.0097}_{-0.011}\, ( 3.055^{+0.022}_{-0.020} ) $ \\
$ H_0  $ & $  67.76\pm 0.61\, ( 67.8^{+1.2}_{-1.2} ) $ & $  67.55\pm 0.58\, ( 67.6^{+1.1}_{-1.1} ) $ \\
\hline \hline
\end{tabular} }
\end{center}
\caption{ Results for $w_0w_a\mathrm{CDM}+\Omega_k$ at $68\%\,(95\%)$ CL. }
\label{ tab:cpl_omk_results }
\end{table*}




\begin{table*}
\begin{center}
\renewcommand{\arraystretch}{1.5}
\resizebox{\textwidth}{!}{
\begin{tabular}{l  c c}
\hline
\textbf{Parameter} & \textbf{ CMB+DESI+PP } & \textbf{ CMB+DESI+DD } \\
\hline\hline

$ \Omega_\mathrm{b} h^2  $ & $  0.022452\pm 0.000092\, ( 0.02245^{+0.00018}_{-0.00018} ) $ & $  0.022448\pm 0.000093\, ( 0.02245^{+0.00018}_{-0.00019} ) $ \\
$ \Omega_\mathrm{c} h^2  $ & $  0.11905\pm 0.00076\, ( 0.1191^{+0.0015}_{-0.0015} ) $ & $  0.11915\pm 0.00077\, ( 0.1191^{+0.0015}_{-0.0015} ) $ \\
$ \tau_\mathrm{reio}  $ & $  0.0605^{+0.0051}_{-0.0064}\, ( 0.060^{+0.012}_{-0.011} ) $ & $  0.0599^{+0.0052}_{-0.0061}\, ( 0.060^{+0.012}_{-0.011} ) $ \\
$ 100\theta_\mathrm{MC}  $ & $  1.04081\pm 0.00023\, ( 1.04081^{+0.00044}_{-0.00045} ) $ & $  1.04080\pm 0.00023\, ( 1.04080^{+0.00045}_{-0.00045} ) $ \\
$ n_\mathrm{s}  $ & $  0.9716\pm 0.0030\, ( 0.9716^{+0.0060}_{-0.0059} ) $ & $  0.9714\pm 0.0031\, ( 0.9714^{+0.0061}_{-0.0060} ) $ \\
$ \log(10^{10} A_\mathrm{s})  $ & $  3.0552^{+0.0096}_{-0.012}\, ( 3.055^{+0.023}_{-0.020} ) $ & $  3.054^{+0.010}_{-0.011}\, ( 3.054^{+0.022}_{-0.020} ) $ \\
$ H_0  $ & $  67.65\pm 0.61\, ( 67.6^{+1.2}_{-1.2} ) $ & $  67.45\pm 0.54\, ( 67.5^{+1.1}_{-1.0} ) $ \\
\hline \hline
\end{tabular} }
\end{center}
\caption{ Results for $w_0w_a\mathrm{CDM}+r$ at $68\%\,(95\%)$ CL. }
\label{ tab:cpl_r_results }
\end{table*}




\begin{table*}
\begin{center}
\renewcommand{\arraystretch}{1.5}
\resizebox{\textwidth}{!}{
\begin{tabular}{l  c c}
\hline
\textbf{Parameter} & \textbf{ CMB+DESI+PP } & \textbf{ CMB+DESI+DD } \\
\hline\hline

$ \Omega_\mathrm{b} h^2  $ & $  0.02241\pm 0.00010\, ( 0.02241^{+0.00020}_{-0.00020} ) $ & $  0.02241\pm 0.00010\, ( 0.02241^{+0.00020}_{-0.00020} ) $ \\
$ \Omega_\mathrm{c} h^2  $ & $  0.11904\pm 0.00074\, ( 0.1190^{+0.0014}_{-0.0015} ) $ & $  0.11915\pm 0.00074\, ( 0.1191^{+0.0014}_{-0.0014} ) $ \\
$ \tau_\mathrm{reio}  $ & $  0.0595^{+0.0053}_{-0.0064}\, ( 0.060^{+0.012}_{-0.011} ) $ & $  0.0587^{+0.0052}_{-0.0063}\, ( 0.059^{+0.012}_{-0.011} ) $ \\
$ 100\theta_\mathrm{MC}  $ & $  1.04081\pm 0.00022\, ( 1.04081^{+0.00043}_{-0.00044} ) $ & $  1.04080\pm 0.00023\, ( 1.04080^{+0.00046}_{-0.00045} ) $ \\
$ n_\mathrm{s}  $ & $  0.9716\pm 0.0031\, ( 0.9716^{+0.0061}_{-0.0060} ) $ & $  0.9714\pm 0.0030\, ( 0.9714^{+0.0059}_{-0.0058} ) $ \\
$ \log(10^{10} A_\mathrm{s})  $ & $  3.052^{+0.010}_{-0.012}\, ( 3.052^{+0.023}_{-0.021} ) $ & $  3.051^{+0.010}_{-0.012}\, ( 3.051^{+0.023}_{-0.021} ) $ \\
$ H_0  $ & $  67.63\pm 0.59\, ( 67.6^{+1.2}_{-1.1} ) $ & $  67.44\pm 0.55\, ( 67.4^{+1.1}_{-1.1} ) $ \\
\hline \hline
\end{tabular} }
\end{center}
\caption{ Results for $w_0w_a\mathrm{CDM}+r+\alpha_s$ at $68\%\,(95\%)$ CL.}
\label{ tab:cpl_r_nrun_results }
\end{table*}




\begin{table*}
\begin{center}
\renewcommand{\arraystretch}{1.5}
\resizebox{\textwidth}{!}{
\begin{tabular}{l  c c}
\hline
\textbf{Parameter} & \textbf{ CMB+DESI+PP } & \textbf{ CMB+DESI+DD } \\
\hline\hline

$ \Omega_\mathrm{b} h^2  $ & $  0.02239\pm 0.00010\, ( 0.02239^{+0.00020}_{-0.00020} ) $ & $  0.02238\pm 0.00010\, ( 0.02238^{+0.00020}_{-0.00020} ) $ \\
$ \Omega_\mathrm{c} h^2  $ & $  0.11914\pm 0.00075\, ( 0.1191^{+0.0015}_{-0.0015} ) $ & $  0.11926\pm 0.00075\, ( 0.1193^{+0.0015}_{-0.0015} ) $ \\
$ \tau_\mathrm{reio}  $ & $  0.0612^{+0.0056}_{-0.0067}\, ( 0.061^{+0.012}_{-0.012} ) $ & $  0.0606^{+0.0054}_{-0.0062}\, ( 0.061^{+0.012}_{-0.011} ) $ \\
$ 100\theta_\mathrm{MC}  $ & $  1.04081\pm 0.00023\, ( 1.04081^{+0.00044}_{-0.00044} ) $ & $  1.04080\pm 0.00023\, ( 1.04080^{+0.00045}_{-0.00045} ) $ \\
$ n_\mathrm{s}  $ & $  0.9681\pm 0.0039\, ( 0.9681^{+0.0075}_{-0.0076} ) $ & $  0.9679\pm 0.0039\, ( 0.9679^{+0.0076}_{-0.0075} ) $ \\
$ \log(10^{10} A_\mathrm{s})  $ & $  3.055^{+0.011}_{-0.012}\, ( 3.055^{+0.024}_{-0.022} ) $ & $  3.054^{+0.010}_{-0.012}\, ( 3.054^{+0.023}_{-0.021} ) $ \\
$ H_0  $ & $  67.59\pm 0.60\, ( 67.6^{+1.2}_{-1.2} ) $ & $  67.42\pm 0.54\, ( 67.4^{+1.1}_{-1.1} ) $ \\
\hline \hline
\end{tabular} }
\end{center}
\caption{ Results for $w_0w_a\mathrm{CDM}+r+\alpha_s+\beta_s$ at $68\%\,(95\%)$ CL. }
\label{ tab:cpl_r_nrun_nrunrun_results }
\end{table*}




\begin{table*}
\begin{center}
\renewcommand{\arraystretch}{1.5}
\resizebox{\textwidth}{!}{
\begin{tabular}{l  c c}
\hline
\textbf{Parameter} & \textbf{ CMB+DESI+PP } & \textbf{ CMB+DESI+DD } \\
\hline\hline

$ \Omega_\mathrm{b} h^2  $ & $  0.02226\pm 0.00017\, ( 0.02226^{+0.00033}_{-0.00032} ) $ & $  0.02225\pm 0.00017\, ( 0.02225^{+0.00033}_{-0.00032} ) $ \\
$ \Omega_\mathrm{c} h^2  $ & $  0.1182^{+0.0027}_{-0.0030}\, ( 0.1182^{+0.0058}_{-0.0054} ) $ & $  0.1179\pm 0.0028\, ( 0.1179^{+0.0059}_{-0.0053} ) $ \\
$ \tau_\mathrm{reio}  $ & $  0.0611^{+0.0055}_{-0.0067}\, ( 0.061^{+0.013}_{-0.011} ) $ & $  0.0608^{+0.0057}_{-0.0069}\, ( 0.061^{+0.014}_{-0.012} ) $ \\
$ 100\theta_\mathrm{MC}  $ & $  1.04090\pm 0.00034\, ( 1.04090^{+0.00067}_{-0.00067} ) $ & $  1.04092\pm 0.00034\, ( 1.04092^{+0.00066}_{-0.00067} ) $ \\
$ n_\mathrm{s}  $ & $  0.9591\pm 0.0094\, ( 0.959^{+0.019}_{-0.018} ) $ & $  0.9580\pm 0.0094\, ( 0.958^{+0.018}_{-0.018} ) $ \\
$ \log(10^{10} A_\mathrm{s})  $ & $  3.053^{+0.013}_{-0.014}\, ( 3.053^{+0.029}_{-0.026} ) $ & $  3.052^{+0.013}_{-0.015}\, ( 3.052^{+0.030}_{-0.027} ) $ \\
$ H_0  $ & $  67.1\pm 1.0\, ( 67.1^{+2.1}_{-2.0} ) $ & $  66.8\pm 1.0\, ( 66.8^{+2.0}_{-1.9} ) $ \\
\hline \hline
\end{tabular} }
\end{center}
\caption{ Results for $w_0w_a\mathrm{CDM}+r+\alpha_s+\beta_s+\Omega_k+\sum m_\nu+N_{\mathrm{eff}}$ at $68\%\,(95\%)$ CL.}
\label{ tab:cpl_all_results }
\end{table*}

\clearpage
\section{Posterior Distributions}
\label{appendix.Plots}

\begin{figure}[h]
    \centering
    \includegraphics[width=1\linewidth]{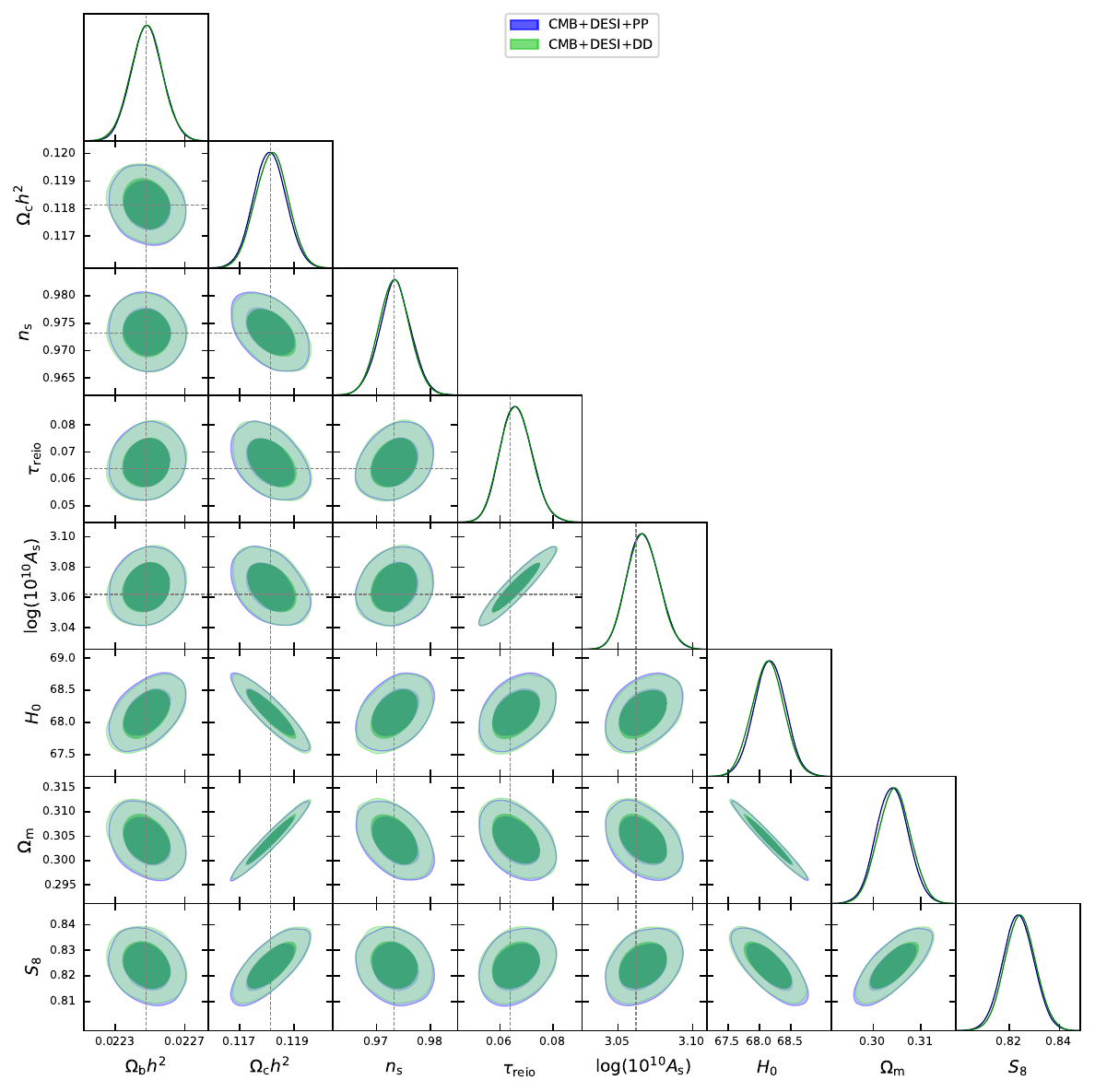}
    \caption{$\Lambda\mathrm{CDM}$ parameter constraints.}
    \label{fig:triangle_LCDM}
\end{figure}

\begin{figure}[h]
    \centering
    \includegraphics[width=1\linewidth]{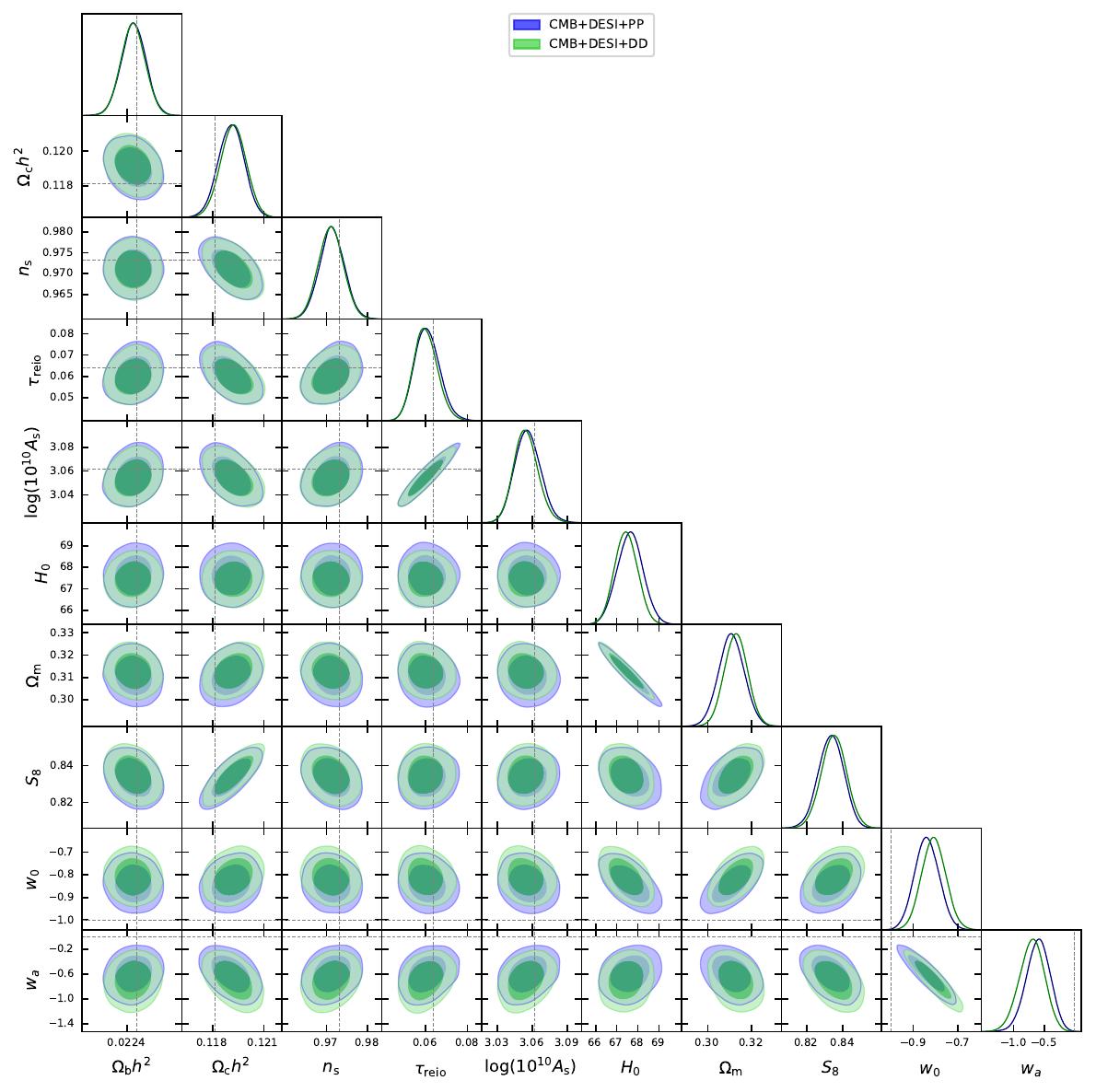}
    \caption{$w_0w_a\mathrm{CDM}$ parameter constraints.}
    \label{fig:triangle_CPL}
\end{figure}

\begin{figure}[h]
    \centering
    \includegraphics[width=1\linewidth]{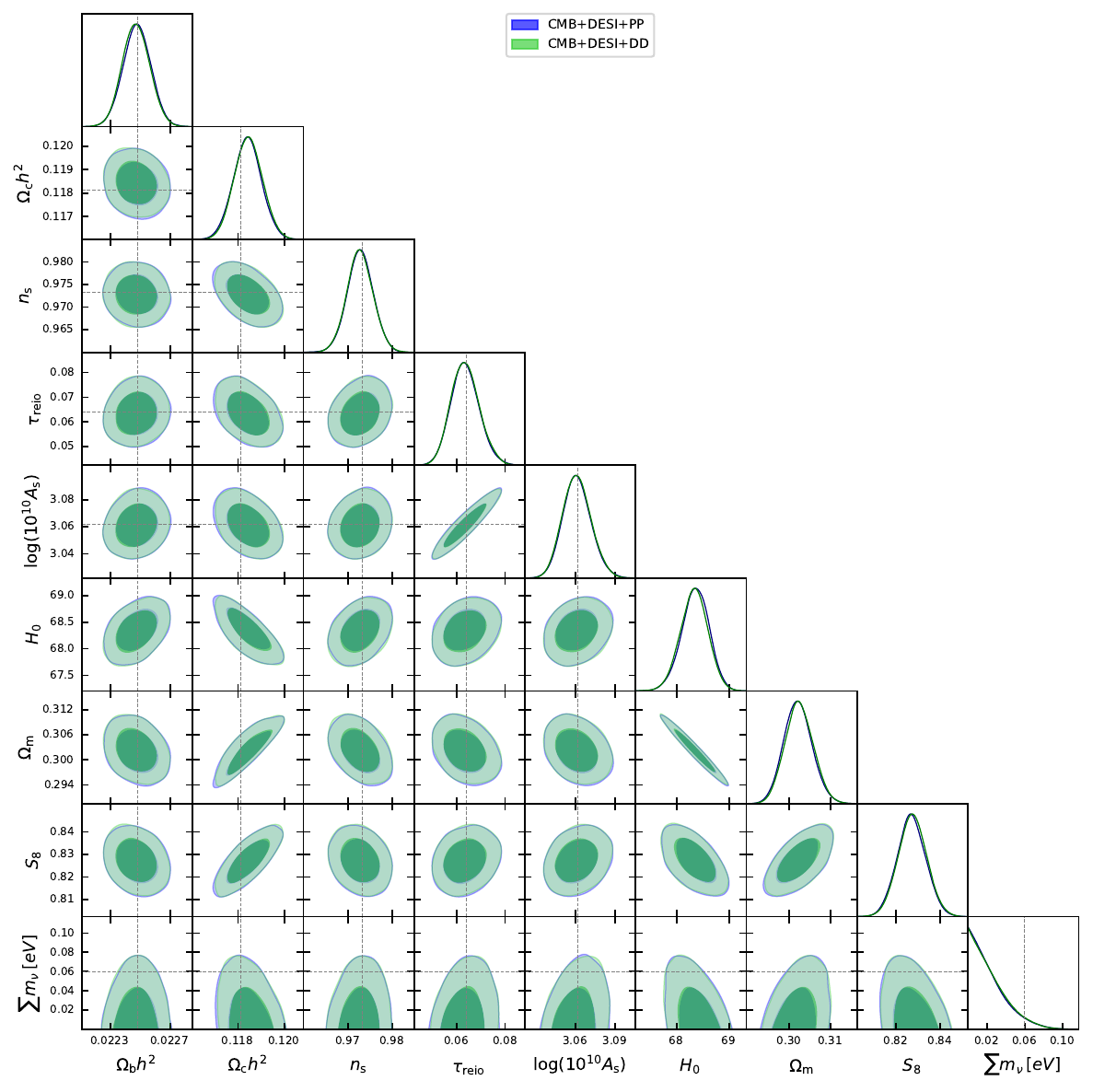}
    \caption{$\Lambda\mathrm{CDM}+\sum m_\nu$ parameter constraints.}
    \label{fig:triangle_LCDM_mnu}
\end{figure}

\begin{figure}[h]
    \centering
    \includegraphics[width=1\linewidth]{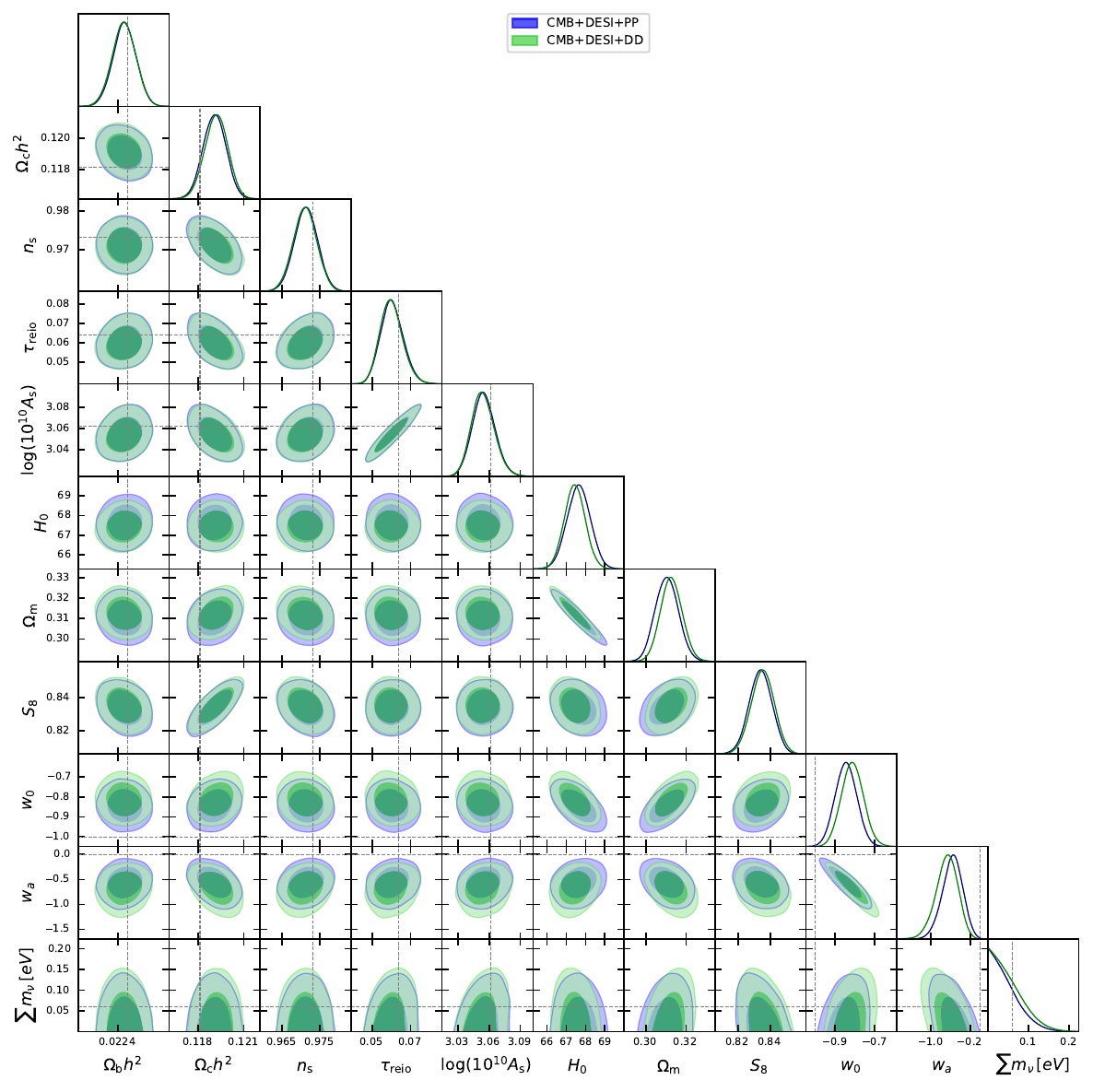}
    \caption{$w_0w_a\mathrm{CDM}+\sum m_\nu$ parameter constraints.}
    \label{fig:triangle_CPL_mnu}
\end{figure}

\begin{figure}[h]
    \centering
    \includegraphics[width=1\linewidth]{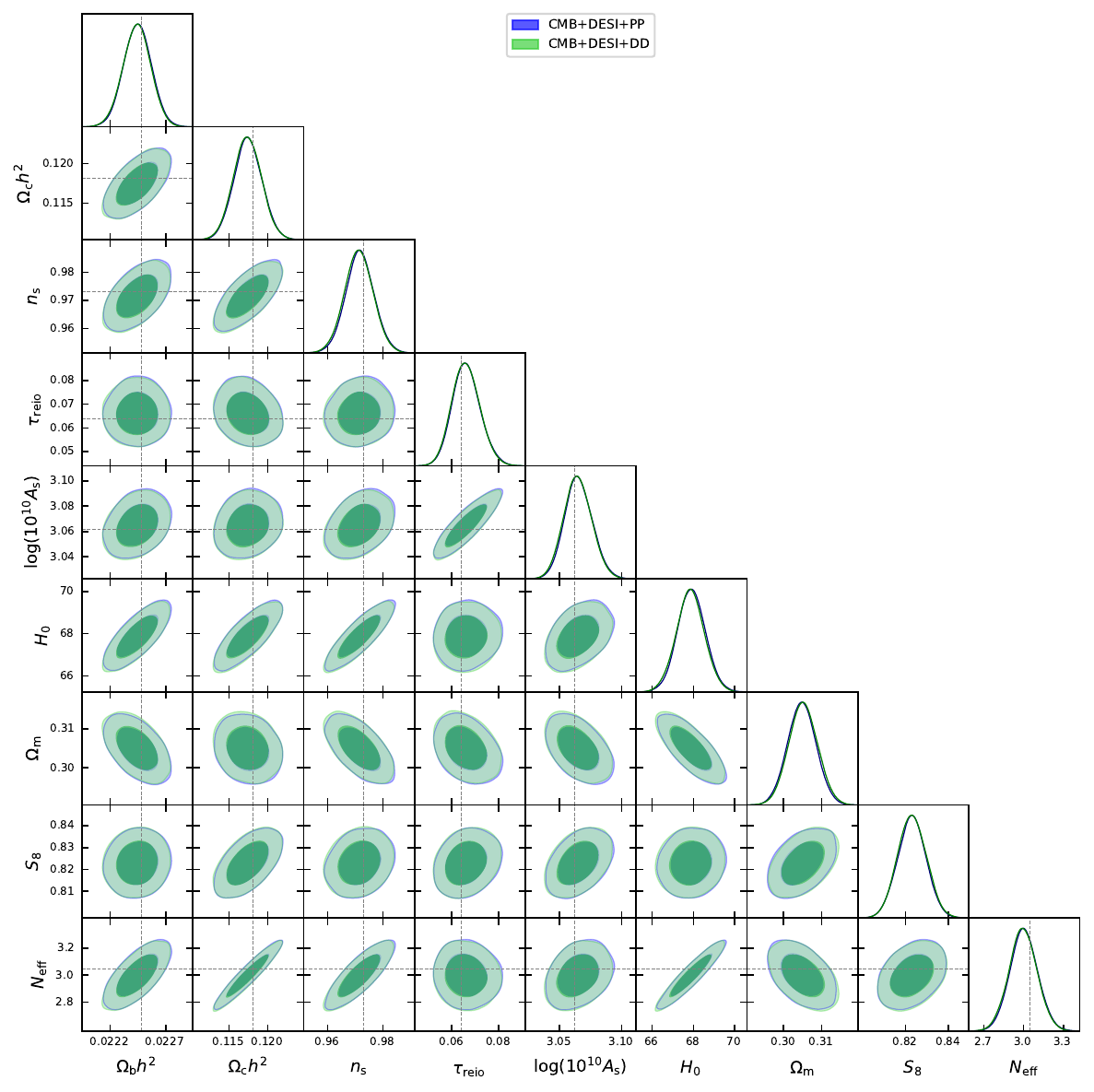}
    \caption{$\Lambda\mathrm{CDM}+N_\mathrm{eff}$ parameter constraints.}
    \label{fig:triangle_LCDM_nnu}
\end{figure}

\begin{figure}[h]
    \centering
    \includegraphics[width=1\linewidth]{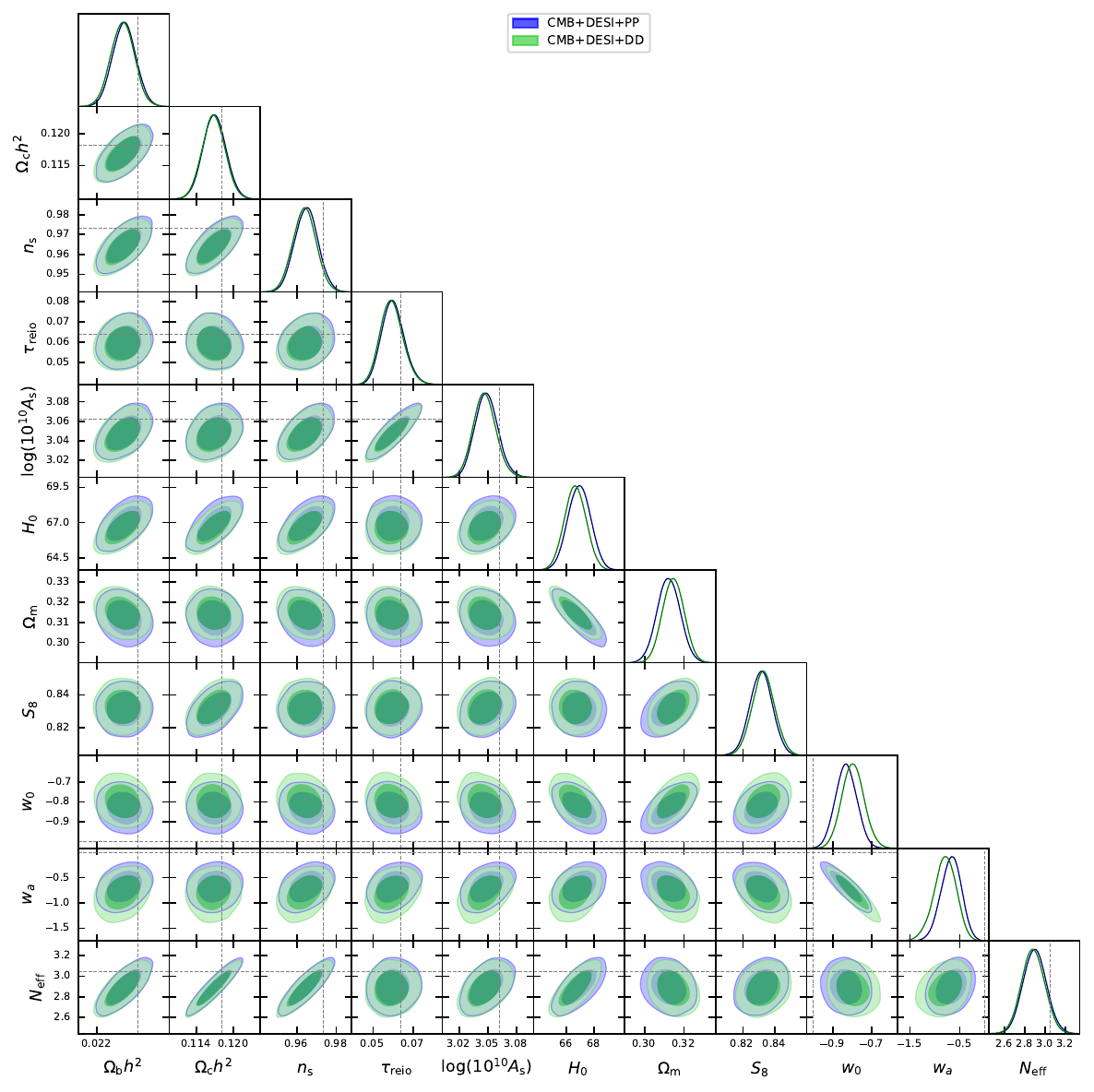}
    \caption{$w_0w_a\mathrm{CDM}+N_\mathrm{eff}$ parameter constraints.}
    \label{fig:triangle_CPL_nnu}
\end{figure}

\begin{figure}[h]
    \centering
    \includegraphics[width=1\linewidth]{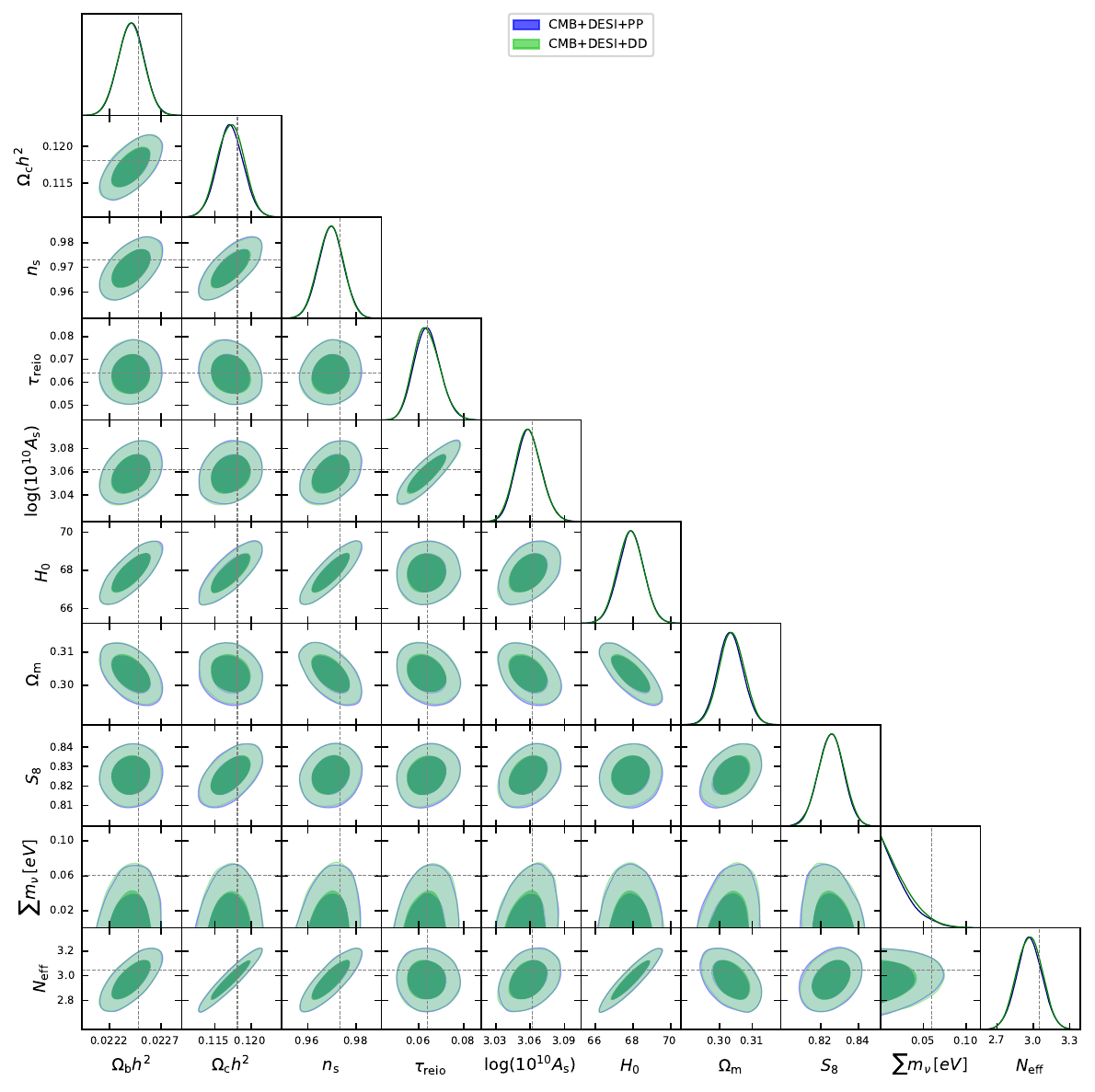}
    \caption{$\Lambda\mathrm{CDM}+\sum m_\nu+N_\mathrm{eff}$ parameter constraints.}
    \label{fig:triangle_LCDM_mnu_nnu}
\end{figure}

\begin{figure}[h]
    \centering
    \includegraphics[width=1\linewidth]{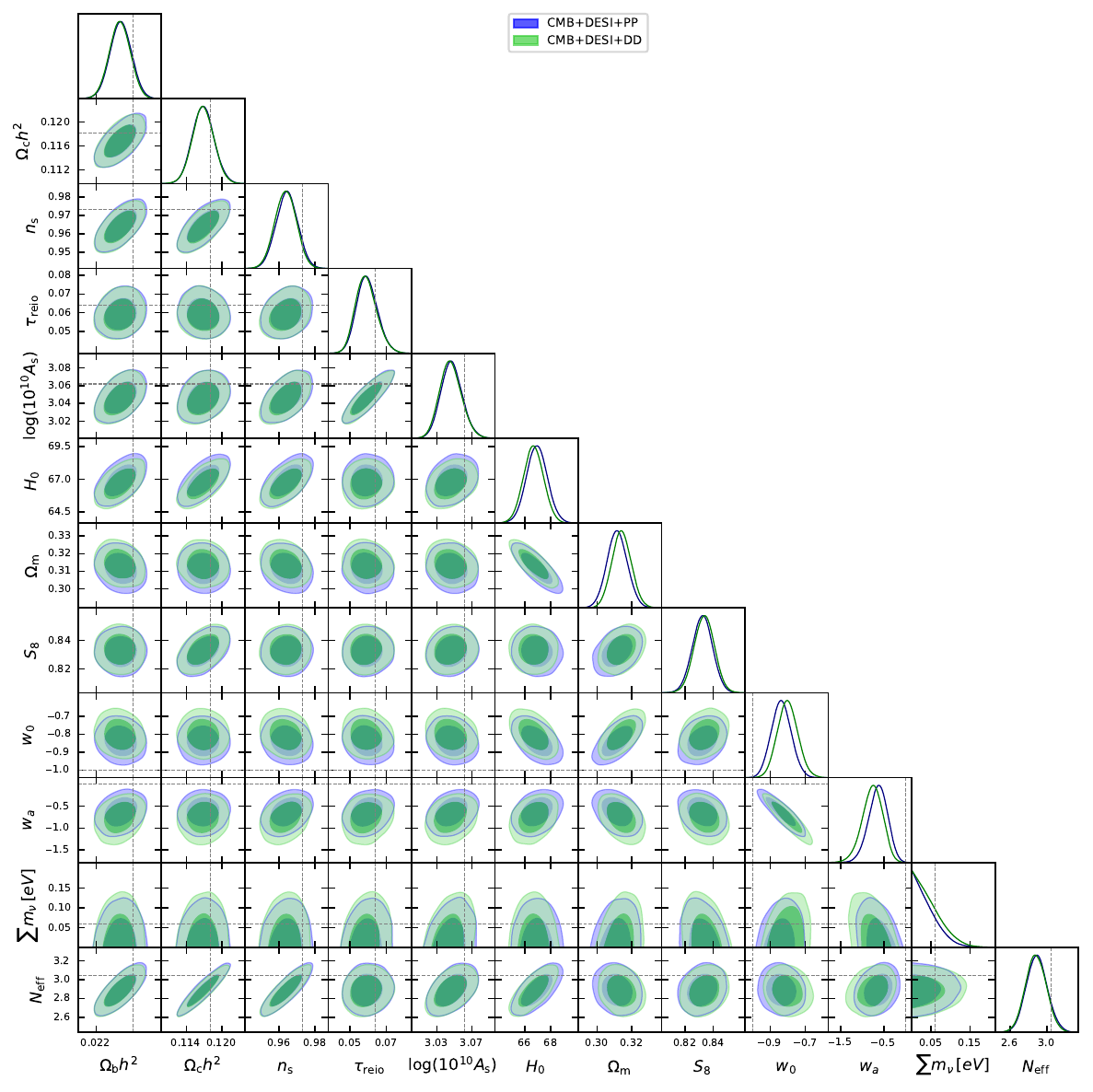}
    \caption{$w_0w_a\mathrm{CDM}+\sum m_\nu+N_\mathrm{eff}$ parameter constraints.}
    \label{fig:triangle_CPL_mnu_nnu}
\end{figure}

\begin{figure}[h]
    \centering
    \includegraphics[width=1\linewidth]{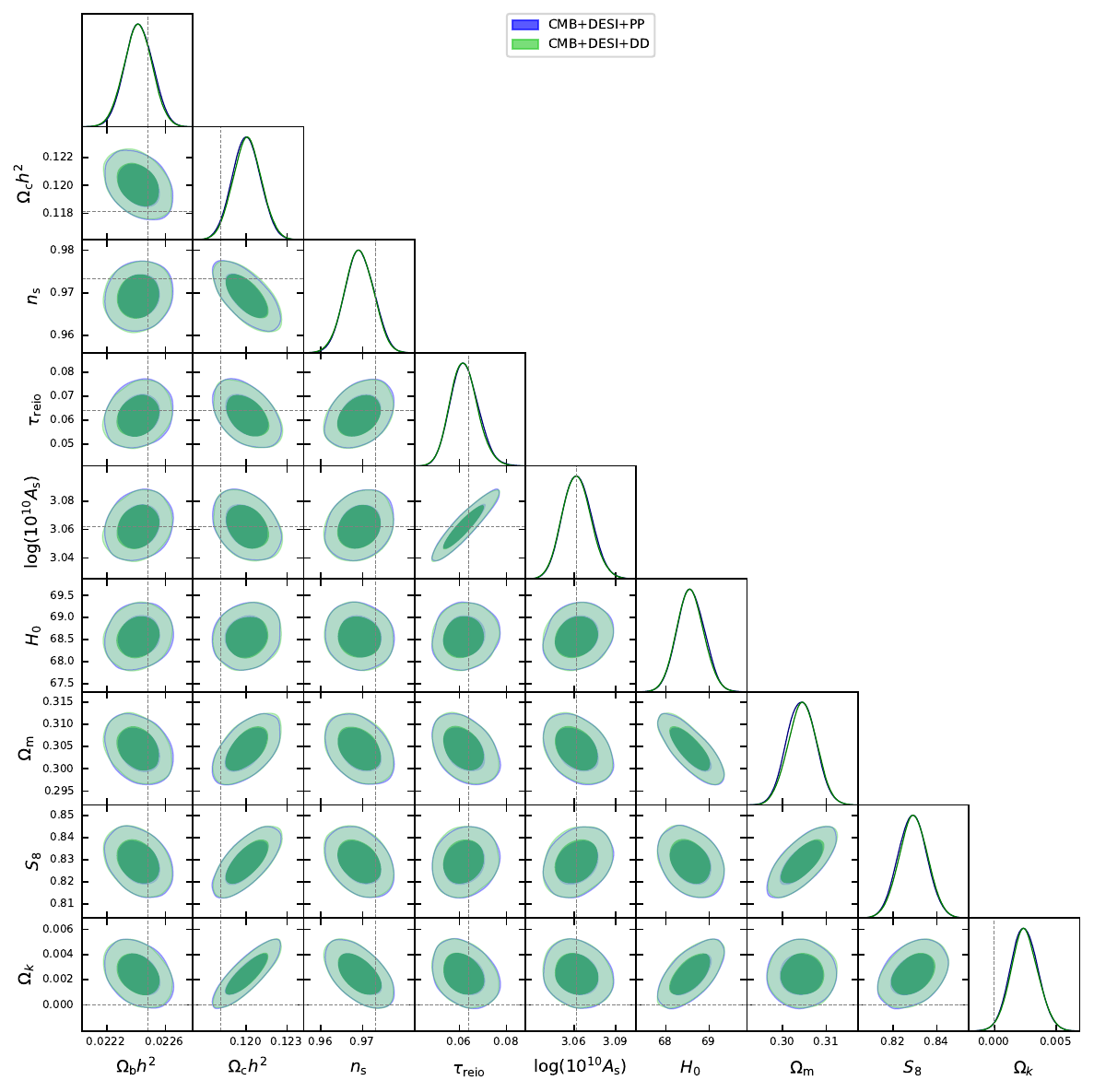}
    \caption{$\Lambda\mathrm{CDM}+\Omega_k$ parameter constraints.}
    \label{fig:triangle_LCDM_omk}
\end{figure}

\begin{figure}[h]
    \centering
    \includegraphics[width=1\linewidth]{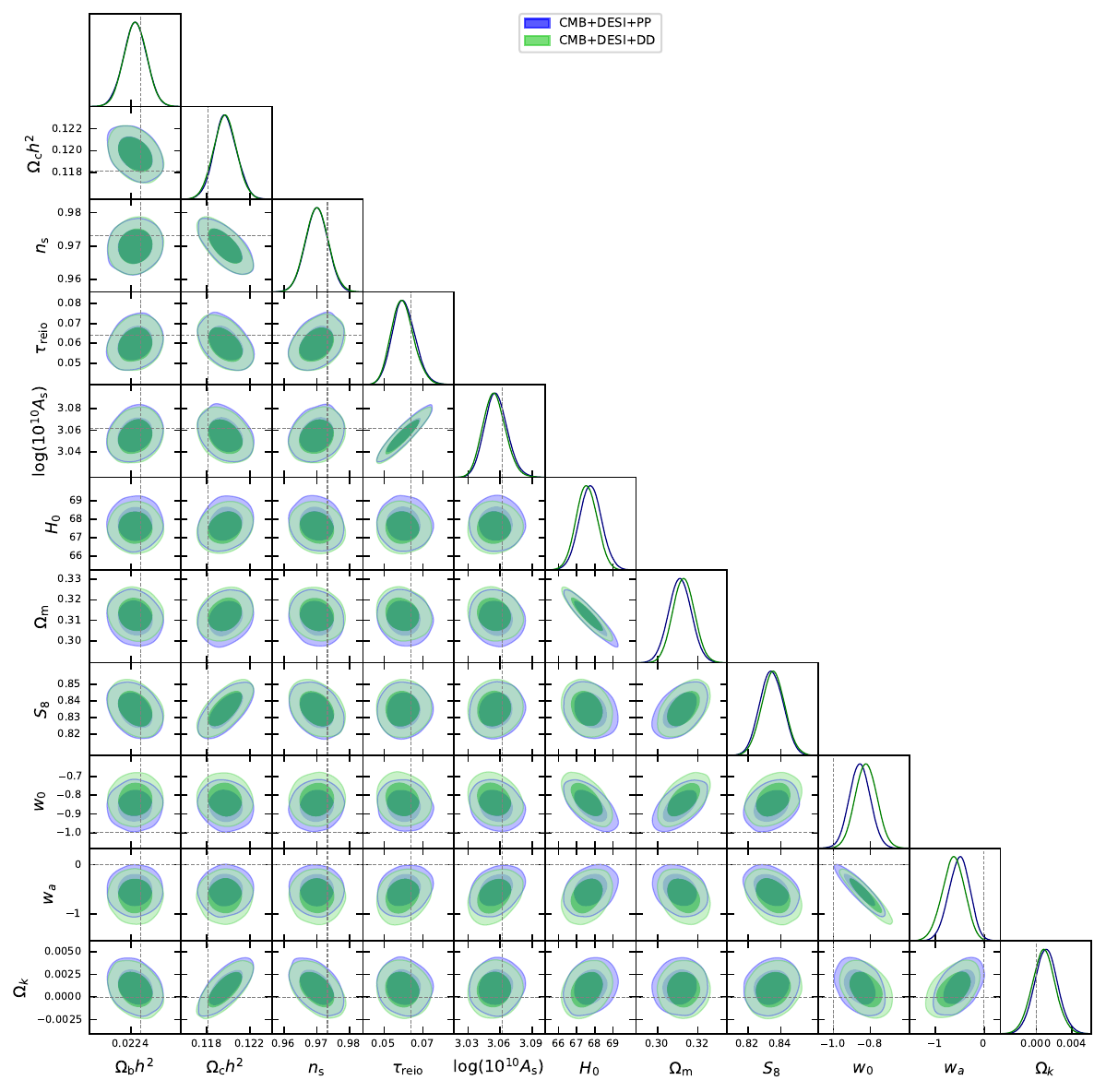}
    \caption{$w_0w_a\mathrm{CDM}+\Omega_k$ parameter constraints.}
    \label{fig:triangle_CPL_omk}
\end{figure}

\begin{figure}[h]
    \centering
    \includegraphics[width=1\linewidth]{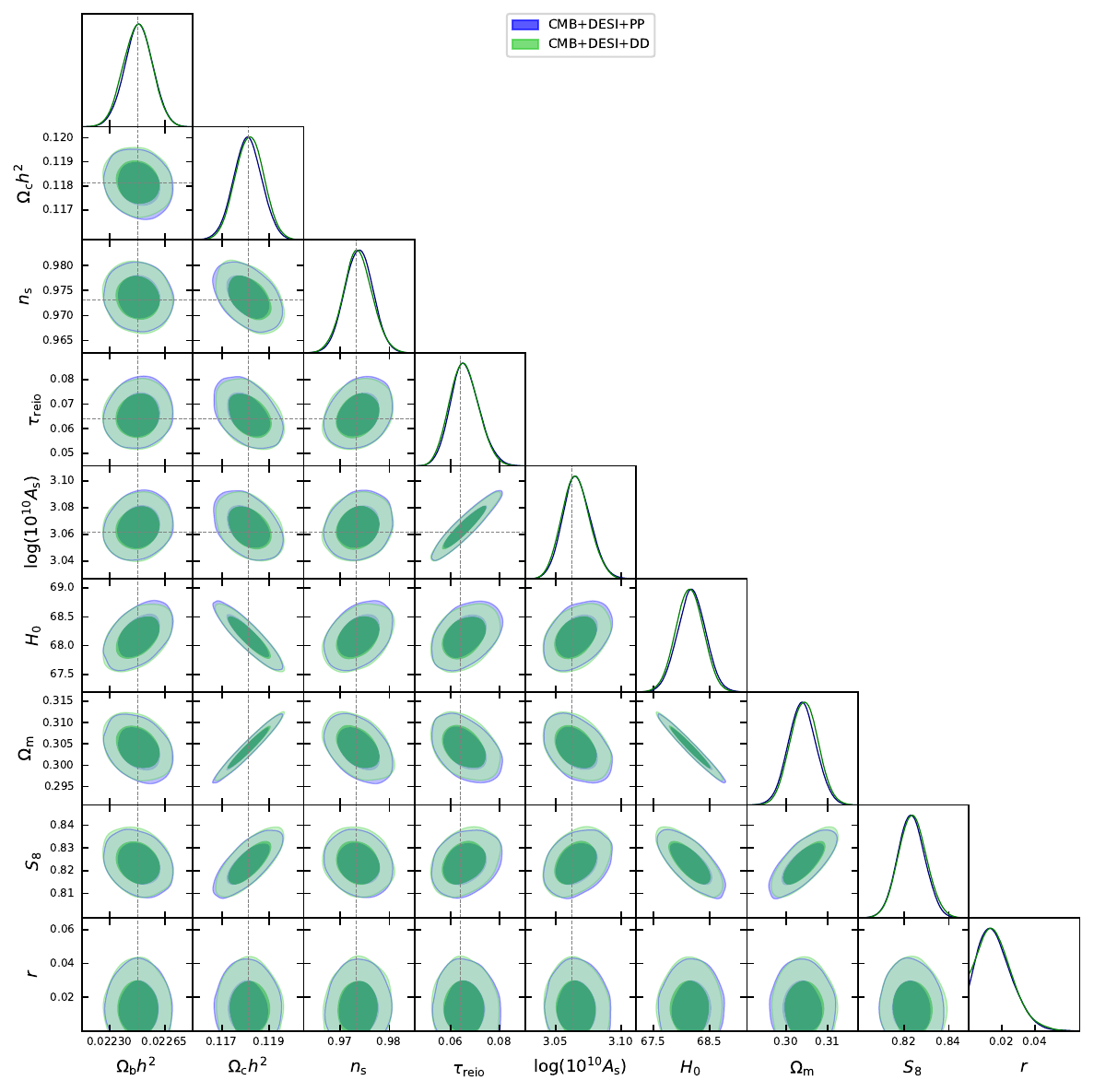}
    \caption{$\Lambda\mathrm{CDM}+r$ parameter constraints.}
    \label{fig:triangle_LCDM_r}
\end{figure}

\begin{figure}[h]
    \centering
    \includegraphics[width=1\linewidth]{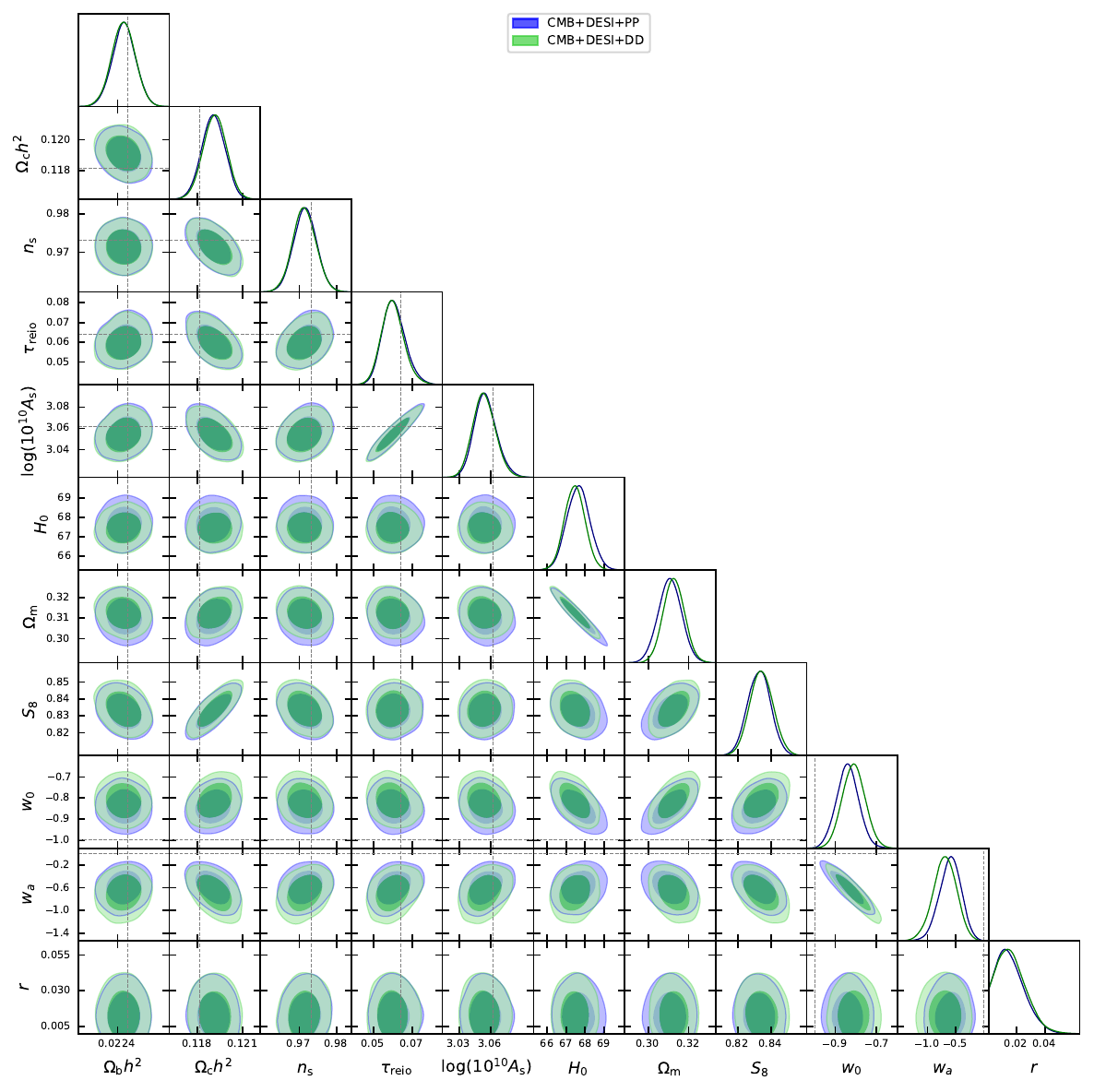}
    \caption{$w_0w_a\mathrm{CDM}+r$ parameter constraints.}
    \label{fig:triangle_CPL_r}
\end{figure}

\begin{figure}[h]
    \centering
    \includegraphics[width=1\linewidth]{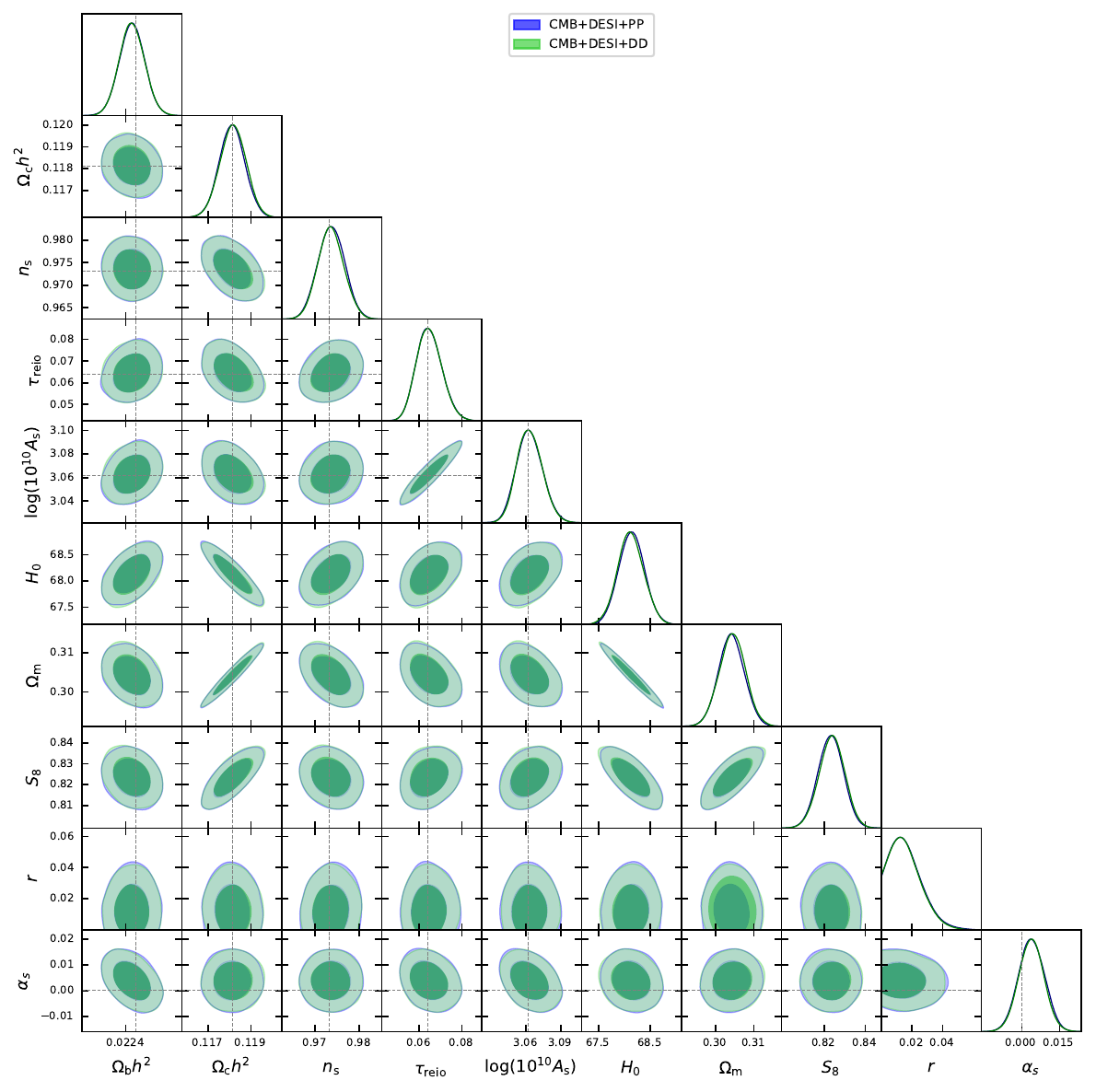}
    \caption{$\Lambda\mathrm{CDM}+r+\alpha_s$ parameter constraints.}
    \label{fig:triangle_LCDM_r_nrun}
\end{figure}

\begin{figure}[h]
    \centering
    \includegraphics[width=1\linewidth]{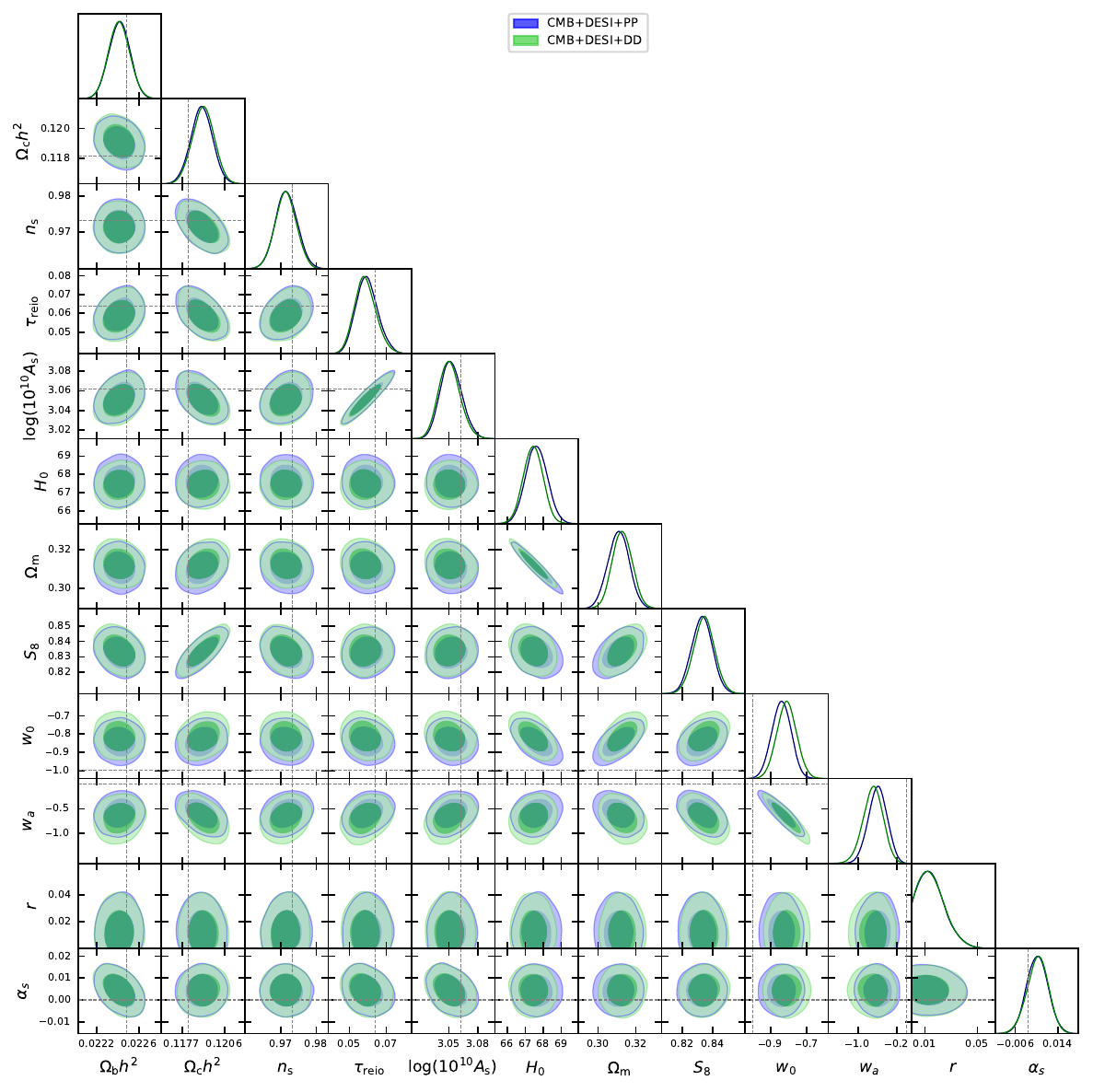}
    \caption{$w_0w_a\mathrm{CDM}+r+\alpha_s$ parameter constraints.}
    \label{fig:triangle_CPL_r_nrun}
\end{figure}

\begin{figure}[h]
    \centering
    \includegraphics[width=1\linewidth]{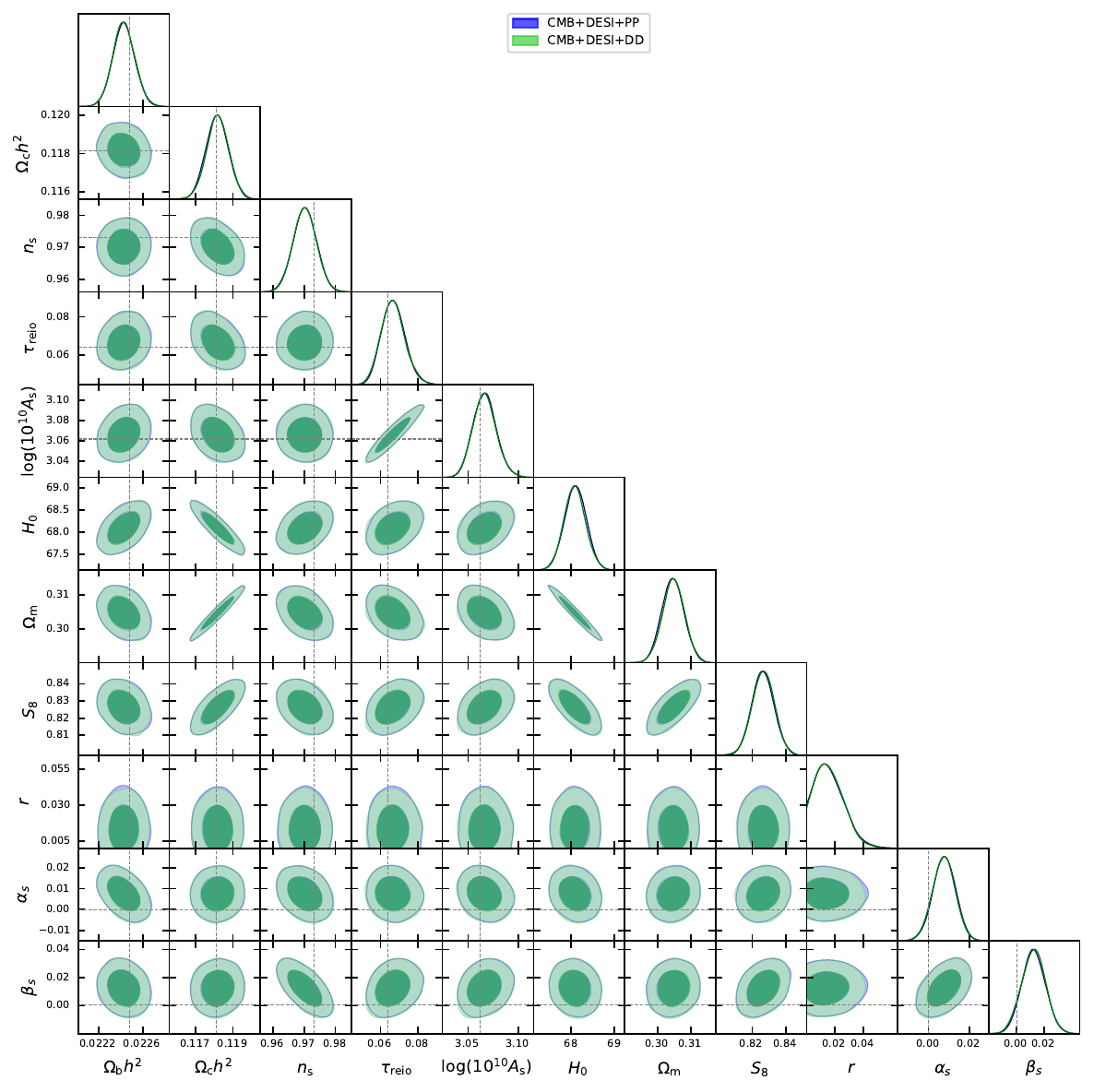}
    \caption{$\Lambda\mathrm{CDM}+r+\alpha_s+\beta_s$ parameter constraints.}
    \label{fig:triangle_LCDM_r_nrun_nrunrun}
\end{figure}

\begin{figure}[h]
    \centering
    \includegraphics[width=1\linewidth]{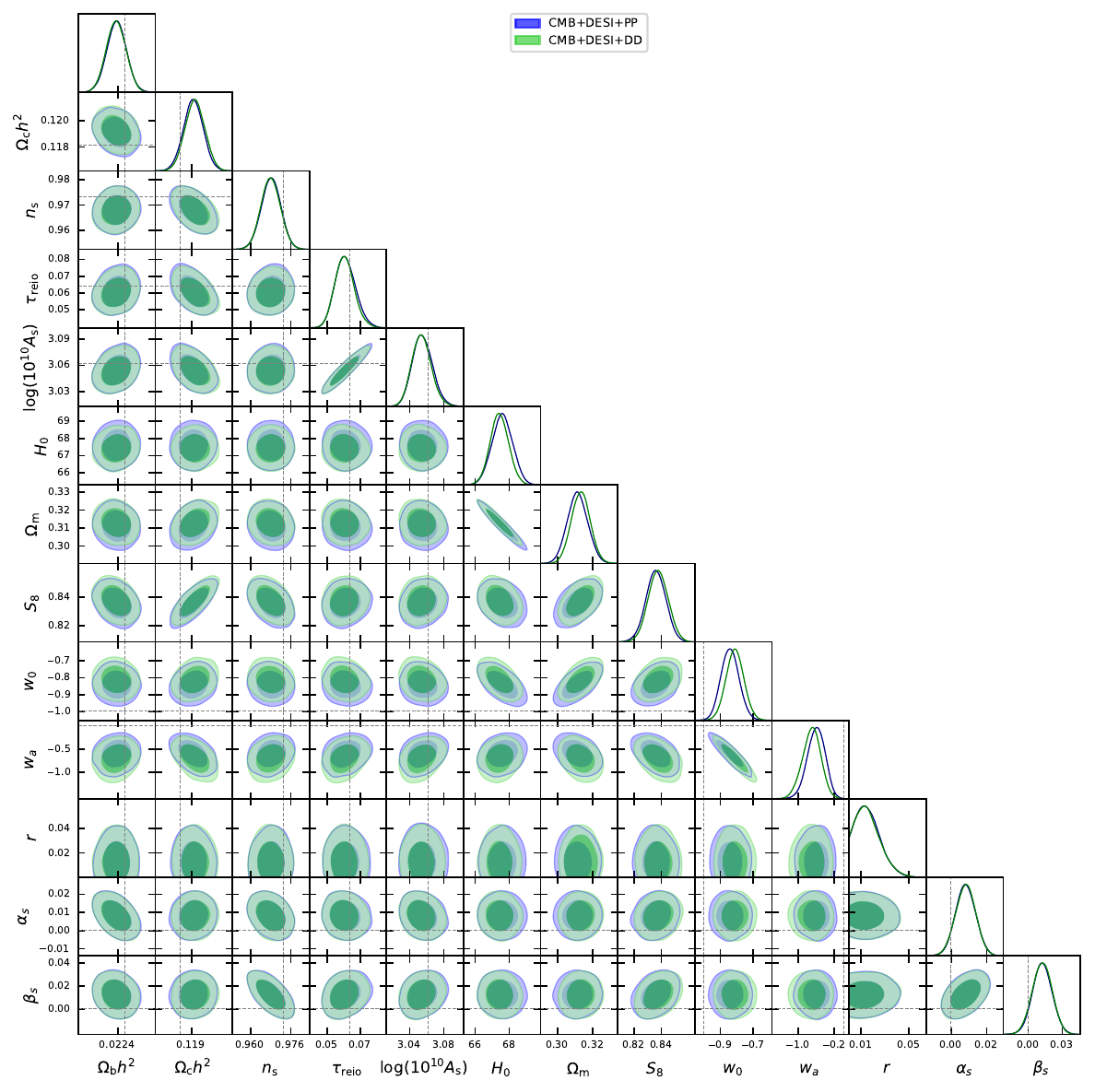}
    \caption{$w_0w_a\mathrm{CDM}+r+\alpha_s+\beta_s$ parameter constraints.}
    \label{fig:triangle_CPL_r_nrun_nrunrun}
\end{figure}

\begin{figure}[h]
    \centering
    \includegraphics[width=1\linewidth]{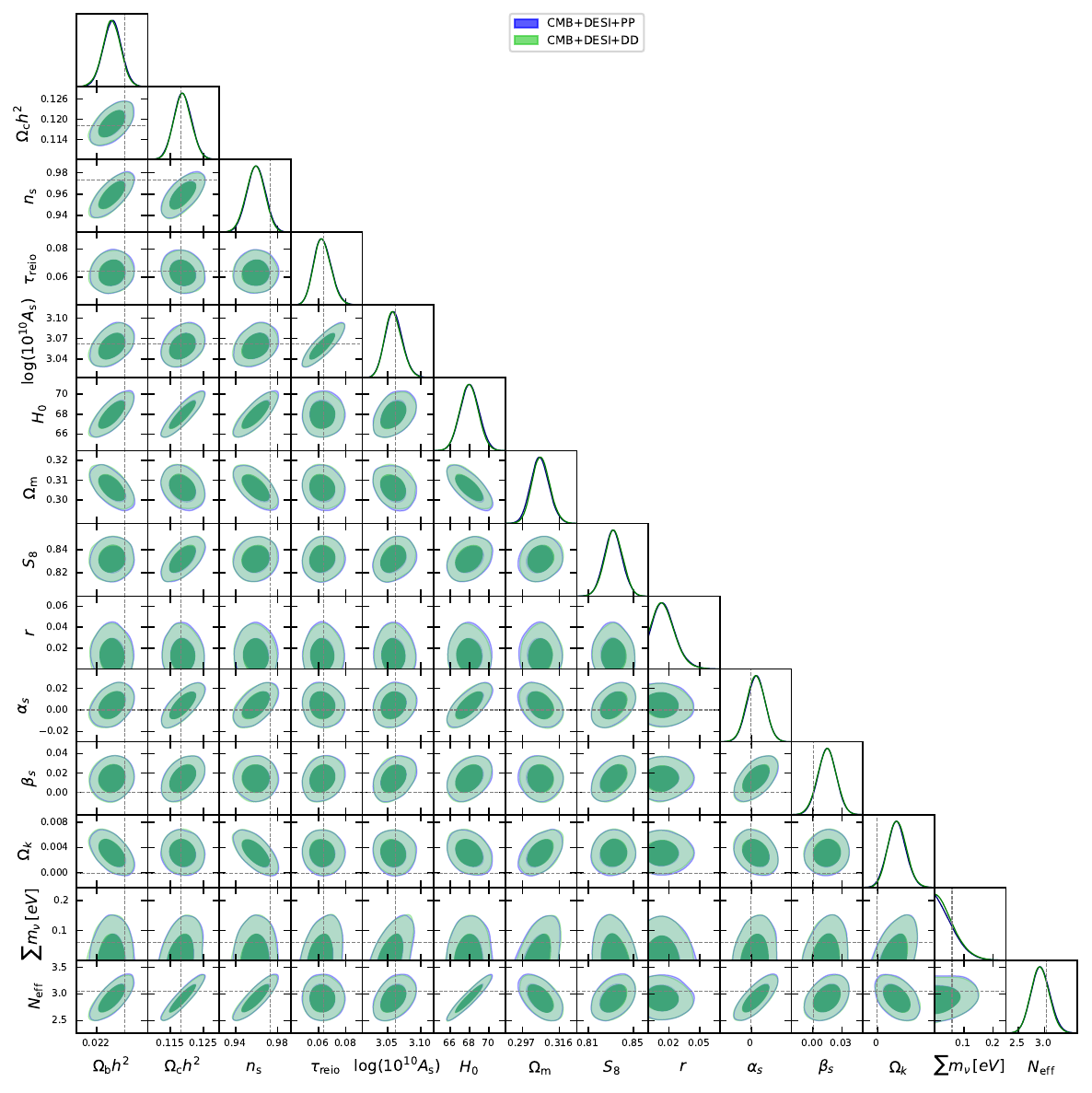}
    \caption{$\Lambda\mathrm{CDM}+r+\alpha_s+\beta_s+\Omega_k+\sum m_\nu+N_\mathrm{eff}$ parameter constraints.}
    \label{fig:triangle_LCDM_all}
\end{figure}

\begin{figure}[h]
    \centering
    \includegraphics[width=1\linewidth]{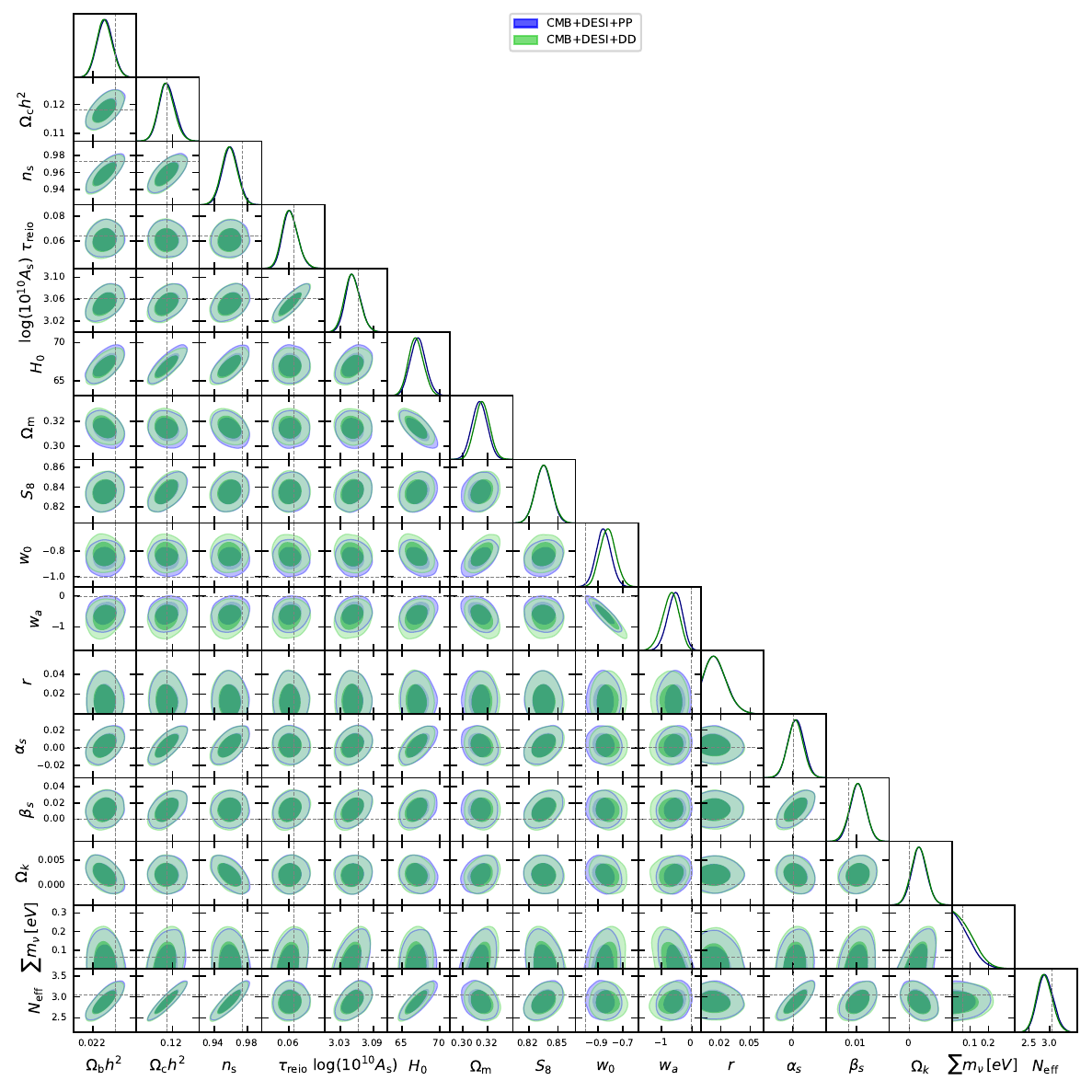}
    \caption{$w_0w_a\mathrm{CDM}+r+\alpha_s+\beta_s+\Omega_k+\sum m_\nu+N_\mathrm{eff}$ parameter constraints.}
    \label{fig:triangle_CPL_all}
\end{figure}

\clearpage


%
\bibliographystyle{JHEP}
\bibliography{biblio}

\end{document}